\documentclass{aa}

\usepackage{graphicx}
\usepackage{txfonts}
\usepackage{lscape}
\usepackage{subfigure}
\usepackage{natbib}
\begin{document}
\newcommand{\blue}{\textcolor{blue}}
\newcommand{\red}{\textcolor{red}}
\newcommand{\oii}{\textrm{[O\,{\sc ii}]}}
\newcommand{\oiii}{\textrm{[O\,{\sc iii}]}}
\newcommand{\nii}{\textrm{[N\,{\sc ii}]}}
\newcommand{\sii}{\textrm{[S\,{\sc ii}]}}
\newcommand{\hii}{\textrm{H\,{\sc ii}}}

\title{SDSS IV MaNGA: Deep observations of extra-planar, diffuse ionized gas around late-type galaxies from stacked  
IFU spectra \thanks{SDSS IV}}

\author{A. Jones
          \inst{1}
\and
          G. Kauffmann\inst{1}
\and
	  R. D'Souza\inst{1}
\and
	  D. Bizyaev\inst{2,3}
\and
	  D. Law\inst{4}
\and
	  L. Haffner\inst{5}
\and
          Y. Bah\'e\inst{1}
\and
          B. Andrews\inst{6}
\and
	  M. Bershady\inst{5}
\and
          J. Brownstein\inst{7}
\and
	  K. Bundy\inst{8}
\and
	  B. Cherinka\inst{9}
\and
	  A. Diamond-Stanic\inst{10}
\and
	  N. Drory\inst{11}
\and	
          R. A. Riffel\inst{12,13}
\and
	  S. F. S\'anchez\inst{14}
\and
          D. Thomas\inst{15}
\and
	  D. Wake\inst{16}
\and
	  R. Yan\inst{17}
\and
	  K. Zhang\inst{17}
          }

   \institute{Max-Planck Institute for Astrophysics, Karl-Schwarzschild-Str 1, 85748 Garching, Germany\\
		\email{ajones@mpa-garching.mpg.de}
   \and
    Apache Point Observatory, P.O. Box 59, Sunspot, NM 88349 
   \and
   Special Astrophysical Observatory of the Russian AS, Nizhnij Arkhyz, Russia
   \and
   Space Telescope Science Institute, 3700 San Martin Drive, Baltimore, MD 21218 USA 
   \and
   Department of Astronomy, University of Wisconsin-Madison, 475N. Charter St., Madison WI 53703, USA
   \and
    PITT PACC, Department of Physics and Astronomy, University of Pittsburgh, Pittsburgh, PA 15260, USA
  \and
  Department of Physics and Astronomy, University of Utah, 115 S. 1400 E., Salt Lake City, UT 84112, USA
  \and
   UCO/Lick Observatory, University of California, Santa Cruz, 1156 High St. Santa Cruz, CA 95064, USA
  \and
   Center for Astrophysical Sciences, Department of Physics and Astronomy, Johns Hopkins University, 3400 North Charles Street, Baltimore, MD 21218, USA
   \and
   Department of Physics and Astronomy, Bates College, Lewiston, ME 04240
   \and
   McDonald Observatory, The University of Texas at Austin, 1 University Station, Austin, TX 78712, USA 
   \and
   Departamento de F\'isica, CCNE, Universidade Federal de Santa Maria, 97105-900, Santa Maria, RS, Brazil
   \and
   Laborat\'orio Interinstitucinal de e-Astronomia-LInea, Rua Gal. Jos\'e Cristino 77, Rio de Janeiro, RJ - 20921-400, Brazil
    \and
   Instituto de Astronomia, Universidad Nacional Autonoma de Mexico, A.P. 70-264, 04510, Mexico, D.F., Mexico
  \and
  Institute of Cosmology \& Gravitation, University of Portsmouth, Dennis Sciama Building, Portsmouth, PO1 3FX, UK
   \and
   School of Physical Sciences, The Open University, Milton Keynes, MK1 15 UK
   \and
   Department of Physics and Astronomy, University of Kentucky, 505 Rose Street, Lexington, KY 40506, USA }
   
   \date{}

   \abstract{We have conducted a study of extra-planar diffuse ionized gas using the first year data from the MaNGA IFU survey.   We have stacked spectra from 49 edge-on, late-type galaxies as a function of distance from the midplane of the galaxy. With this technique we can detect the bright emission lines H$\alpha$, H$\beta$, \oii$\lambda\lambda$3726, 3729, \oiii$\lambda$5007, \nii$\lambda\lambda$6549, 6584, and \sii$\lambda\lambda$6717, 6731 out to about 4 kpc above the midplane.  With 16 galaxies we can extend this analysis out to about 9 kpc, i.e. a distance of $\sim 2$\;R$_e$, vertically from the midplane.  In the halo, the surface brightnesses of the \oii\ and H$\alpha$ emission lines are comparable, unlike in the disk where H$\alpha$ dominates.  When we split the sample by specific star formation rate, concentration index, and stellar mass, each subsample's emission line surface brightness profiles and ratios differ, indicating that extra-planar gas properties can vary.  The emission line surface brightnesses of the gas around high specific star formation rate galaxies are higher at all distances,  and the line ratios are closer to ratios characteristic of \hii\ regions compared with low specific star formation rate galaxies. The less concentrated and lower stellar mass samples exhibit line ratios that are more like \hii\ regions at larger distances than their more concentrated and higher stellar mass counterparts. The largest difference between different subsamples occurs when the galaxies are split by stellar mass. We additionally infer that gas far from the midplane in more massive galaxies has the highest temperatures and steepest radial temperature gradients based on their \nii/H$\alpha$ and \oii/H$\alpha$ ratios between the disk and the halo.}

\authorrunning{Jones}
\titlerunning{SDSS IV MaNGA: eDIG around late-type galaxies}
\maketitle

\section{Introduction}

Ionized gas in the outskirts of galaxies has been the subject of study for several decades. 
In the Milky Way (MW), a layer of diffuse ionized gas (DIG) a few kpc above the plane of the galaxy (sometimes also
referred to as a  warm ionized medium (WIM)  was discovered in the early 1970s 
and is commonly called the Reynolds Layer \citep{1971PhDT.........1R,1973ApJ...185..869R}. 
This layer contains most of the ionized gas in the MW, 
which is comparable  ($\sim$ 30\%) to the total  mass of neutral hydrogen
in the Galaxy \citep{1990IAUS..139..157R,1991IAUS..144...67R}. 
More recently, several wide-angle surveys of MW H$\alpha$ emission were combined by \citet{2003ApJS..146..407F} to produce an all-sky map of H$\alpha$ emission in the MW.  
H$\alpha$ emission is present practically everywhere, and it contains lots of 
structure, such as loops, filaments, and blobs.  Forbidden optical emission lines 
have also been studied. 
Similar to H$\alpha$, the luminosities of \sii\ and \nii\ vary spatially \citep[e.g.][]{2006ApJ...652..401M}.  
The ratios of  \sii\ and \nii\ with respect to H$\alpha$ are higher 
in the DIG compared to classic \hii\ regions, indicating that the properties of
the DIG differ from those in  \hii\ regions \citep[e.g.][]{2009RvMP...81..969H}.  

Diffuse ionized gas has also been detected around other galaxies.  Some of the first observations
of diffuse gas in external galaxies were those of NGC 891 by \citet{1990ApJ...352L...1R,1990A&A...232L..15D}.  \citet{1999ApJ...522..669H} compared 
narrow-band H$\alpha$ imaging of four nearby galaxies.  They found that the H$\alpha$ emission 
had substructure and that the H$\alpha$ luminosity varied from galaxy to galaxy in their sample,
with the more star-forming galaxies exhibiting stronger H$\alpha$ emission.  \citet{1994ApJ...426L..27L} found that DIG was common in starforming galaxies.  In face-on galaxies it was also seen that the DIG was correlated with \hii\ regions \citep[e.g.][]{2000A&A...363....9Z,2002A&A...386..801Z}.  
This work was followed up with a survey of 74 edge-on galaxies by \citet{2003A&A...406..493R,2003A&A...406..505R} with a goal to
understand how common H$\alpha$ halos are around galaxies, as well as how the
properties of these halos depend on star formation rate.  These authors found that H$\alpha$ could be detected
if the star formation rate per unit area was above a threshold of 
$3.2\pm0.5\times10^{40}\mathrm{erg\;s}^{-1}\;\mathrm{kpc}^{-2}$, with an estimated mean sensitivity for the galaxies observed with DFOSC at La Silla around $6\times10^{-18}$erg\;s$^{-1}$cm$^{-2}$arcsec$^{-2}$ \citep{1996ApJ...462..712R}. 

In addition to narrow-band imaging, there have also been spectroscopic studies 
of nearby galaxies \citep{1997ApJ...483..666G,2003ApJ...586..902H}.  
The detection of multiple emission lines allows better constraints on a possible additional heating 
source, as well as the physical conditions of the extra-planar diffuse ionized gas (eDIG), such as its density and temperature.   
Radiation from OB stars escaping from the disk has been traditionally
considered to be the main source of ionization \citep{2009RvMP...81..969H}, however in some cases there is a need for an additional heating source beyond photoionization.  
 \citet{2001ApJ...560..207O} studied the 
optical emission lines from \oii$\lambda$3727 to \sii$\lambda$6717 for three galaxies.  All three galaxies had an \oii/H$\alpha$ ratio
that increased with distance from the galaxy midplane.  Keeping the oxygen abundance constant 
while increasing the temperature as a function of radius yielded results in 
good agreement with the data. It is difficult to accommodate these results if only radiation from
OB stars is considered; OB stars are confined to the disk and 
heating effects are expected to drop as a function of distance from the disk plane (however, see \citet{2010ApJ...721.1397W,2014MNRAS.440.3027B} for a contrasting view).
Another possible additional heating source are hot, 
low-mass evolved stars (HOLMES) in the thick disk and stellar halo, which have been proposed to explain the 
emission line ratios in NGC 891 \citep{2011MNRAS.415.2182F}. 
Planetary nebulae and white dwarfs are hot and have a harder ionizing spectrum that could
explain the observed emission line ratios, but it is unclear whether the density of
such sources is high enough to maintain these temperatures at the required level.  Other possible additional sources
of heat in the diffuse halo gas  are shocks \citep[e.g.][]{2001ASPC..240..392C}, photoelectric heating \citep{1992ApJ...400L..33R}, 
turbulence \citep[e.g.][]{2009A&A...500..817B}, and magnetic reconnection \citep[e.g.][]{1999ApJ...525L..21R}.  

One of the main challenges in understanding the eDIG is to disentangle ionization from hot stars in the disk and other possible additional heating sources.  
This additional heating source may also depend on the galaxy mass or type,
as seen in a study of irregular galaxies which have higher \oiii/H$\beta$ and lower \nii/H$\alpha$ compared to spiral galaxies \citep{2006AJ....131.2078H}. 
Finally, it is not yet understood whether the eDIG is inflowing, outflowing or simply in pressure-supported equilibrium
around the galaxy. 
Inflowing gas from the circumgalactic medium (CGM) surrounding galaxies has been proposed 
to explain ongoing star formation in spiral galaxies like the MW \citep{1978MNRAS.183..341W}.  
Inflowing gas is thought to come from gas that is cooling from the surrounding
halo, or it may be recycled gas that is first ejected and then falls back onto the disk in the form of galactic fountains or chimneys \citep[][and references therein]{2012ARA&A..50..491P}.  
Outflowing gas can be from galactic winds produced in supernovae explosions or from
active galactic nuclei (AGN). Understanding the kinematics of the inflowing and outflowing gas around
galaxies is key to understanding the processes that regulate star formation in these systems.  
In a few studies \citep{2006ApJ...647.1018H,2006ApJ...636..181H,2007A&A...468..951K,2007ApJ...663..933H}, the outer regions of several galaxies, NGC 5775, NGC 891, and NGC 4302, 
were found to have negative velocity gradients with distance in the eDIG.  These gradients were inconsistent with the ballistic model of \citet{2002ApJ...578...98C} for a star formation driven disk-halo flow.

In a recent study, \citet{2016MNRAS.457.1257H} used the Sydney-AAO Multi-object Integral field spectrograph (SAMI) galaxy survey \citep{2015MNRAS.447.2857B} to study the eDIG of 40 galaxies.  They only considered galaxies with a clear H$\alpha$ detection in the outskirts, which limited their sample to galaxies with higher star formation rates.  They were interested in the connection between star formation rates and galactic winds and they found higher amounts of eDIG in galaxies with a recent star formation burst.  They also detected emission lines (H$\alpha$, H$\beta$, [OI], \oii, \oiii, \nii, and \sii) out to $\sim$10 kpc, however their galaxies are on average twice as large as the galaxies in our sample due to their higher median stellar mass.  In our study we focus on how the average properties of the eDIG vary as a function of height above the disk and for galaxies with a range of properties.

Large IFU spectroscopic surveys, such as the Mapping Nearby Galaxies at
Apache Point (MaNGA) survey, allow us to gain new and different perspectives on the eDIG. 
MaNGA will eventually obtain spectroscopic observations for 10\;000 nearby galaxies, with 1392 galaxies observed already in the first year.
As we will show, a small, but significant fraction of these galaxies are edge-on, late-type 
systems for which it is possible to study the eDIG out to between about 4 and 9 kpc 
above and below the galactic plane. By stacking together spectra from similar galaxies, we 
can clearly detect emission lines farther out into the halo than previously possible due to the increased signal to noise for an average late-type galaxy.  Previous very deep observations of NGC 891 and NGC 5775 have detections out to about 10\;kpc \citep{1997ApJ...474..129R,2000ApJ...537L..13R}.  We can study how the properties of the emission lines depend
on galaxy mass, morphological type, and specific star formation rate. In this paper, 
we present results from the first year of observations.  
In Section 2 we describe the method for selecting our galaxy sample,
our method for stacking the spectra, and our techniques for measuring emission line 
surface brightnesses and their associated errors. In Section 3, we present our main results for the full sample as well as 
for a sample where it is possible to measure the radial profiles of the H$\alpha$ and
\oii\ lines out to $\sim$\;9\;kpc. We also present results for several subsamples of galaxies which were
selected according to specific star formation rate, concentration index, and stellar mass. 
Finally, we conclude and summarize in Section 4.

For the rest of the paper, we refer to the ionized gas in the outskirts of galaxies as eDIG for simplicity, even though this ionized gas could also originate from outflows or shocks.  DIG refers to the diffuse ionized gas that is present anywhere in the galaxy.  We assume a Hubble constant $H_0=70$\;km\;s$^{-1}$Mpc$^{-1}$ throughout.

    
\section{Method}

In this section, we will first discuss the MaNGA survey and how we selected 
our galaxies (Section 2.1).  We also show that our sample is representative of the 
late-type galaxies currently observed by MaNGA.  We then discuss our stacking technique, including corrections and normalizations for each 
spectrum that is included into a stack, in Section 2.2.  We also demonstrate the improvement to the achieved S/N by stacking.  
Lastly, in Section 2.3, we discuss the spectral fitting of the stacked spectra for
extracting the emission line surface brightnesses. 

\subsection{Data Set}
 
We use the data from the SDSS-IV MaNGA survey \citep[see Blanton et al. 2016 for an overview of SDSS-IV and for MaNGA][]{2015ApJ...798....7B}.  MaNGA began operations in 2014 and is a six 
year IFU survey of nearby galaxies.
MaNGA uses the BOSS spectrographs, with a spectral range from 
3622 to 10354 $\AA$  and a resolution of R$\sim2000$ \citep{2013AJ....146...32S}.  For this work we only use the wavelength
range shortwards of 7000 $\AA$, because sky residuals from the OH skylines limit the achievable depths in the outer low surface brightness regions
of the galaxies at longer wavelengths (see Law et al. 2016).  

The observations were taken with the 2.5 meter Sloan Foundation 
telescope at Apache Point Observatory in New Mexico, USA \citep{2006AJ....131.2332G}. 
Each observation is a plate containing 1423 fibers that are bundled into 
different sized units \citep{2015AJ....149...77D}.  For science purposes, there are five 
127, two 91, four 61, four 37, and two 19 fiber bundles (Wake et al., in prep).  There are 
twelve 7 fiber bundles for observing spectrophotometric standard stars and 92 
sky fibers for the sky background subtraction \citep{2016AJ....151....8Y}.  For the analysis presented in this paper, 
we mostly use the larger science bundles, because they typically have more fibers 
that extend to larger angular radii.  The bundles are hexagonal in shape with the galaxy center at the origin (except in a few test cases which are not part of our sample).  The list of plates and IFU bundles 
for each galaxy in our sample is provided in Table \ref{gal_prop}. 
Note, that the IFU design number is the IFU bundle size followed by 
two digits signifying which IFU bundle of that size was used 
(so 12705 would be the fifth IFU bundle with 127 fibers).  

The survey observes 2/3 of the galaxies out to a minimum radius of 1.5 effective radii (R$_e$), known as the primary sample, and the remaining 1/3 out to 2.5 R$_e$ (secondary sample).  The observational strategy is given in \citet{2015AJ....150...19L} and the first year survey data is described in Yan et al. (2016).  This targeting strategy enables a more comprehensive study of the outskirts of galaxies 
compared to other IFU surveys that do not have such a requirement  
\citep[e.g. CALIFA;][]{2012A&A...538A...8S}.  Each plate is observed in three different dithering positions multiple times, which are then reconstructed into a datacube.  For our analysis we use the row stacked spectra instead of the data cube, which provides the spectra from each dithered position prior to the resampling.  The data reduction pipeline was improved since SDSS III with better sky subtraction and less systemic residual flux which allows the stacking of low surface brightness spectra of many individual fibers without becoming dominated by background noise (Law et al. 2016).

We choose to study edge-on, late type galaxies.  The constraint that the 
galaxies must be edge-on is valuable for two reasons.  First, there is a greater 
area of the IFU that is in the halo of the galaxy. 
This also means that we are not limited to using galaxies only from the secondary sample.  Second, studying
the emission perpendicular to the plane of the disk minimizes contamination from gas and stars within
the disk itself. We require the ratio of the 
semi-minor to semi-major axis (b/a) to be less than 0.3.  This cutoff is similar to the one in \citet{2016MNRAS.457.1257H} with a cutoff of b/a<0.26. 

In this study we are interested in studying late-type systems with as wide a range of 
stellar masses (M$_{star}$), star formation rates, and morphologies as possible. 
We define a late-type galaxy to have a concentration index 
C (defined as the ratio $R_{90}/R_{50}$, where
$R_{90}$ and $R_{50}$ are the radii enclosing 90\% and 50\% of the total $r$-band
light from the galaxy) less than 2.6 \citep{2001AJ....122.1238S}.  \citet{2001AJ....122.1238S} show that this is a robust way to separate late and early type galaxies even for highly inclined systems.  With this definition there are 81 galaxies with b/a<0.3 in the entire first year sample. 
Six of these galaxies have an effective radius (R$_e$) that appears to be wrongly measured or overly influenced by a bright bulge, 
one is a galaxy merger, 17 have another object (e.g. a star) in the field of view, and five show asymmetries along the disk, so they are discarded.
We also required that there must be fibers out to at least 4\;kpc above the disk midplane, 
which excludes another three galaxies.  This leaves us with a total sample of 49 galaxies.  We call this the full sample. 
Another sample, called the large-$z$ sample, consists of galaxies with fibers simultaneously 
out to at least 9 kpc and 2 R$_e$ along the minor axis. We relaxed the b/a restriction to b/a<0.4, which added four more galaxies.  We also required that b/a*R$_e$ be less than 2 kpc, which limits the apparent height of the disk, to ensure that the outer stacks are coming from the halo and not a mixture of disk and halo.  In total, there are 16 galaxies in this sample.  
As the survey continues, the number of galaxies suitable for this study will increase, 
allowing for a more detailed analysis of the extra-planar, diffuse ionized gas.

Table \ref{gal_prop} lists the galaxies used in this study, 
their right ascension (RA) and declination (dec), redshift, S\'{e}rsic effective radius (R$_e$), 
b/a, M$_{star}$, concentration C,  and specific star formation rate sSFR (defined as SFR/M$_{star}$). 
All of these values, except  sSFR are taken directly from the 
NASA-Sloan Atlas\footnote{http://www.nsatlas.org} (NSA) catalog.  
The sSFR is calculated using the luminosity of H$\alpha$
enclosed within the central bin (see Sec 2.2), taking the area of the galaxy as the area of the ellipse with semi-major axis equal to 1\;R$_e$ to convert from surface brightness to luminosity.  
The H$\alpha$ luminosity is dust-corrected using a Calzetti extinction law and assuming
a Case-B recombination rate value H$\alpha$/H$\beta=2.83$ in the absence of dust \citep{2001PASP..113.1449C}. 
Using the conversion SFR=L(H$\alpha$)$*7.9\times10^{-42}\mathrm{M}_\odot\mathrm{yr}^{-1}$ from \citet{1998ARA&A..36..189K}, we arrive at a rough estimate for the sSFR for each galaxy.  The sSFR
for the galaxies in our sample ranges from 7.85$\times10^{-11}$ to 
5.21$\times10^{-9}$yr$^{-1}$, with a median of 5.81$\times10^{-10}$yr$^{-1}$.
In the table we have also marked the galaxies which are part of the large-$z$ sample, including the additional four galaxies with 0.3<b/a<0.4 below the line.

\begin{figure*}[!ht]
   \centering   
  \includegraphics[width=0.8\textwidth]{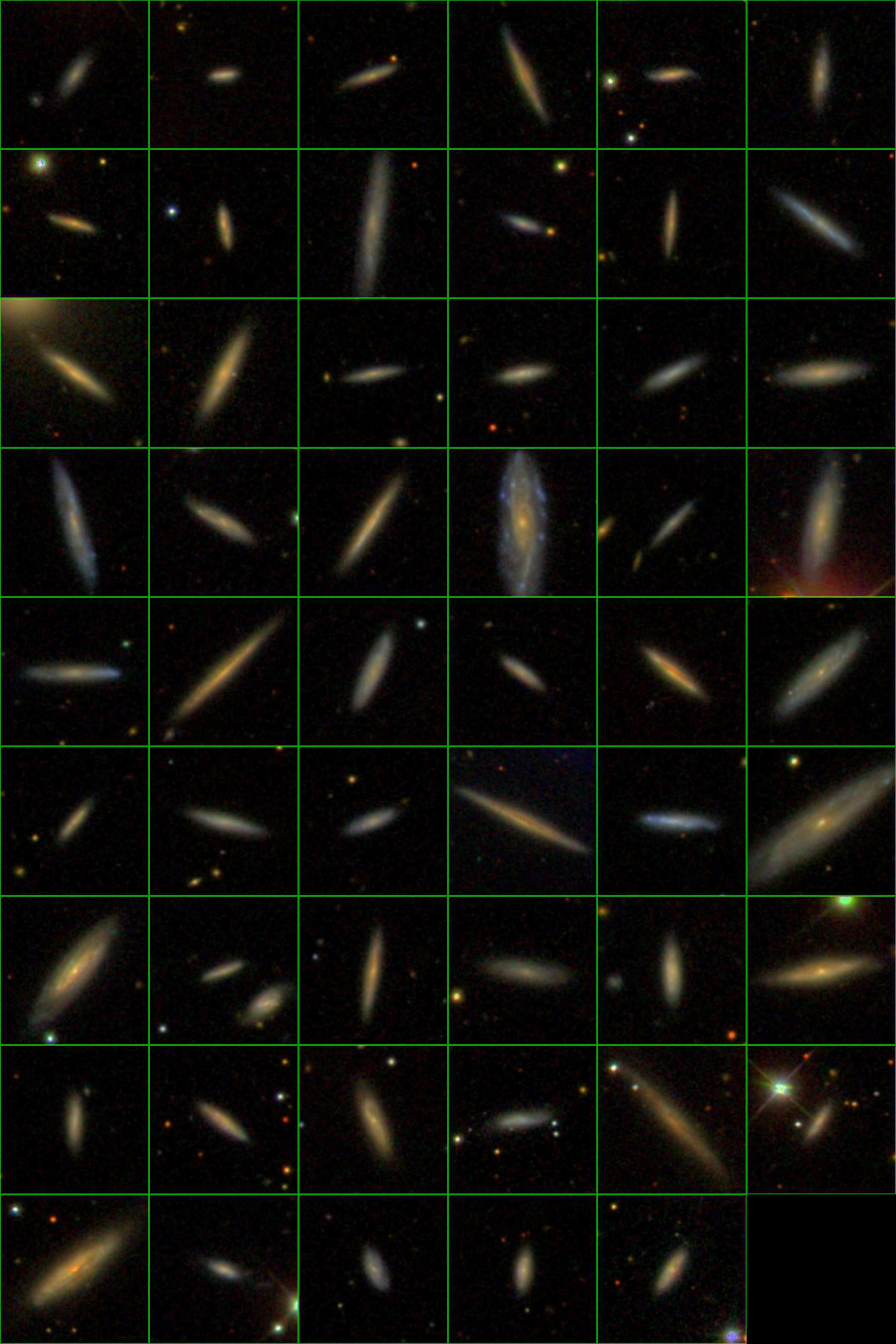}
  \caption{SDSS images of all the galaxies used in the analysis.  The images are in the same order as Table \ref{gal_prop} (left to right, then top to bottom).  Each box is 60 x 60 arcsecs with the centers corresponding to the coordinates given in Table \ref{gal_prop}.  The four additional galaxies for the large-$z$ sample are also included as the last four galaxies.}
\label{mosaic}
    \end{figure*}

A mosaic of our sample (49 plus four galaxies for the large-$z$ sample) is shown in Fig \ref{mosaic}. As can be seen, the majority of our galaxies are not truly edge-on, defined as showing no visible spiral structure and a clear dust lane.  We chose a b/a < 0.3 to ensure a decent sample size.  A discussion of the effect of different inclinations is in Appendix A.  Despite some detailed differences, our qualitative conclusions are not affected by our inclusion of not truly edge-on galaxies.  One of the galaxies, 8257-12705 (fourth column, fourth row of Fig \ref{mosaic}) shows extended emission, including extended H$\alpha$ emission in the outskirts.  We have made stacks excluding this galaxy and the change is negligible.  The additional galaxies for the large-$z$ sample (last four galaxies in Fig \ref{mosaic}) appear smaller compared to the rest of the galaxies.  These galaxies are at a higher redshift, between 0.039 and 0.043, compared to the average redshift (z=0.03) of the rest of the sample.  This is most likely due to the constraint on the width of the minor axis for the large-$z$ galaxies. 

\begin{figure*}[!ht]
   \centering   
   \includegraphics[width=0.49\textwidth]{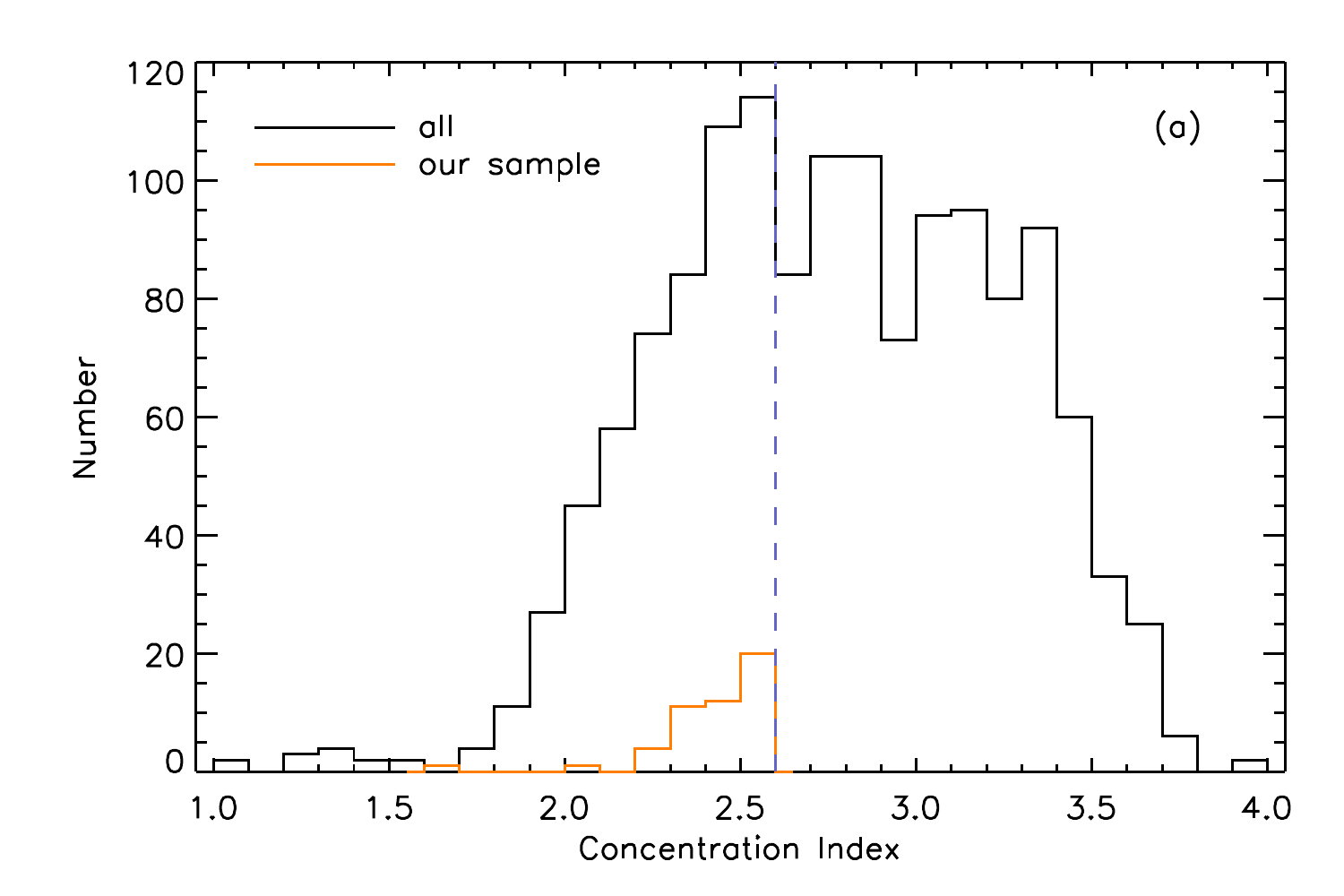}
\includegraphics[width=0.49\textwidth]{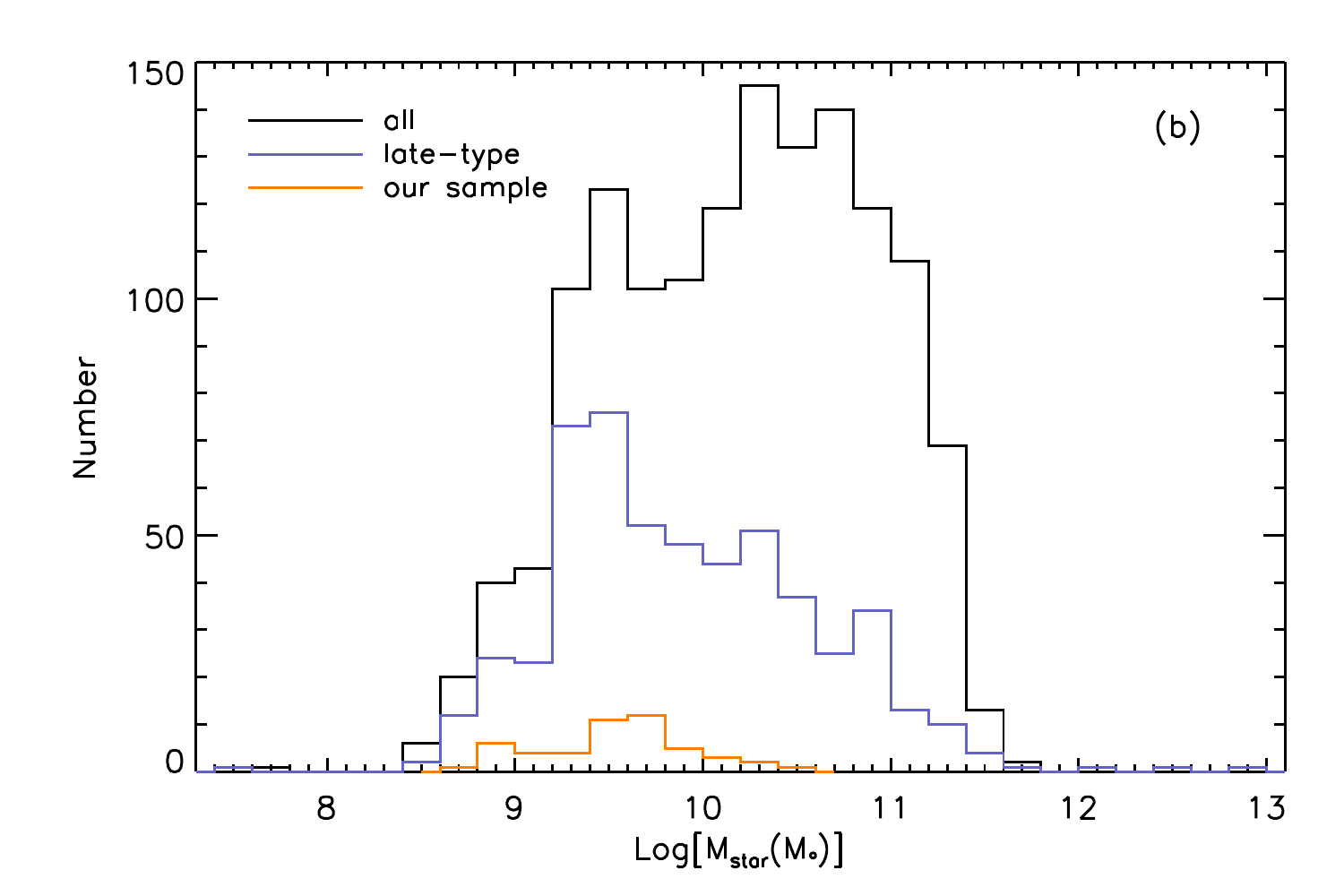}
   \includegraphics[width=0.49\textwidth]{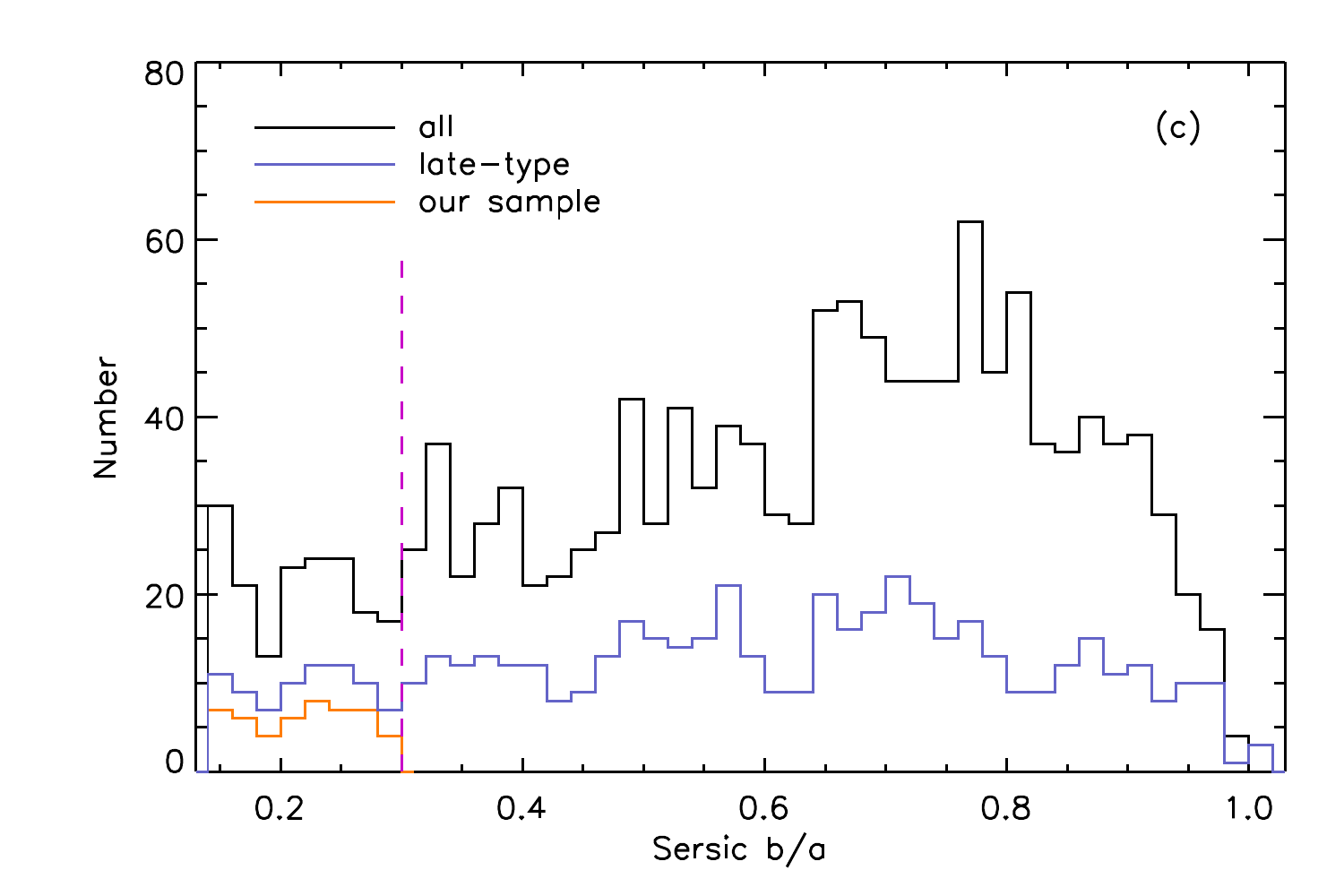}
\includegraphics[width=0.49\textwidth]{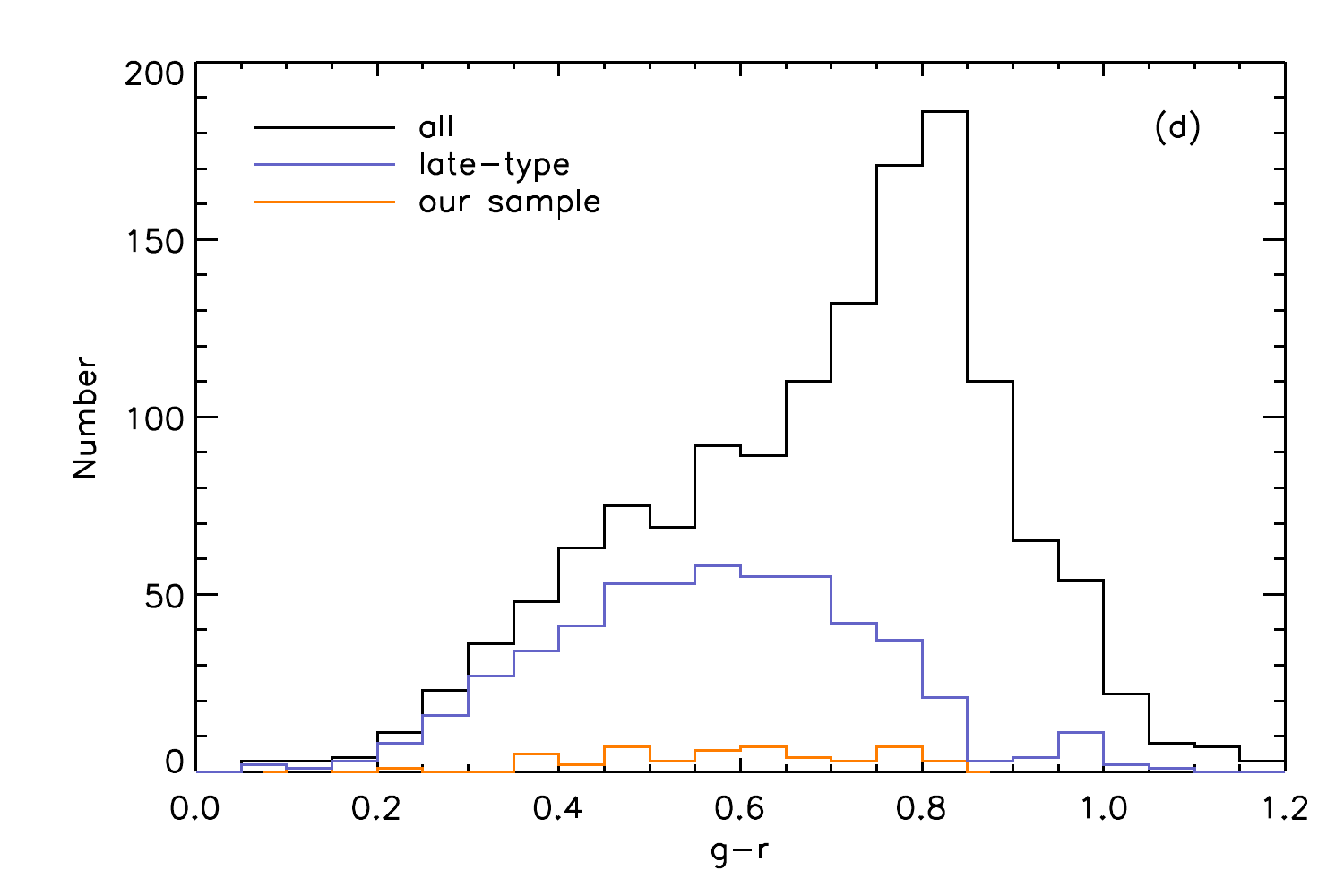}
   \caption{Histograms showing the distributions of the MaNGA first year sample (black) and our sample of galaxies (orange) for various galactic properties.  Panel \textit{a} shows the distribution of concentration index C with the vertical dashed purple line showing the division between late-type and early-type galaxies at C=2.6.  Panel \textit{b} gives the M$_{star}$ for the full sample (black), late-type galaxies (blue) defined by C<2.6, and our sample (orange).  
Panel \textit{c} shows the inclination distribution (given by b/a) for the same sets of galaxies with the vertical dashed line at our cutoff b/a=0.3. Panel \textit{d} shows the distribution of g-r colors.}
\label{manga_sample}
    \end{figure*}

In Fig \ref{manga_sample}, we show distributions of a variety of properties for the MaNGA 1st year, fourth MaNGA Product Launch MPL-4 (i.e. the fourth product launch as an internal
release) parent sample, as well as for our sample of 49 galaxies. 
Figure \ref{manga_sample}\;\textit{a} shows the Concentration index C with the vertical line 
showing our adopted separation between late-type and early-type galaxies at C=2.60.  In our sample, C ranges from 1.68 to 2.60 with a median value of 2.44.   
Panel \textit{b} shows the distribution of M$_{star}$.  Late-type galaxies in the parent sample (blue histogram) 
tend to have smaller M$_{star}$ compared to the early-types as expected.  Our sample roughly follows the mass distribution 
of the late-type galaxies of the MaNGA parent sample, with a minimum, maximum and median mass of $5.36\times10^8$, $2.79\times10^{10}$, and $3.73\times10^{9}$\;M$_\odot$, respectively.  Figure \ref{manga_sample}\;\textit{c} 
shows the distribution of inclination (approximated by b/a) for the three samples.  
Since we selected our sample to have b/a<0.3 (vertical dashed line), our sample lies on the extreme low end, with a median b/a=0.22, 
of the distributions for both the parent sample and the late-type galaxies 
of the parent sample.  In panel \textit{d} we plot the g-r color of 
the full parent sample, the late-type galaxy subsample, and our sample. 
The g and r magnitudes are derived from the Petrosian fluxes provided in the 
NSA catalog.  The late-type galaxies have bluer colors, as expected, compared to the full 
parent sample and our sample roughly follows the distribution of the late-type galaxies.  In addition (not shown), our sample follows the redshift distribution of the MaNGA sample with an average redshift of 0.03.
In summary, we conclude that our sample of 49 edge-on galaxies is a fair 
representation of the late-type galaxies observed in the first year of the MaNGA survey.

\subsection{Stacking Procedure}

We are interested in studying the average properties of the extra-planar diffuse ionized gas as a function of distance from the galactic plane.  Therefore, we need to create a set of stacks (with a narrow range in distance) along the minor axis from multiple galaxies.  
Since each galaxy is a different size and at a different redshift, we must normalize and scale the distances from the midplane for each galaxy.  We have done this in three different ways: by the minor axis effective radius (Method 1), major axis effective radius (Method 2), and by the physical distance from the midplane (Method 3).  In all cases we define $z$ as the distance along the minor axis.  The S\'{e}rsic effective radius R$_e$ and b/a are taken from the NSA catalog.

For Method 1, scaling each galaxy by its minor axis R$_e$, we define the minor axis R$_e$ ($b_e$) as $b_e=\mathrm{R}_e*b/a$, where b/a is the ratio of the minor to major axis given in the NSA catalog.  We are using the b/a ratio as an approximation for inclination. The less inclined galaxies tend to have a larger $b_e$ compared to the more inclined galaxies.  This allows one to probe the disk-halo boundary across galaxies which have small differences in inclination. In Method 2 we scale by the \textit{major axis} R$_e$ along the vertical height which takes into account that normally larger galaxies reside in larger halos.  The last method, by physical distance, assumes that the gaseous halo should behave similarly at similar physical heights above/below the midplane.  Since all three methods are useful for probing different information about the eDIG and the gas across the disk-halo boundary, we consider all three methods in this paper. Our approach follows that of \citet{2015ApJ...807...11G} who analyzed trends in stellar halo properties of early-type galaxies both as a function of physical radius and radius scaled by R$_e$.

For each galaxy, we use the row stacked spectra (RSS) output from the MaNGA data release pipeline (DRP), 
which contains all the individual spectra from the different observations that go into the final cube.  
Each RSS file contains the flux, wavelength, error, mask, and fiber position for all 
the spectra.  We prefer using the RSS instead of the cube files because the spectra have 
not been resampled, thus each spectrum is still independent.  This allows 
us to calculate errors for line fluxes in the stacked spectra more easily.

We stack fibers into different bins along the minor axis.  If we define $z$ as the distance along the minor axis and $x$ as the distance along the major axis (perpendicular to $z$), 
then for each bin, fibers that have $z$ between  $d_{min}$ and  $d_{max}$,
and $x$ within 75\% of the major axis R$_e$ are stacked.

\begin{equation}
\begin{split}
d_{min}<&|z|<d_{max}\\
&|x|<0.75*\mathrm{R}_e.
\end{split}
\label{fib_loc}
\end{equation}

The parameter $d$ is in units of $b_e$ for Method 1, R$_e$ for Method 2, and kpc for Method 3. 
For each minor axis bin, we consider fibers both above and below the disk.  We use the \texttt{xpos} 
and \texttt{ypos} fits extensions in the RSS file at the wavelength of 5000 $\AA$ to 
determine the location of each fiber.  The position of the fiber can shift as a function of wavelength due to differential atmospheric refraction (DAR).  Since each galaxy can have a different size, 
shape and IFU bundle size, the number of fibers that are located in a given minor 
axis bin varies from galaxy to galaxy.  However, we have selected our sample 
so that every galaxy contributes some fibers to each bin.

The spectra from each fiber were first 
corrected using the masks and error information provided in the RSS files.  Any wavelength pixel that was flagged 
to be masked was considered as a bad pixel and the flux was interpolated
over it.  If a given fiber had less than 100 good pixels, the fiber was removed from the stack, which occurred in $\lesssim0.5$\% of the fibers.  
The wavelength regions dominated by sky line residuals were also removed and interpolated over.  
Since we are considering the spectral range from 3600 to 7000 $\AA$, there are only 
three dominant sky lines at 5578.5, 5894.6, and 6301.7 $\AA$.  The flux from each spectrum was 
converted into surface brightness using the redshift for each galaxy.
   
Before stacking, the spectra were also corrected for galactic rotation.  Since 
most of the individual spectra have very low S/N, we could not measure 
directly the velocity shift from the emission lines.  Instead, we derive the 
rotation curve from the H$\alpha$ emission line measured along the major axis 
\citep{2010ApJ...720.1126M}.  We fit a cubic function 
to $V_c(R)$, where $R$ is the distance along the major axis and we assume that the rotation curve does not 
vary as a function distance from the galactic midplane.  This velocity 
rotation correction made only a slight difference to our measurements of the widths 
of the stacked emission lines. As noted in Section 3, some of the broadening of the 
emission lines seen in the outer minor axis bins may be attributed to inaccuracies 
in this procedure and the assumption that the velocity is constant with distance along the minor axis.

Once all the spectra for the different bins were corrected, shifted to rest frame 
and put on the same wavelength grid, they were stacked by the following procedure.  At each wavelength pixel, we took the mean surface brightness of all the fibers excluding the lowest and highest 10\% \citep[e.g.][]{2014MNRAS.443.1433D},
with the errors for each spectrum propagated appropriately to a stacked error.  We chose a clipped mean instead of a weighted mean to better exclude outliers and be less biased by extreme regions (like an off axis \hii\ region or an unseen satellite) or fibers with extreme negative or positive values.  Since we are interested in DIG in the outskirts and this should be a relatively low surface brightness, diffuse feature, taking a clipped mean over just taking a mean was less biased and closer to the median averaged stacked spectrum, with a difference between the clipped mean and median of $\lesssim$10\%.  We chose the clipped mean over a median to have a better handle on propagating the errors to have a stacked error spectrum as well as the stacked surface brightness spectrum.  Regardless of whether we took the median, mean, or clipped mean, the general results and trends were the same, only the value of the surface brightnesses changed.
We did not perform any additional sigma clipping because most of the spectra had very 
low S/N, usually $\lesssim 0.5$ and so this would greatly limit the number of fibers to stack. 
    
\begin{figure}[!ht]
   \centering   
   \includegraphics[width=0.49\textwidth]{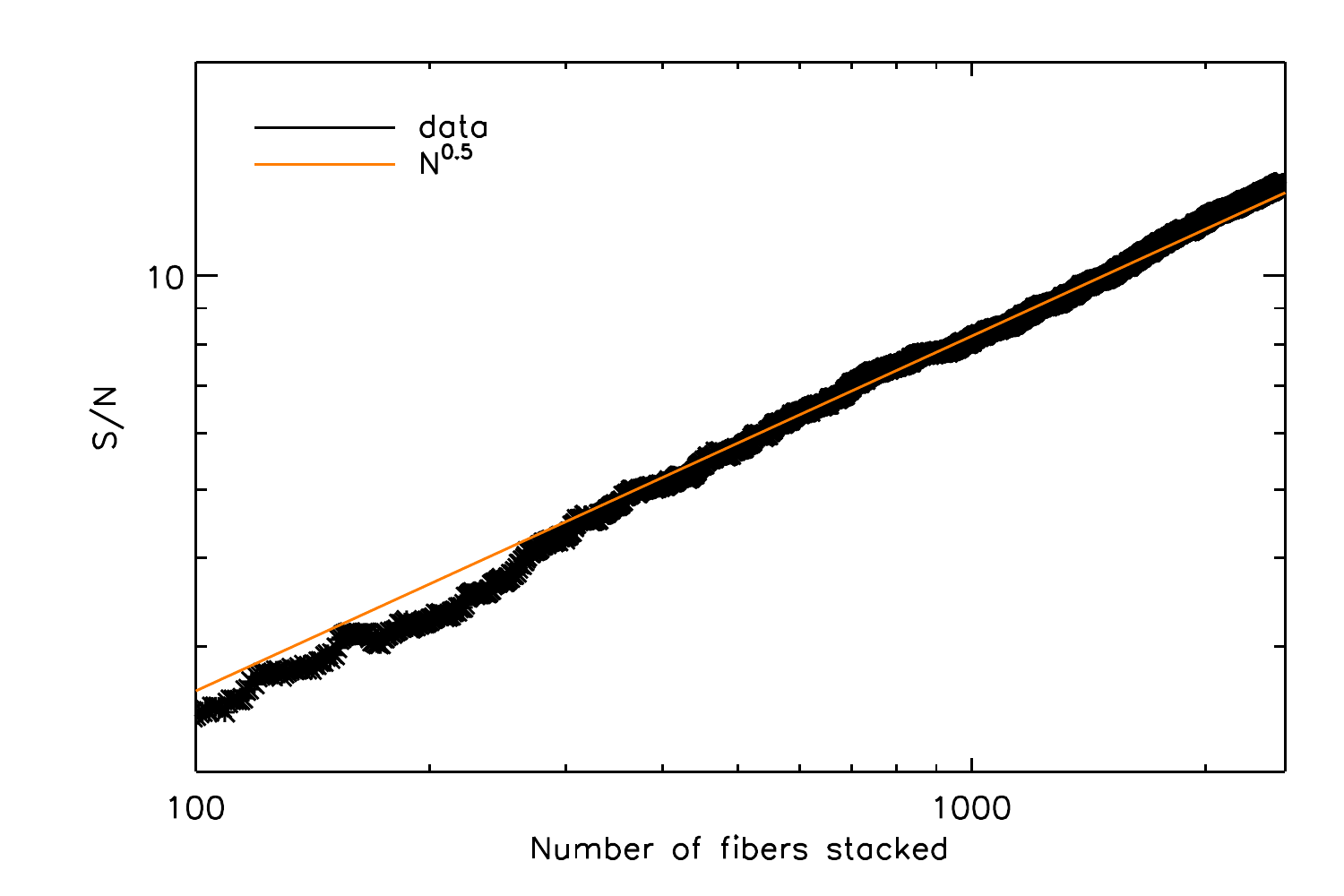}
\includegraphics[width=0.49\textwidth]{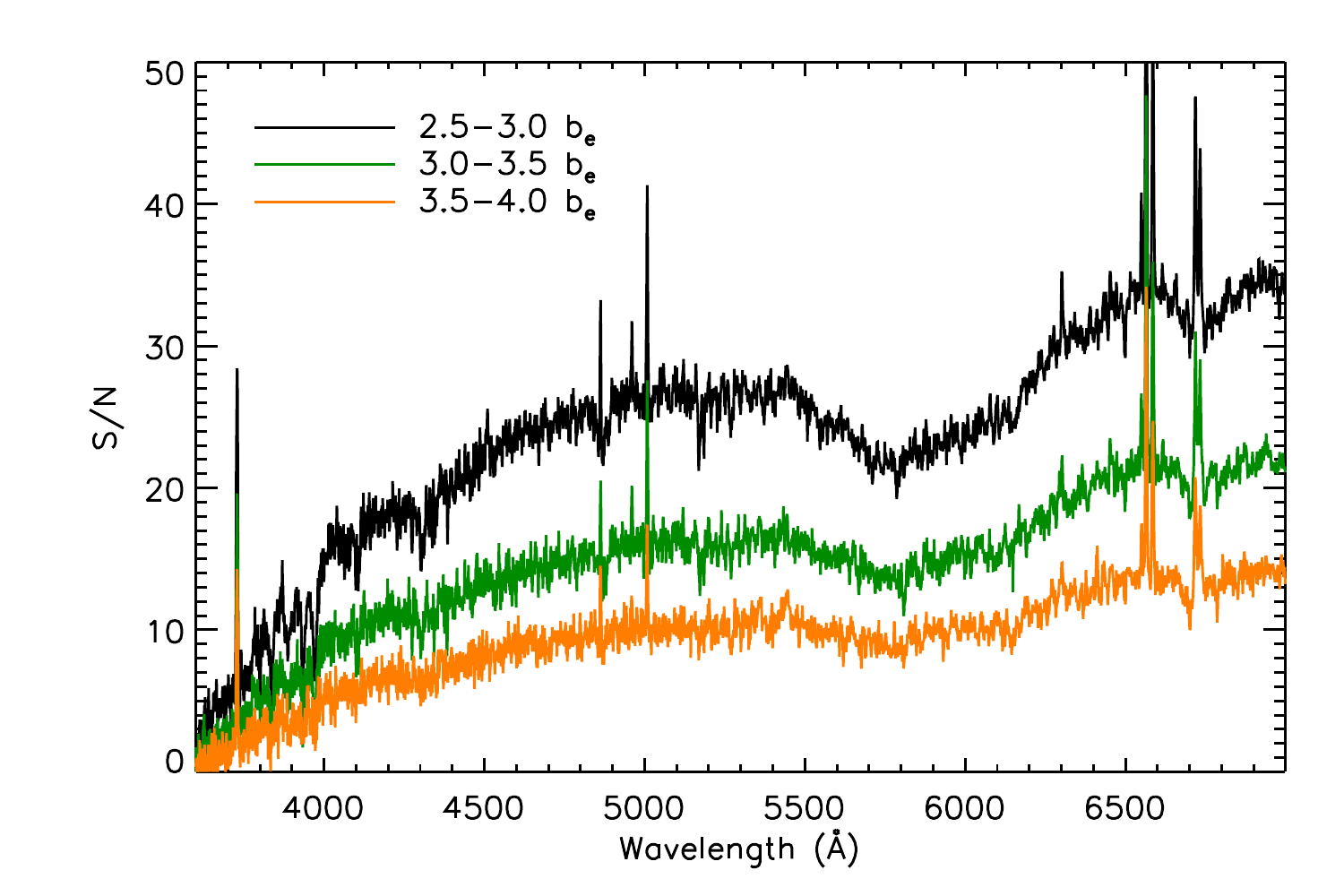}
   \caption{\textit{Top}: S/N ratio in the blue (4000-5500 $\AA$) as a function of the number of fibers stacked.
   The S/N ratio scales close to the theoretical expectation of $N^{0.5}$ overplotted in orange. 
   This demonstrates that we are not currently limited by the background or systematics in this wavelength range.
   \textit{Bottom}: S/N ratio for three different $b_e$ bins as a function of wavelength for the full sample. The S/N is higher for the emission lines.
   There is a dip $\sim 5700\;\AA$ around where the two spectrographs are joined and an increase in background noise from high pressure sodium streetlamps.}
\label{snr}
    \end{figure}

    In Fig \ref{snr}, the top panel shows that the S/N in the blue part of the spectrum (4000-5500 $\AA$) 
increases roughly as the square root of the number of fibers stacked, showing that we are not
limited by residual sky background subtraction errors in this wavelength range. 
The example shown in this figure is from the 3.0-3.5 $b_e$ bin of the full stack of 49
galaxies using Method 1.  The other minor axis bins and methods give similar results. 
Note that for the first few hundred fibers, the trend is not as smooth because 
each fiber was chosen randomly and each galaxy has a different surface brightness 
profile and therefore different average S/N ratio.  
The bottom panel of Fig \ref{snr} shows the S/N ratio as a function of 
wavelength (in rest frame) for the stacked spectrum in three different minor axis bins. There is a dip in the S/N around 5700\;$\AA$ which could be from where the two spectrographs are joined, between 5900-6300\;$\AA$ in observer wavelength frame, and/or by high pressure sodium from streetlamps which is a broad feature around 5900\;$\AA$ in the observers frame causing an increase in the background noise. The S/N also decreases shortwards of 4500\;$\AA$ where the throughput
of the spectrographs is lower.  As expected, the average S/N shows spikes where there are strong emission lines. 
 
\begin{figure*}[!ht]
   \centering   
   \includegraphics[width=0.89\textwidth]{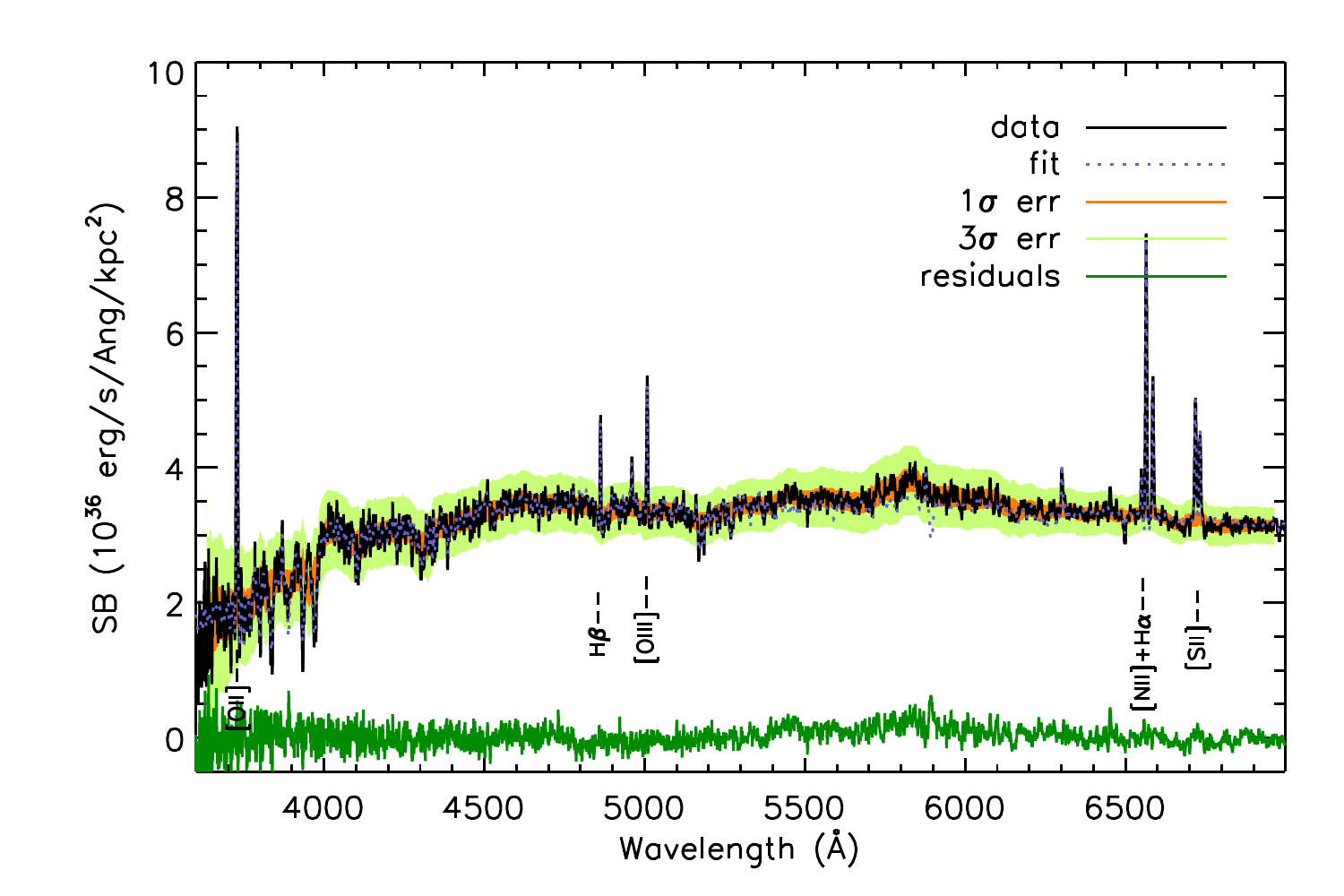}
   \caption{Example of the spectrum (in rest frame) obtained by stacking fibers from the full sample between 2.5 and 3.0 $b_e$ (black solid).  
The 1 and 3 $\sigma$ errors on the continuum are shown in the orange and yellow bands. 
The best-fit model spectrum (blue dotted) is overplotted along with the residuals from the fit (green). The bright emission lines H$\alpha$, H$\beta$, \oii, \oiii, \nii, and \sii, discussed in this paper, are labeled.}
\label{eg_spectra}
    \end{figure*}

The total number of galaxies, number of fibers, surface brightness estimate of the continuum in the blue, S/N in the blue, and the S/N near H$\alpha$ 
for the outer minor axis bins are provided in Table \ref{bin_info} for
each stack with $b_e > 2$, and $z>4$\;kpc for the large-$z$ sample.  We provide results from Method 3 for the large-$z$ sample instead of Method 1.
For the full sample with 49 galaxies, 
the number of fibers stacked ranges from 2981 in the inner bins to 2281 in the outermost bins and for the large-$z$ sample it ranges between 331 and 1871 fibers.  The outermost bins for the large-$z$ sample are larger because more fibers are needed in order to increase the S/N to a high enough level to be able to detect emission lines. 
The continuum S/N in the blue drops from 40 in the disk to $\sim5$ in 
the outer regions, while the S/N of H$\alpha$ drops from 72 to 8 for the samples with only 24 galaxies.  
The S/N of the emission lines are always about twice that of the continuum.  For the large-$z$ bins the S/N in the continuum can be below one and for H$\alpha$ the S/N is around four.  This demonstrates the need for stacking many galaxies to detect the eDIG at >5\;kpc (or >1.5\;R$_e$) from the galactic plane.   
An example of a stacked spectrum between 2.5 and 3.0 $b_e$ is shown in Fig \ref{eg_spectra}. 
The 1 and 3 $\sigma$ error contours are also plotted. As can be seen the main
strong emission lines H$\alpha$, H$\beta$, \oii$\lambda\lambda$3726, 3729, \oiii$\lambda$5007, \nii$\lambda\lambda$6549, 6584, and \sii$\lambda\lambda$6717, 6731
are easily detected at greater than 3 $\sigma$ confidence.  A small bump in the spectrum is visible near the
region where the two spectrographs are joined.

\begin{figure*}[!ht]
   \centering     
\includegraphics[width=0.89\textwidth]{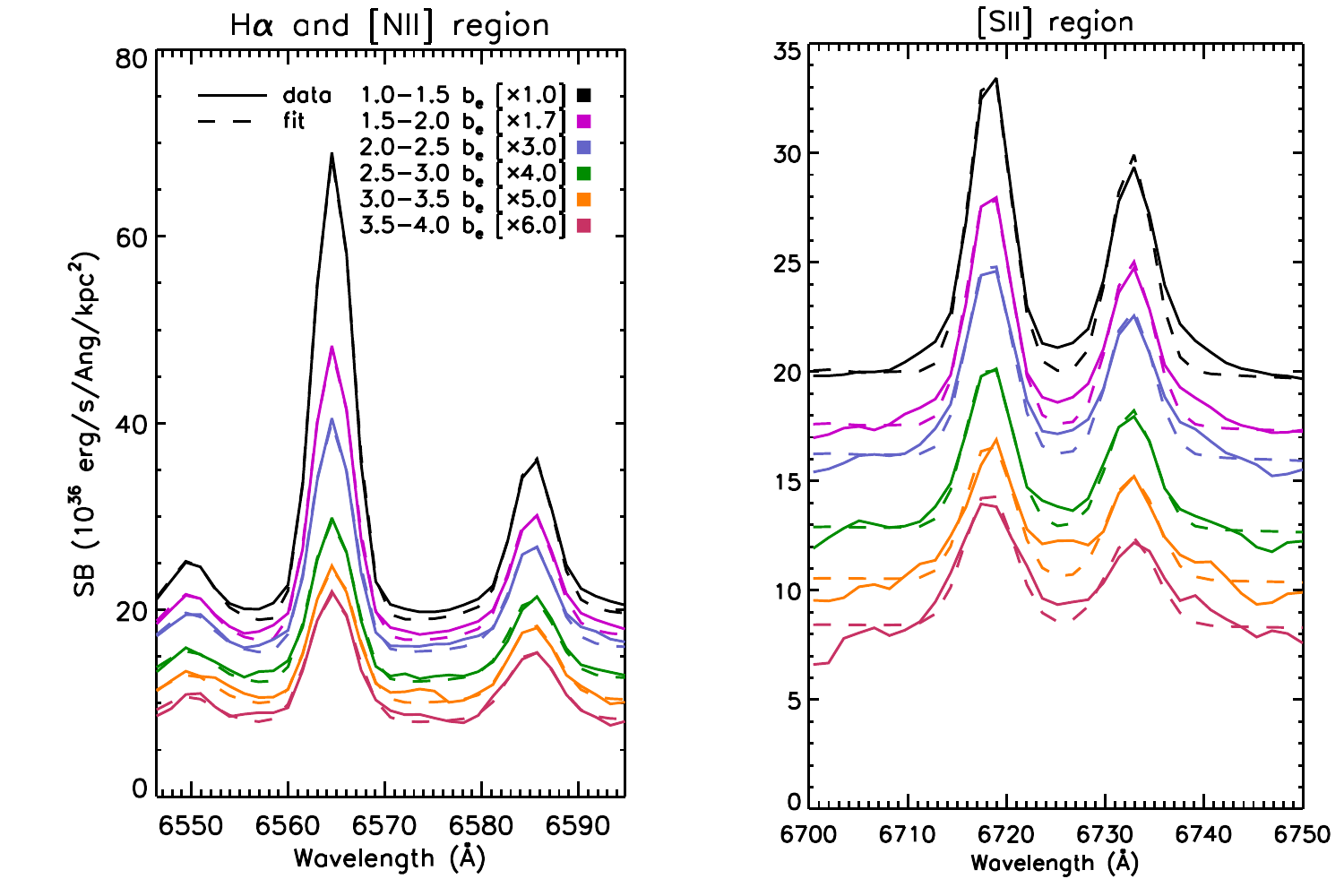}
   \caption{Zoom-in on the emission lines for the full sample for various 
$b_e$ bins using Method 1.  The data is shown as the solid line and the emission
line fits are the dashed lines, with the different colors corresponding to different  $b_e$ bins. 
The spectra have been scaled in the vertical direction for clarity, with the amount shown in the legend.}
\label{zoom_36g}
    \end{figure*}

\subsection{Spectral Fitting}

For finding the emission line surface brightnesses, we used a version of the MPA-JHU spectral fitting code 
\citep{2004ApJ...613..898T} that has been  
modified by C.Tremonti for the MaNGA spectra.  
The MPA-JHU spectral fitting method uses the models from 
\citet{2003MNRAS.344.1000B}, with three metalicities and 11 different ages to fit the stellar continuum 
with the emission lines 
masked, and then fits the emission lines as Gaussians.
Line surface brightnesses and widths with errors are output by the code. 
We note that the subsolar metalicity models consistently had a lower $\chi^2$; 
this is not unexpected since almost
all the galaxies in our sample have stellar masses less than $10^{10} \mathrm{M}_{\odot}$, below
the "knee" in the stellar mass-metallicity relation.

As shown in Fig \ref{eg_spectra} with the blue dotted line, the best-fit model spectrum agrees well with the data  (the residuals (data-fit) shown in green
in the plot are always near zero).  Figure \ref{zoom_36g} presents a closer look at some of the main
emission lines. We plot the wavelength regions around H$\alpha$ and \nii, as well as around \sii, 
for the minor axis bins from 0 to 4 $b_e$.  
Even though the S/N decreases significantly in the outermost bins, the emission lines are
well fit by simple Gaussians and the emission lines are clearly detected.

\section{Results}
 
We present our results of the surface brightnesses of the bright optical emission lines from the stacked spectra.  First, we discuss the results from the full sample of 49 galaxies in Section 3.1. 
For the full sample, the minor axis bins for all three methods are given in Table \ref{em_values}.  We have bins for the central areas of the galaxies
within the main disk to aid in understanding how the halo gas differs from interstellar medium
gas.  We checked the central bins for active galactic nucleus (AGN) contamination by stacking from 0.0-0.2 and 0.2-1.0 $b_e$, 
and there was no noticeable difference in the line ratios.  In 0.0-0.2 $b_e$ stacks, we found three galaxies that are consistent with having an AGN.
To extend farther into the halo, the large-$z$ sample contains 16 galaxies that have fibers 
out to at least 9\;kpc and we discuss these results in Section 3.2.  The minor axis bins are the same as the full sample with additional bins for Method 2 between 1.0-1.2, 1.2-1.4, 1.4-1.7, 1.6-1.9, and 1.7-2.0\;R$_e$, and for Method 3 between 4.0-4.5, 4.5-5.0, 5.0-6.0, 6.0-7.0, 7.0-9.0, and 7.0-10.0\;kpc.  We do not include the analysis done with Method 1 for the large-$z$ stack, but the results from Method 1 are similar to Method 2 for the large-$z$ sample.
Then in Section 3.3 we present subsamples, where we split the full sample into two halves, 
by sSFR (at $5.8\times10^{-10}\mathrm{yr}^{-1}$), by concentration index (at 2.44), and
by stellar mass (at $3.73\times10^9\mathrm{M}_\odot$) with the same minor axis 
bins as the full sample.  As shown in Table \ref{bin_info}, the S/N in the H$\alpha$ region $\gtrsim 3$ 
in all bins with detections.  For the large-$z$ sample, many of the
other emission lines are no longer detected in the outer bins, however \oii\ and H$\alpha$ can still be detected in most of them.
A full discussion of the detection limits is given in Sec 3.2.  
Table \ref{em_values} gives the surface brightness values for the bright emission lines at each minor axis bin for the three stacking methods for the full sample and subsamples split by various galactic properties.

\subsection{Full Sample}

 \begin{figure}[!ht]
   \centering    
\includegraphics[width=0.49\textwidth]{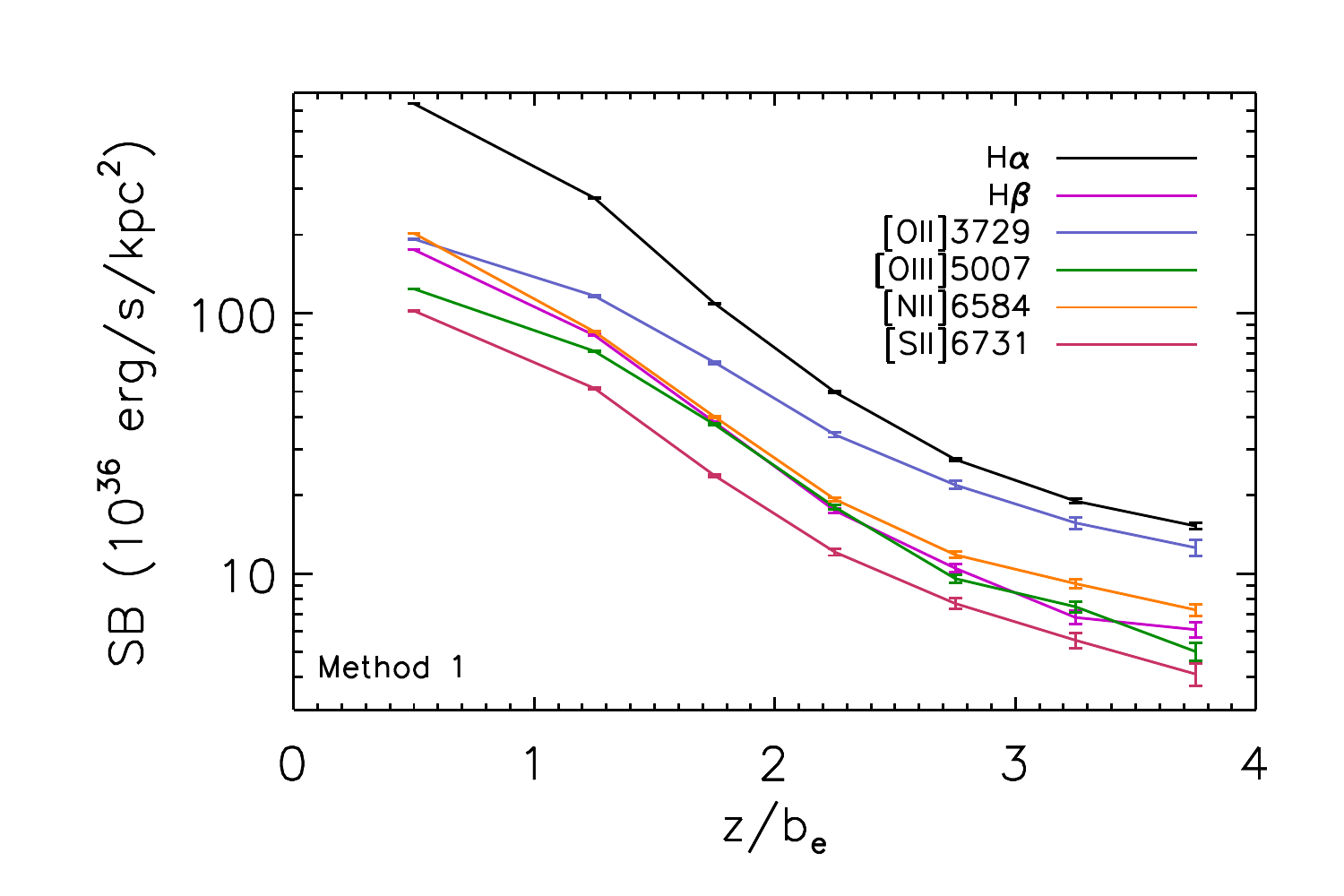}
\includegraphics[width=0.49\textwidth]{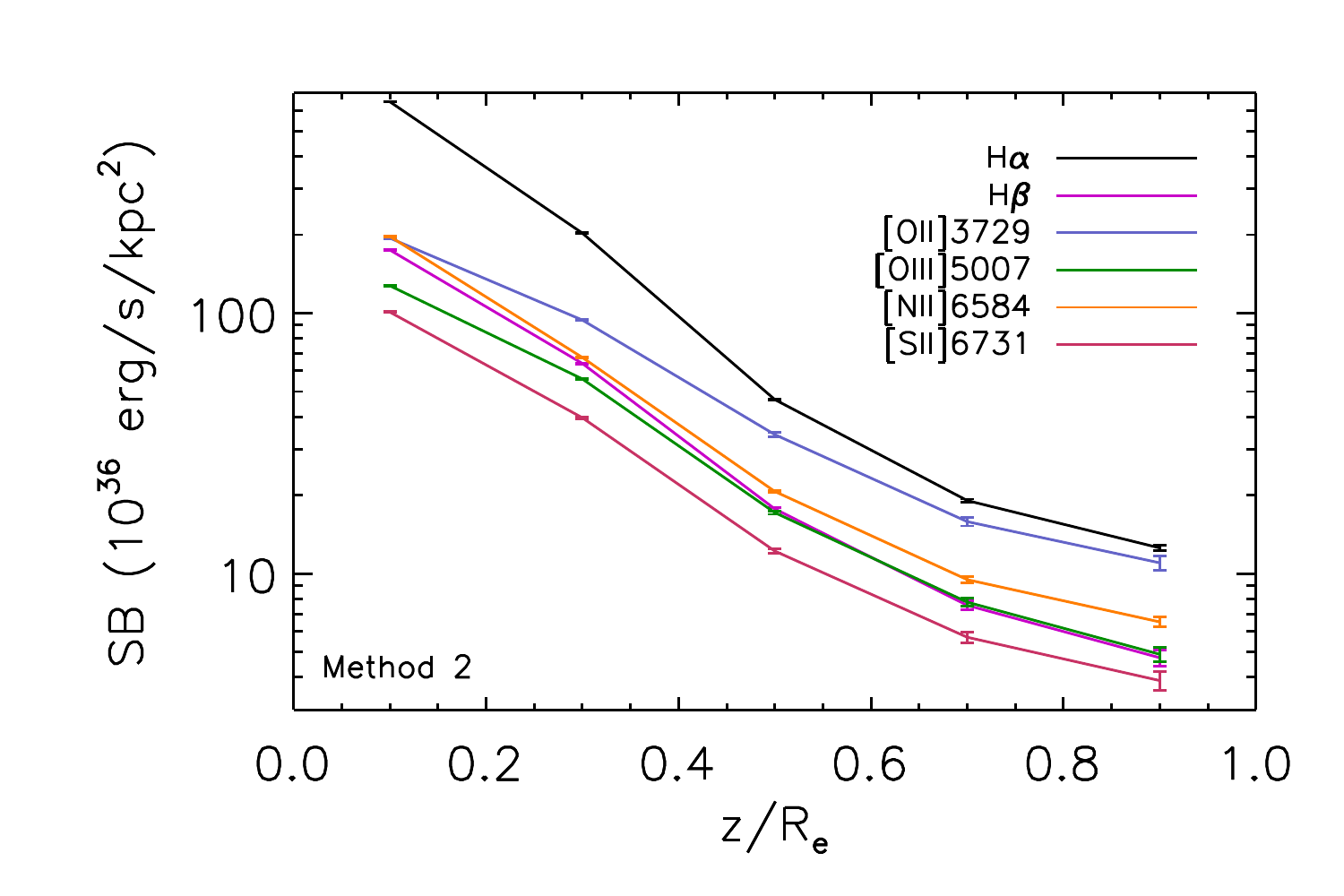}
\includegraphics[width=0.49\textwidth]{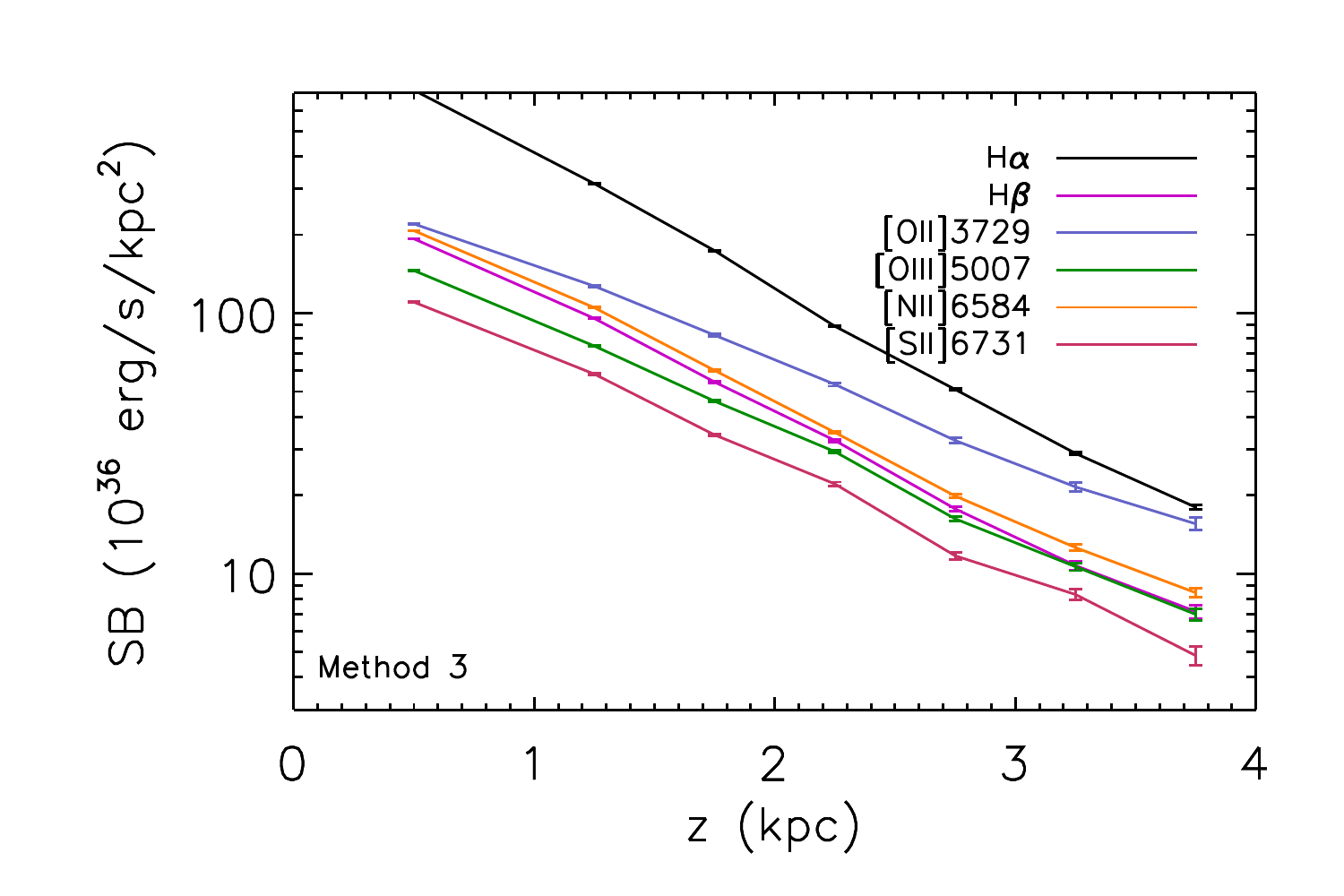}
   \caption{Surface brightness profiles of the six brightest emission lines (different colors) as a function of distance from the midplane for the full sample with the three different stacking methods.  The errors are from the spectral fitting.  \textit{Top} panel is with Method 1, \textit{middle} with Method 2, and the \textit{bottom} is with Method 3, see Section 2.2 for details about these methods.}
\label{em_36g}
    \end{figure}
    

        \begin{figure}[!ht]
   \centering     
\includegraphics[width=0.49\textwidth]{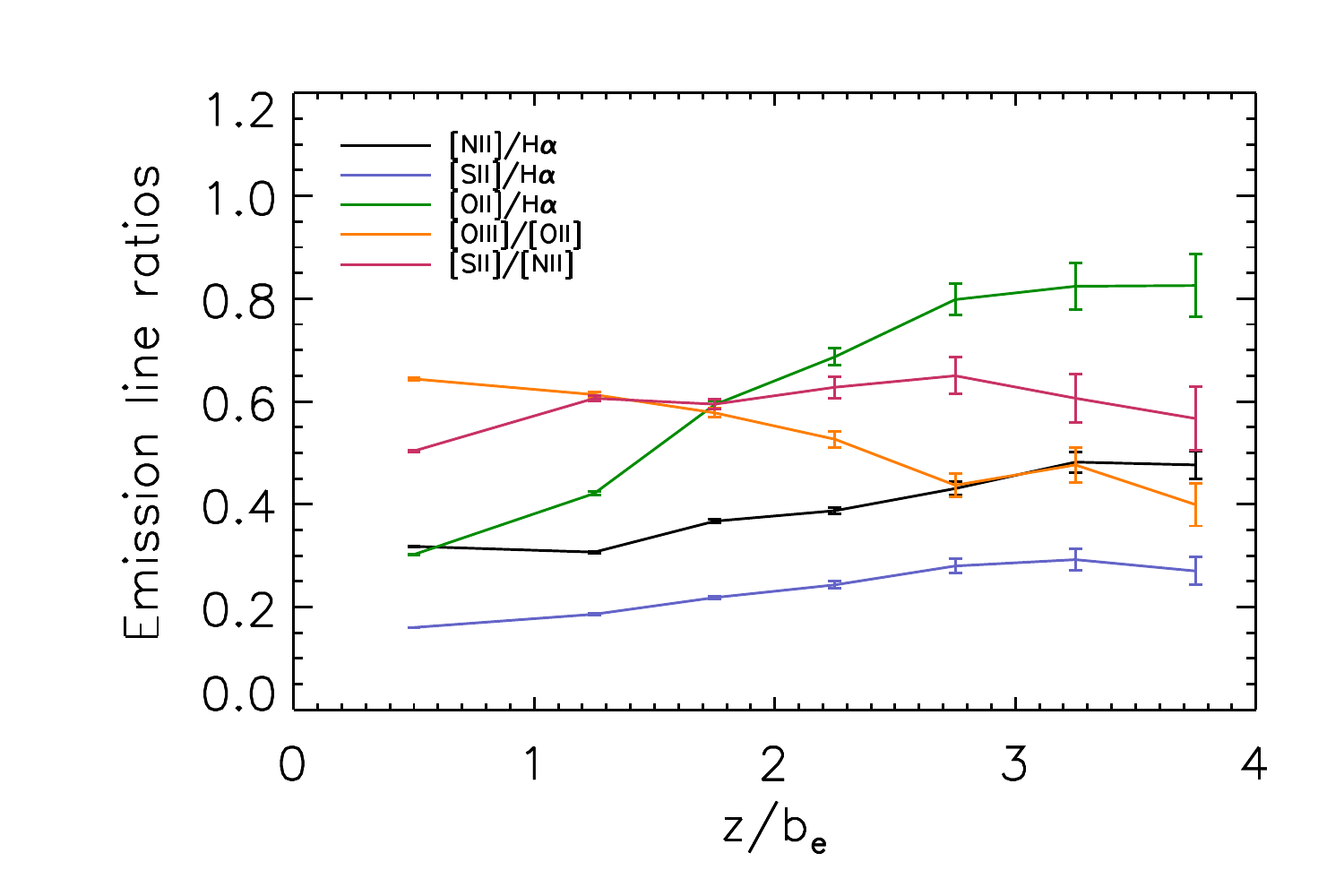}
   \caption{Ratios of the bright emission lines for the full sample using Method 1 as a 
function of $z$.}
\label{emrat_36g}
    \end{figure}


    \begin{figure}[!ht]
   \centering    
\includegraphics[width=0.49\textwidth]{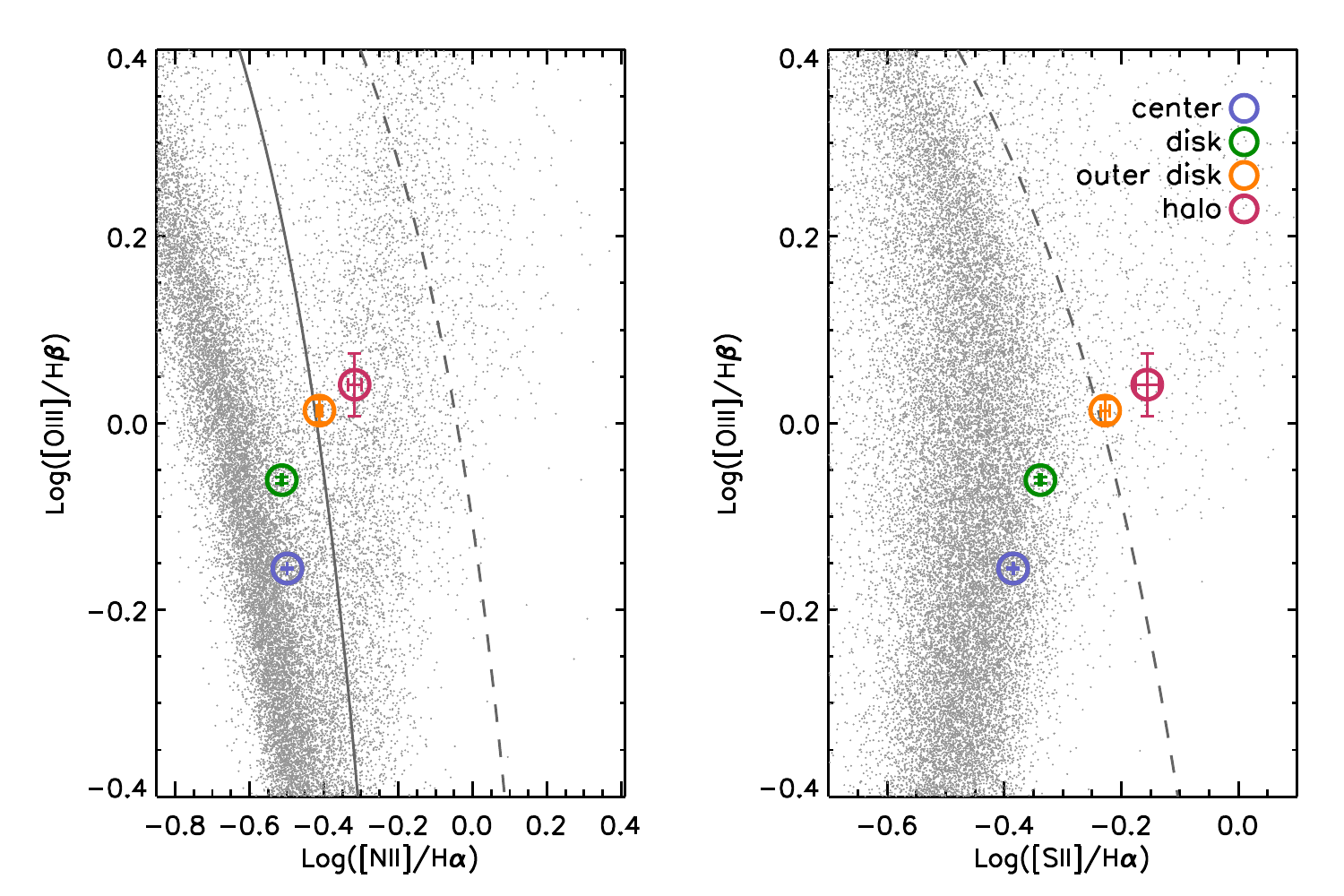}
   \caption{BPT diagram of the different galactic regions from the full sample with 
the K03 (solid) and K01 (dashed) demarcation lines.  The central, disk, outer disk, 
and halo regions correspond to the 0.0-1.0, 1.0-1.5, 2.0-2.5, 3.0-3.5 $b_e$ 
bins, respectively. The error bars correspond to the errors from the spectral fitting, propagated appropriately.  Emission line ratios measured from SDSS DR7 fiber spectra for
the main galaxy sample are plotted as gray dots.}
\label{bpt}
    \end{figure}


The surface brightness profiles of the bright emission lines, H$\alpha$, H$\beta$, \oii$\lambda$3729, \oiii$\lambda$5007, \nii$\lambda$6584, and \sii$\lambda$6731, as a function of distance along the minor axis for the
full sample of 49 late-type galaxies are shown for the three stacking methods
in Fig \ref{em_36g}.  The three methods give similar profiles and all have clear detections of the six emission lines, which exponentially decrease with distance.  H$\alpha$ is the dominant emission line near the disk, but \oii$\lambda$3729
is nearly as strong as H$\alpha$ at distances greater than $\sim 3\;b_e$.  \sii$\lambda$6731 is the weakest. 
The errors shown in the plot in color are the ones provided by the emission line spectral fitting code.  In general there is not a huge difference amongst the different stacking methods.  However, Method 1 seems to have breaks in the emission line profiles whereas Method 3 is smoother and more shallow.  This is probably because we are stacking across galaxies with a range of sizes leading to a smoothing of the profiles in Method 3.  The relative strengths of the emission lines do not vary significantly for the different methods.  All of the emission line surface brightnesses are provided in Table \ref{em_values}.

In Fig \ref{emrat_36g} we show the radial dependence of ratios of some of the bright emission lines, for simplicity only for Method 1 (z/$b_e$). 
\oiii$\lambda$5007/\oii$\lambda$3729 decreases with increasing distance from the mid-plane, which is expected as the gas 
is farther away from the main ionization source (OB stars in the disk). 
\oii$\lambda$3729/H$\alpha$ shows a strong increase with distance.  This could be indicative of gas temperatures increasing
at large distances from the disk \citep[e.g.][]{2009RvMP...81..969H}.  In the outermost bin it almost approaches unity.
The ratios \nii$\lambda$6584/H$\alpha$, \sii$\lambda$6731/H$\alpha$ increase by about 50\% between the inner- and outermost bins, whereas the ratio of \sii$\lambda$6731/\nii$\lambda$6584 remains roughly constant with height.  Here \sii$\lambda$6731/H$\alpha$ varies from 0.2 in the center to 0.3 in the outskirts and \nii$\lambda$6584/H$\alpha$ from 0.4 to 0.5.  These results agree with previous studies of the DIG in the MW and other galaxies \citep[e.g.][Zhang et al, 2016]{2009RvMP...81..969H}, who found that the ratios \nii/H$\alpha$ and \sii/H$\alpha$ in DIG are usually 
enhanced compared to classic \hii\ regions.

To gain further insight, in Fig \ref{bpt} we have also examined correlations between different
line ratios in a series of "BPT" diagrams similar to those introduced by \citet{1981PASP...93....5B}
to distinguish between gas that is being photo-ionized by young stars in \hii\ regions, and gas 
that is being photo-ionized by the central source in an active galactic nucleus (AGN).  For all the BPT diagrams \sii\ refers to the sum of two \sii$\lambda\lambda$6717, 6731 emission lines.
For clarity, we have only plotted about half the $b_e$ bins, sampling different regions of the galaxy.  
The chosen bins are 0.0-1.0 $b_e$ for the center of the galaxy, 1.0-1.5 $b_e$ for the disk, 
2.0-2.5 $b_e$ as the outer disk, and 3.0-3.5 $b_e$ for the region outside the disk (which we term the halo) with errors from the spectral fitting code. For reference, emission line ratios from 
galaxies from the SDSS DR7 main spectroscopic sample 
are plotted in the background as gray small dots. These emission line ratios are obtained from 3 arcsecond
diameter single fiber spectra that typically sample light from the inner 1-3 kpc of the galaxy, i.e.
these ratios often pertain to emission line gas present in galactic bulges rather than disks.
As a consequence, many emission line ratios characteristic of ionization from an AGN are found.

The empirical demarcation curve separating star-forming galaxies and AGN from \citet{2003MNRAS.346.1055K} (K03) 
and the theoretical curve proposed by \citet{2001ApJ...556..121K} (K01) are plotted as solid and dashed lines, 
respectively. As can be seen, the emission line ratios for the central stack lie in the region of the 
diagram appropriate for photo-ionization from young stars. 
As we move away from the center, there is a clear trend with distance $z$ for the emission
line ratios to move into the so-called "composite" region of the diagram. As discussed in K03,
emission line gas sampled by the single fiber SDSS spectra likely lie in this region of the diagram
because photo-ionization is from a mixture of young stars and a central AGN. By comparing
emission line ratios from similar galaxies at different redshifts in the SDSS main sample, K03
demonstrated that the average emission line ratios shifted systematically from the AGN to the
star-forming sections of the BPT diagrams as distance (redshift) increased and the physical aperture
spanned by the SDSS fiber became larger. In this study, we are sampling regions of the galaxy very
far away from any central supermassive black hole, so the physical explanation must be different.
As discussed in K01, gas which is heated by shocks is also expected to produce emission lines with ratios
lying in the "composite" part of the diagram. Shock-heating of gas falling into dark matter halos
is a fundamental physical process in galaxy formation and a proper
accounting of such heating process is likely critical to the interpretation of our observations.  Another phenomenon that should be considered when interpreting these diagrams is the effect of a diluted radiation field which would also cause a shift in the line ratios.  We leave the full investigation of the differences and trends of the emission lines and ratios for a future paper. 

\subsection{Large-$z$ Sample}  

    \begin{figure}[!ht]
   \centering    
\includegraphics[width=0.49\textwidth]{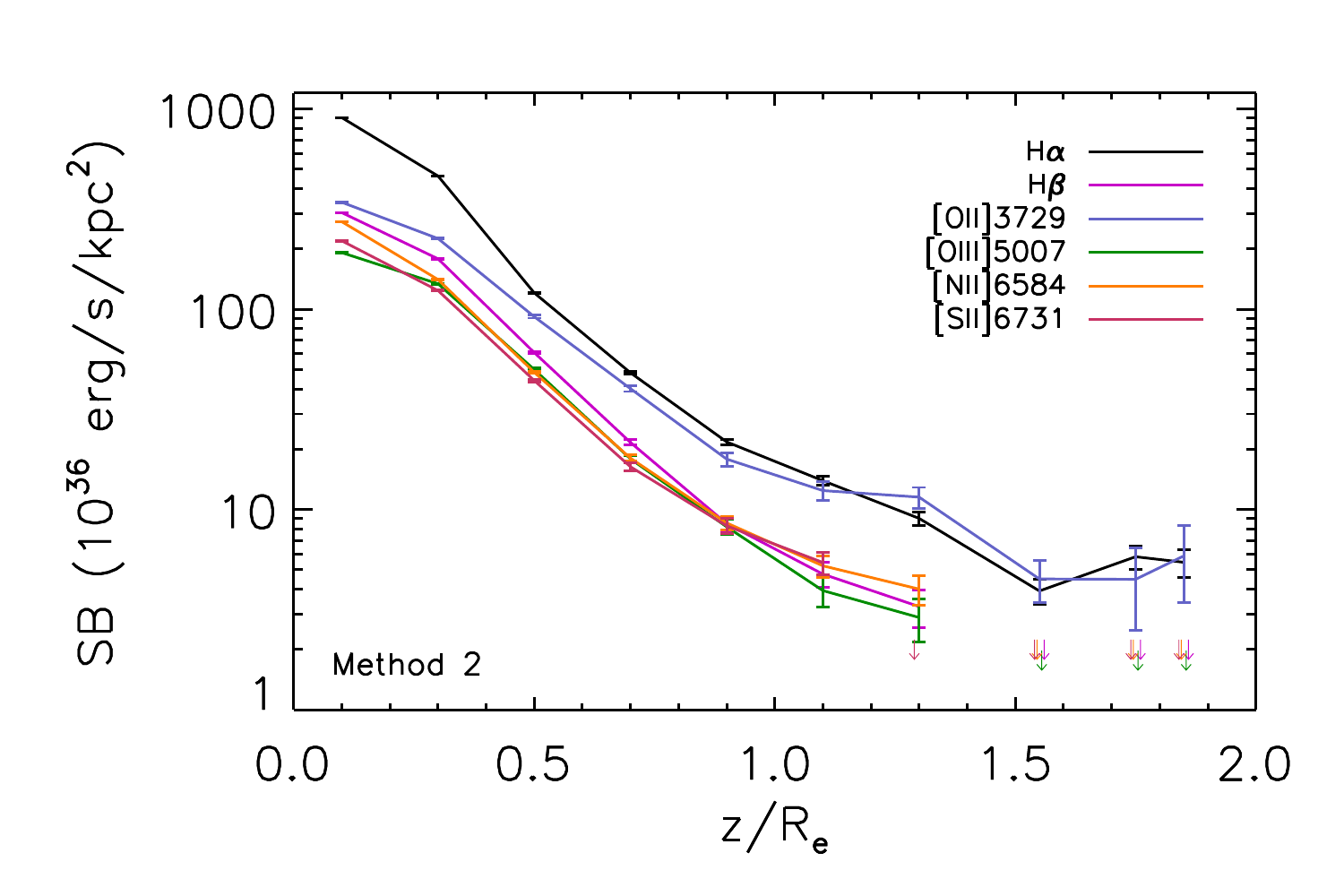}
\includegraphics[width=0.49\textwidth]{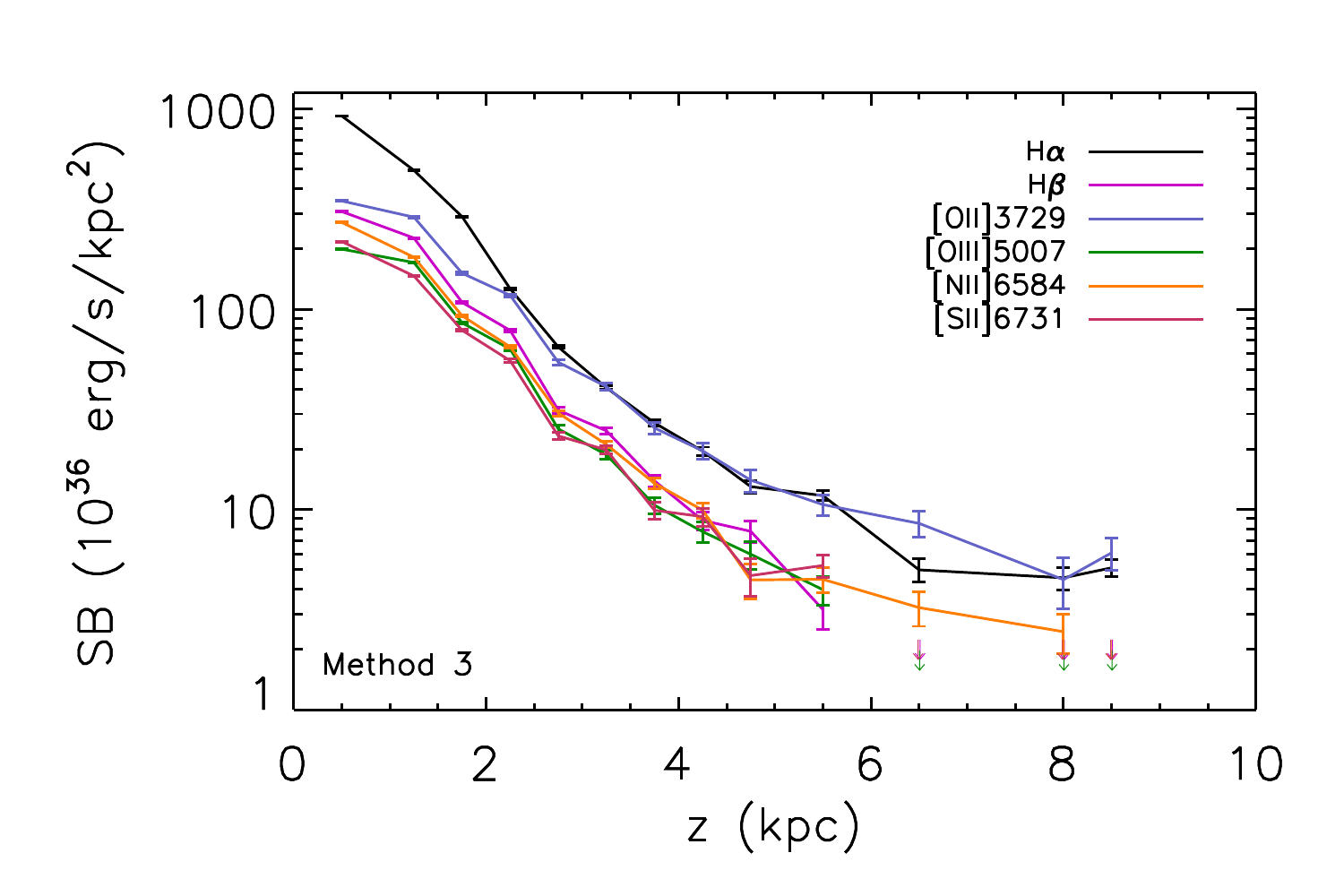}
   \caption{Surface brightnesses of the bright emission lines for the large-$z$ stack as a 
function of $z$ for stacking methods 2 (\textit{top}) and 3 (\textit{bottom}).  Upper limits are indicated by the arrows.}
\label{em_13g}
    \end{figure}

With our stack of 16 galaxies extending past 9\;kpc, we can study emission line profiles out to larger distances from the disk.  This sample was chosen so each galaxy simultaneously had fibers that extended out to at least 2\;R$_e$ and 9\;kpc so they would be part of both stacking methods 2 and 3.  We relaxed the constraint on b/a so each galaxy must have b/a<0.4.  This increased the sample by another four galaxies.  Because the galaxies have a larger range of inclinations and we wanted to ensure that the detections in the outer bins for each galaxy were from the halo and not from the outer disk, we made an additional requirement on the size of $b_e$. Each galaxy must have $b_e\leq2$\;kpc, so at heights $\gtrsim5$\;kpc (or about 1 R$_e$) we are probing the halo.  The galaxies that were included in this sample are marked in Table \ref{gal_prop} and the four additional galaxies with 0.3<b/a<0.4 are shown below the line.  

In Fig \ref{em_13g} we show the emission line surface brightness profiles for the large-$z$ sample with stacking methods 2 and 3.  
We provide upper limits where the signal is not reliably detected.  
This is done by combining between 1000 and 1500 sky spectra into 30 independent 
stacks and adding in artificial emission lines of known flux with Gaussian profiles
and widths the same as that observed in the stacked spectra with 
detections. We decreased the strength of the emission lines until the error on the line fluxes are
greater than 20\% of the true value. This determines the values of the upper limits
plotted in Fig \ref{em_13g} and any value that was less than this is considered to not be reliably detected.  Due to the wavelength-dependence of the S/N (see Fig \ref{snr}) 
these limits are at slightly different levels for different lines.
At large distances, \oii\ becomes comparable with H$\alpha$ beyond 1.0 R$_e$ and 3\;kpc in the stack.
Intriguingly, with both stacking methods all detectable lines in the halo, \oii, H$\alpha$, and for Method 3 also \nii, the line surface brightnesses appear to flatten at these large distances.  This could indicate a change in the heating source at these distances.  
Future analysis with larger samples from the complete MaNGA survey will be useful for solidifying the analysis of
line ratios at large distances from the disk. 

\subsection{Results split by sSFR, C, and M$_{star}$}

With a sample of 49 galaxies we can split the sample in half ($2\times24$) and still have enough S/N for reliable detections of the bright emission lines (see Table \ref{bin_info}).  The amount of ionized gas and possible additional heating source(s) could depend on galactic properties.  For example, the amount of eDIG present is known to depend on the star formation rate of the galaxy \citep[e.g.][]{2003A&A...406..493R}.  We have, therefore, split the full sample in half according to sSFR (at $5.8\times10^{-10}\mathrm{yr}^{-1}$), C (at 2.44), and M$_{star}$ (at $3.7\times10^{9}\mathrm{M}_\odot$).  These quantities for the individual galaxies are
provided in Table \ref{gal_prop}.  The surface brightnesses of the bright emission lines for these subsamples are also given in Table \ref{em_values}.

\begin{figure}[!ht]
   \centering    
\includegraphics[width=0.49\textwidth]{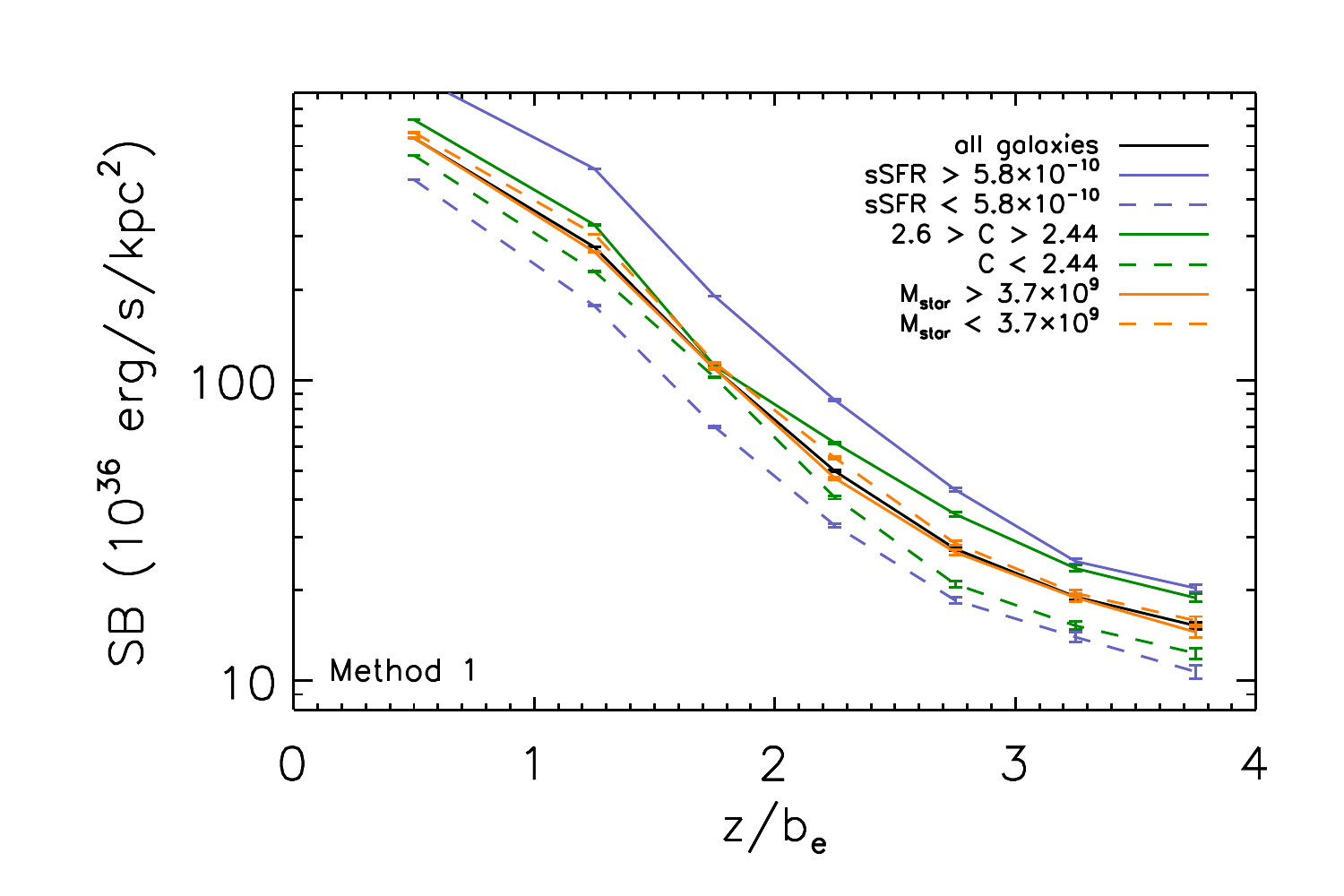}
\includegraphics[width=0.49\textwidth]{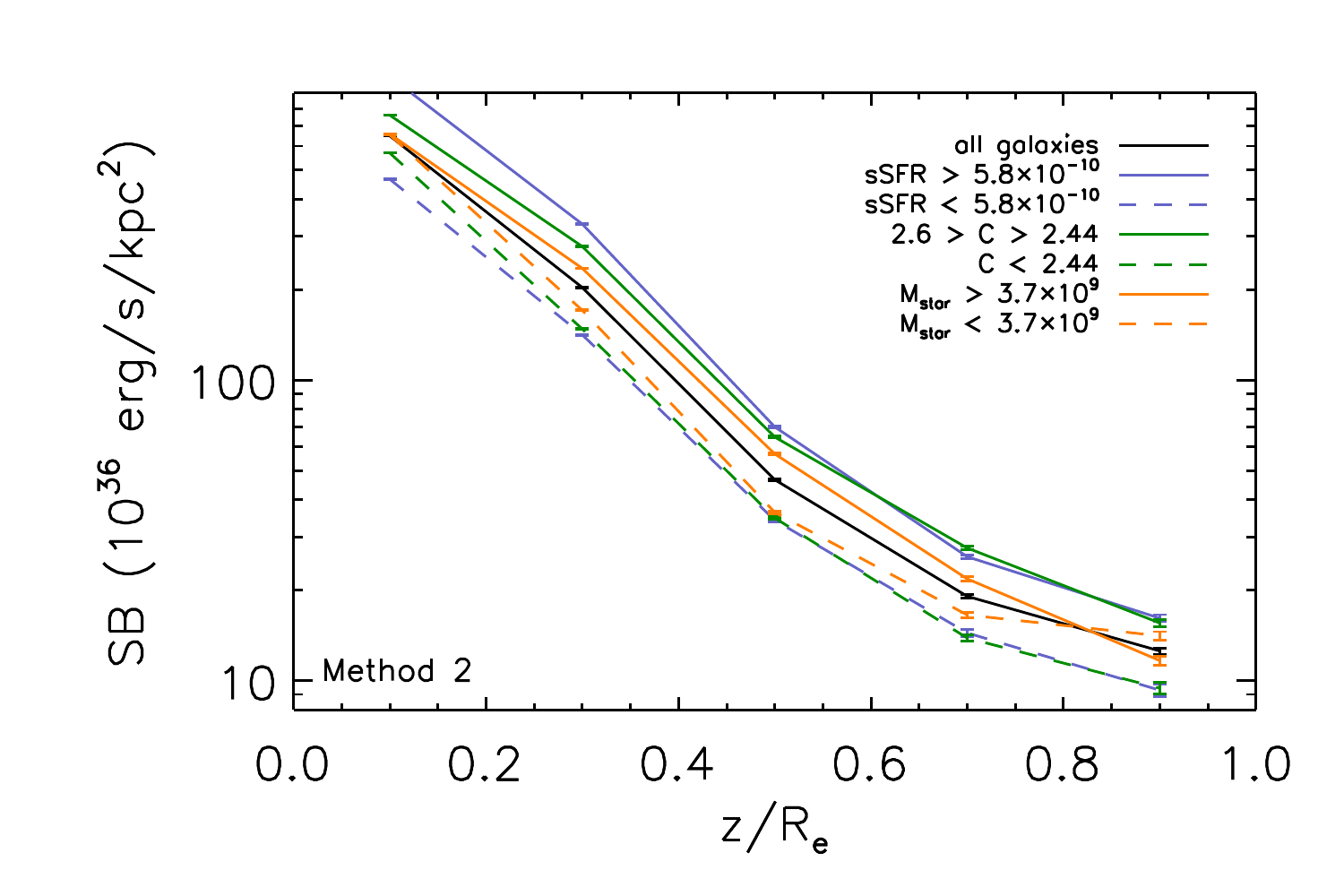}
\includegraphics[width=0.49\textwidth]{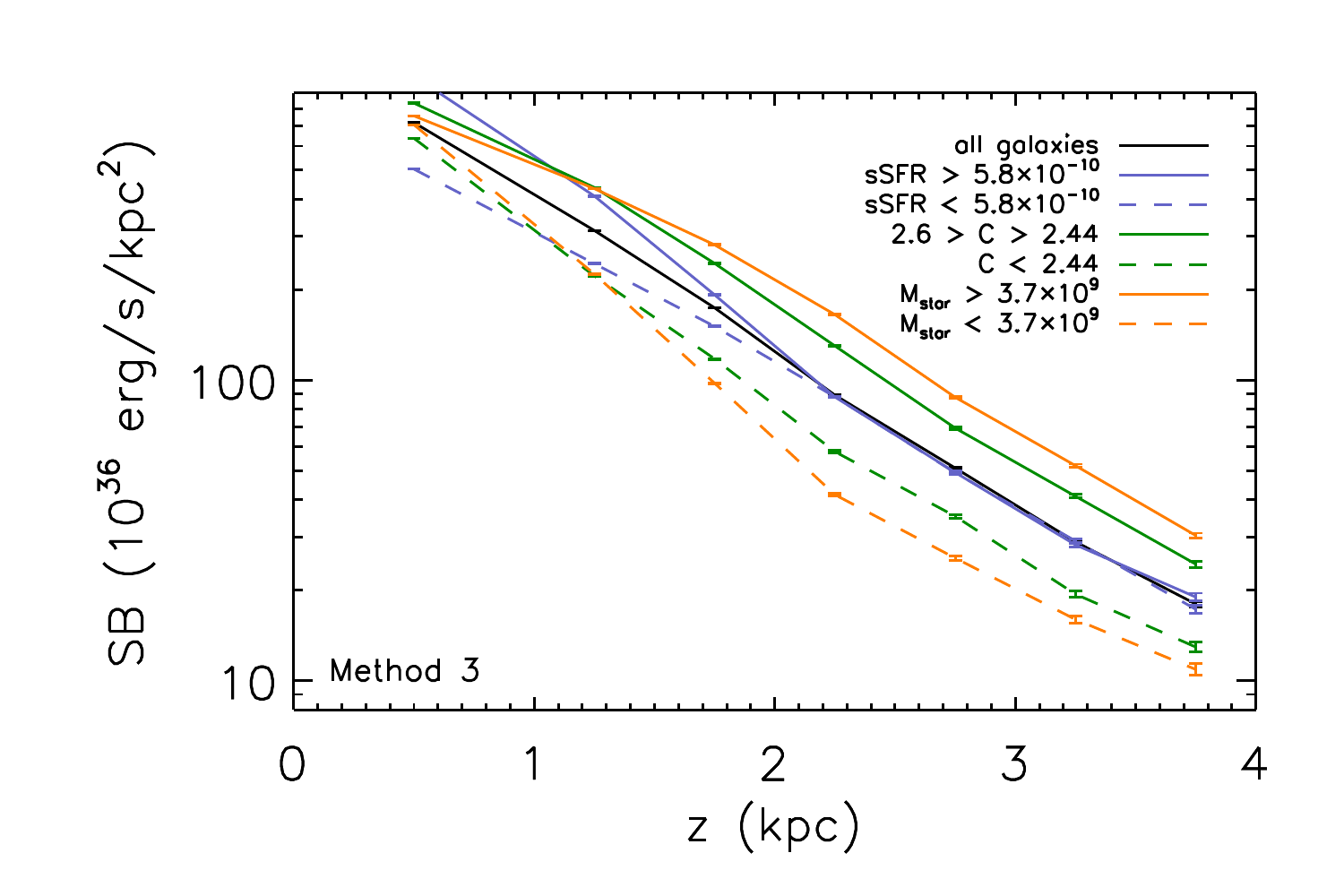}
   \caption{H$\alpha$ surface brightness profile as a function of $z$ for various subsamples with the three different stacking methods, \textit{top} panel is with Method 1, \textit{middle} with Method 2, and the \textit{bottom} is with Method 3.}
\label{ha}
    \end{figure}

\begin{figure}[!ht]
   \centering    
\includegraphics[width=0.49\textwidth]{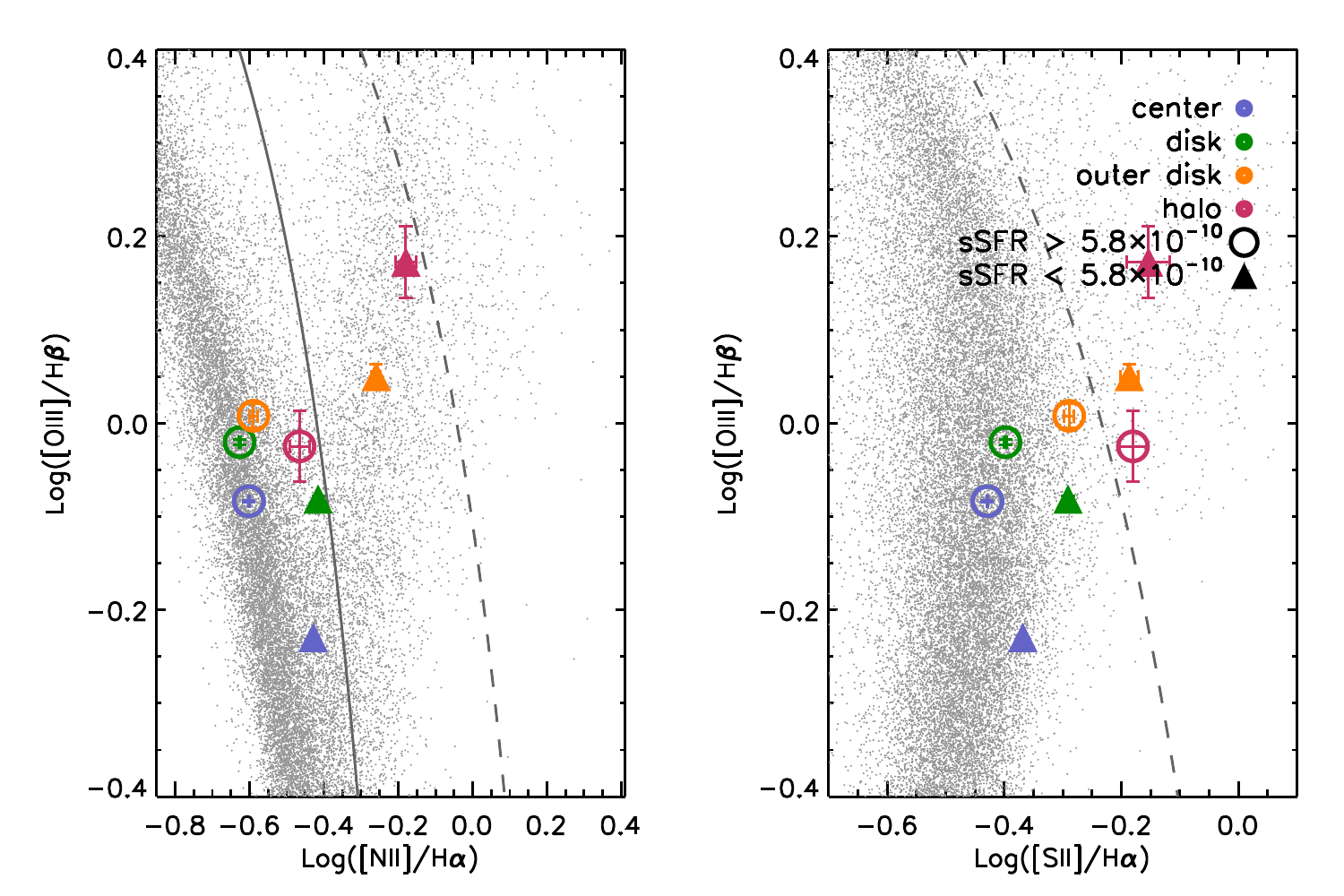}
   \caption{Similar to Fig \ref{bpt} but showing the subsamples that were split by sSFR.  The high sSFR galaxies are shown as open circles and the low sSFR galaxies as filled triangles with the colors corresponding to regions along the minor axis like in Fig. \ref{bpt}.}
\label{bpt_ssfr}
    \end{figure}

Figure \ref{ha} is similar to Fig \ref{em_36g} but only for H$\alpha$ of the different subsamples for the three stacking methods.  The full sample (black solid line) is also given for reference.
It is clearly seen that the largest differences in H$\alpha$ line surface brightness are between the high and low sSFR subsamples for Method 1 and 2, which is expected due to the relationship between eDIG and star formation.  All three stacking methods show very similar trends, except for Method 3 for the high sSFR subsample.  For Methods 1 and 2 the high sSFR subsample has the highest H$\alpha$ surface brightness at all distances, but for Method 3 this is only true at z$\lesssim1$\;kpc and then the high M$_{star}$ dominates.  This is probably because the higher M$_{star}$ galaxies also have larger disks compared to the other subsamples, so at the same physical distance, the high M$_{star}$ sample is at a smaller R$_e$ compared to the other samples.
Interestingly, the difference in line surface brightness for the sSFR subsamples is roughly the same at all distances with Methods 1 and 2. 
The favored interpretation for higher eDIG surface brightnesses is that galactic winds have driven out more gas in 
galaxies with stronger star formation. Galactic winds generally occur as bi-conical outflows from
the central region of the galaxy with opening angles of $\sim 60$ degrees.  With a larger sample, we will test for an azimuthal dependence on the H$\alpha$ surface brightness.   
Figure \ref{bpt_ssfr} shows differences in emission line ratios for the high sSFR and low sSFR subsamples.  Like in Fig \ref{bpt}, we have only plotted a fraction of the minor axis bins for clarity.  The central, disk, outer disk, and halo regions again correspond to the 0.0-1.0, 1.0-1.5, 2.0-2.5, 3.0-3.5 $b_e$ bins, respectively.   
As can be seen, the high sSFR data points are concentrated closer to the star-forming region of the BPT
diagrams than the low sSFR data points. This is consistent with the hypothesis that a larger
fraction of the ionizing photons responsible for exciting the gas arise from young stars in the
actively star-forming systems.  To ensure that these differences are significant and not due to a small sample size, we also split the full sample randomly in half six times and compared the emission line ratios for the halo bin, where the differences are the largest in the BPT diagram.  The mean and standard deviation of the absolute value of the differences for the randomly split samples between $\log{(\oiii/\mathrm{H}\beta)}$, $\log{(\nii/\mathrm{H}\alpha)}$, and $\log{(\sii/\mathrm{H}\alpha)}$ are $0.04\pm0.02$, $0.09\pm0.06$, and $0.03\pm0.02$, respectively.  The changes we see in the two sSFR populations are greater than this.

\begin{figure}[!ht]
   \centering   
\includegraphics[width=0.49\textwidth]{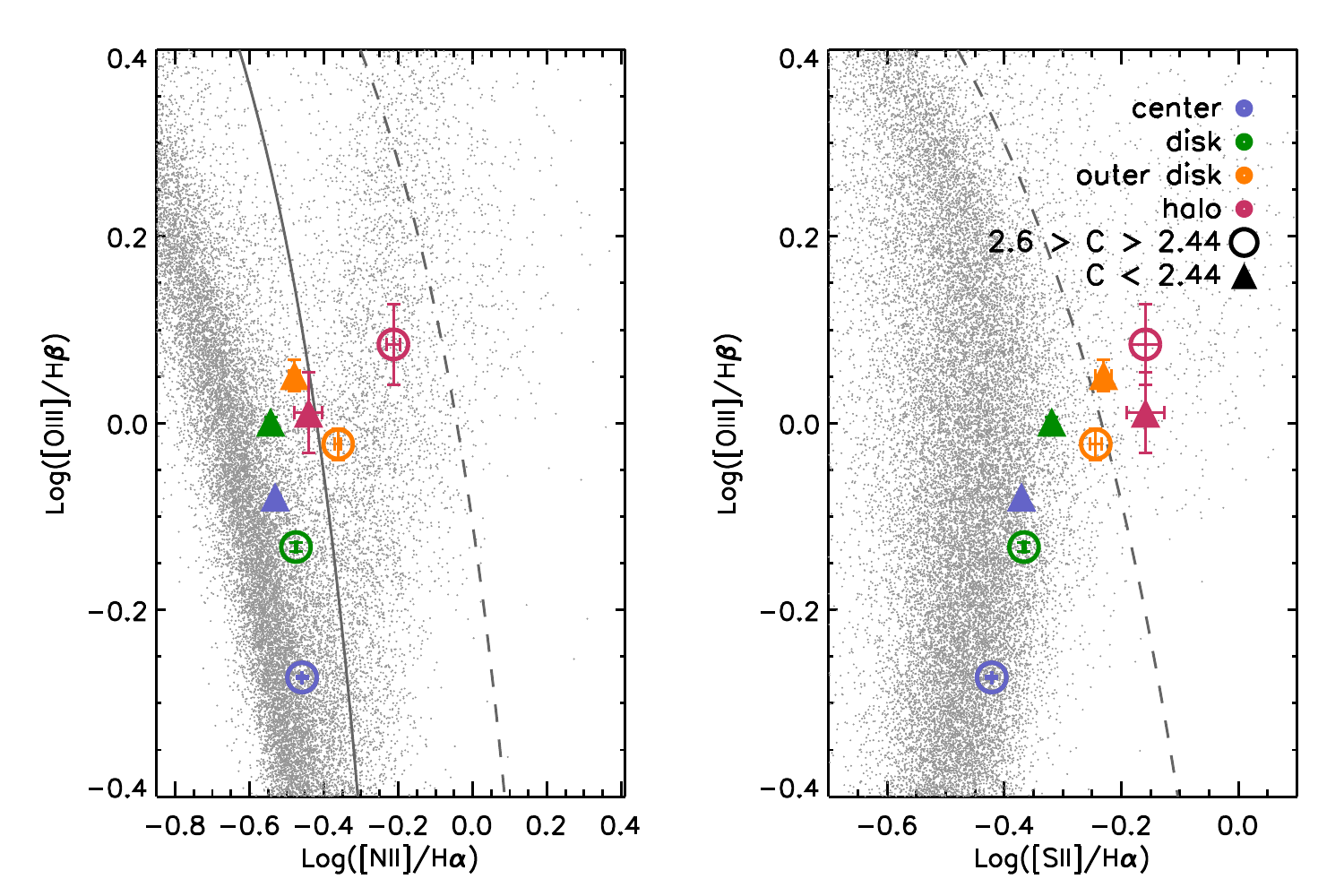}
   \caption{Same as Fig \ref{bpt_ssfr} but split by C, with high C galaxies as open circles and low C galaxies as filled triangles.}
\label{bpt_conc}
    \end{figure}

When split by M$_{star}$ or C, differences in H$\alpha$ surface brightness profiles are much smaller (typically
less than a factor of 1.5) for Methods 1 and 2 (Fig \ref{ha}), but 
significant differences in emission line ratios between the sub-samples remain.
When split by concentration, Fig \ref{bpt_conc} shows that the low C points are more clustered towards the star formation region. 
Interestingly, the outermost "halo point" for the low C sample regresses back to the star forming area, 
particularly in the diagram with \nii/H$\alpha$ on the x-axis. In contrast, there is a 
0.4 dex increase in log \nii/H$\alpha$ from the inner disk to the outer halo for the high C sample. 
When split by $M_{star}$, similar trends are seen -- low mass galaxies cluster towards the star-forming region
of the diagram, whereas high-mass galaxies are generally in the composite region, particularly
at a large distance from the disk plane. The trend for \oiii/H$\beta$, \nii/H$\alpha$, and  \sii/H$\alpha$ to increase as a function of distance
from the disk is strongest for high $M_{star}$ galaxies. It is well established that dark matter halo mass
scales most strongly with stellar mass, and thus the temperature of shock-heated, virialized halo gas could
also scale most strongly with stellar mass. In future work we plan to investigate whether these trends
can be caused by such scalings.   

\begin{figure}[!ht]
   \centering    
\includegraphics[width=0.49\textwidth]{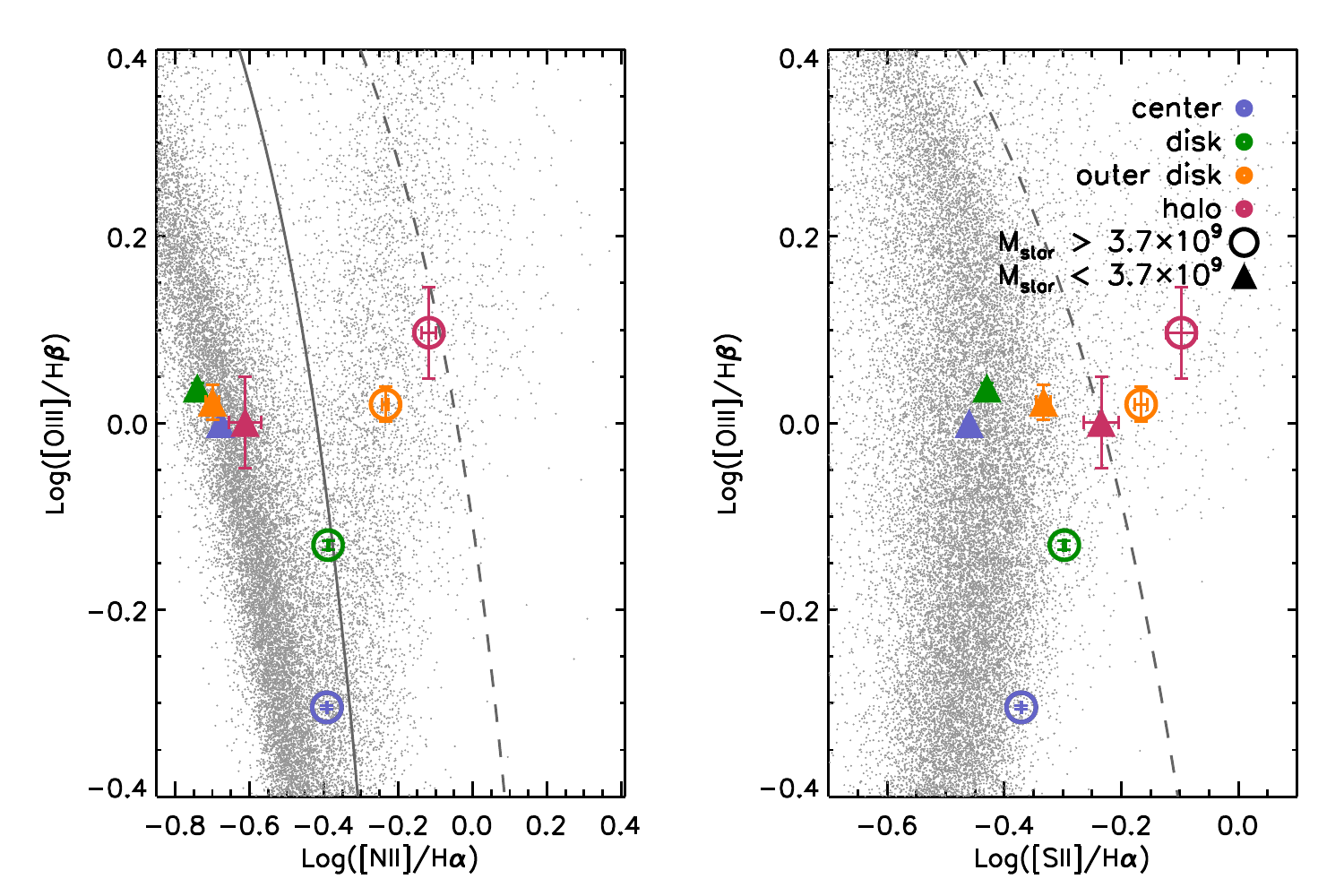}
   \caption{Same as Fig \ref{bpt_ssfr} but split by M$_{star}$, with high M$_{star}$ galaxies as open circles and low M$_{star}$ galaxies as filled triangles.}
\label{bpt_mstar}
    \end{figure}

\begin{figure}[!ht]
   \centering    
\includegraphics[width=0.49\textwidth]{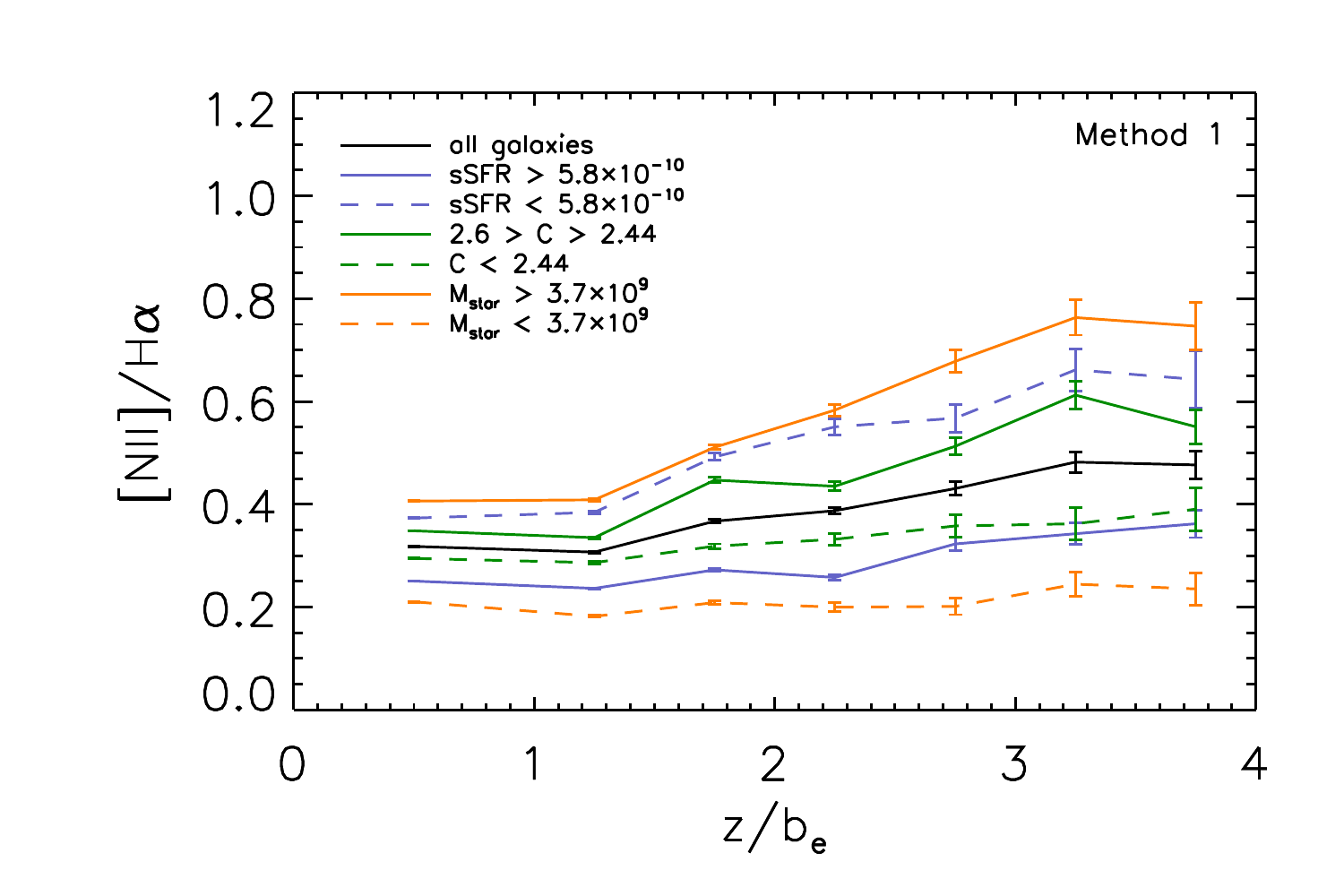}
\includegraphics[width=0.49\textwidth]{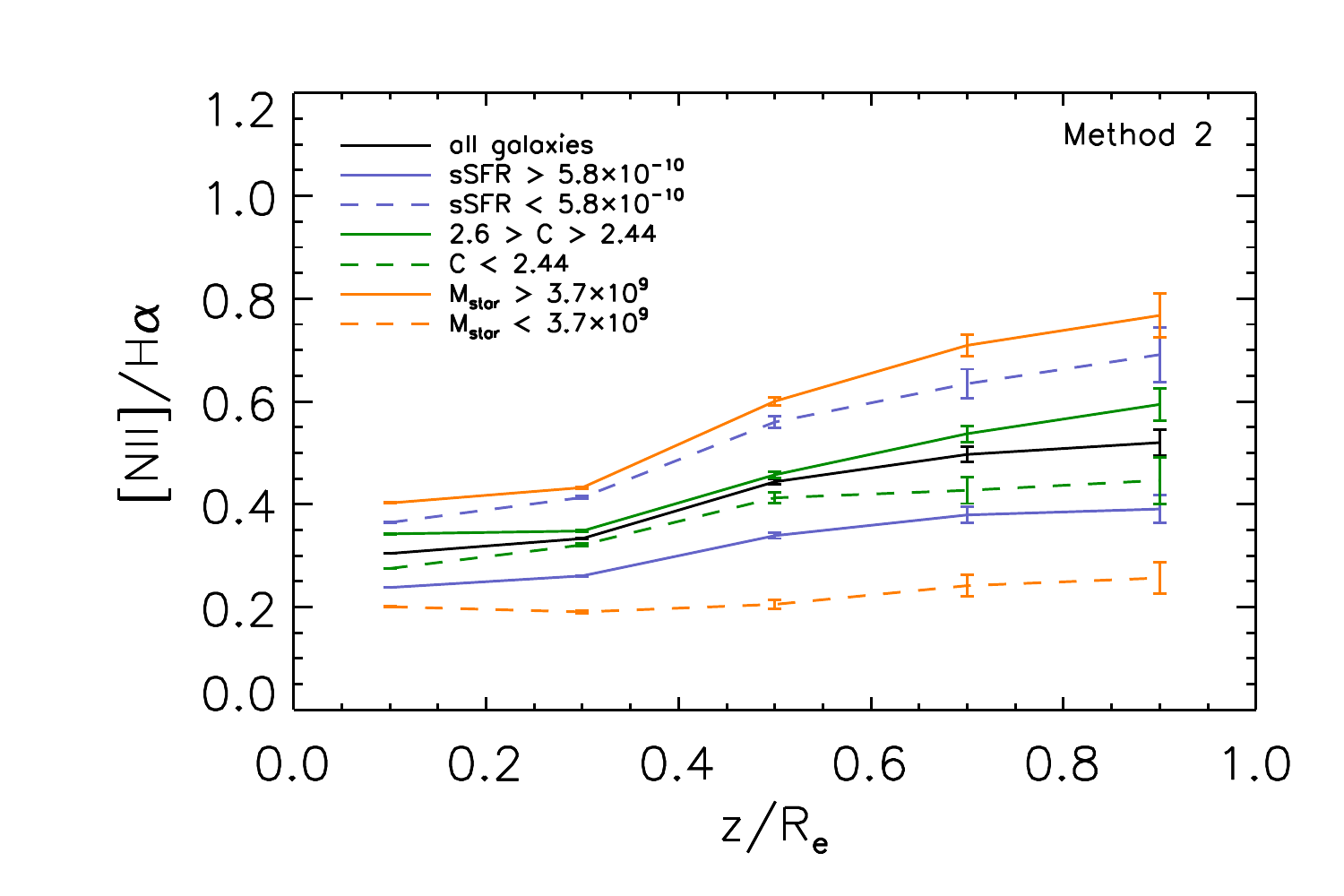}
\includegraphics[width=0.49\textwidth]{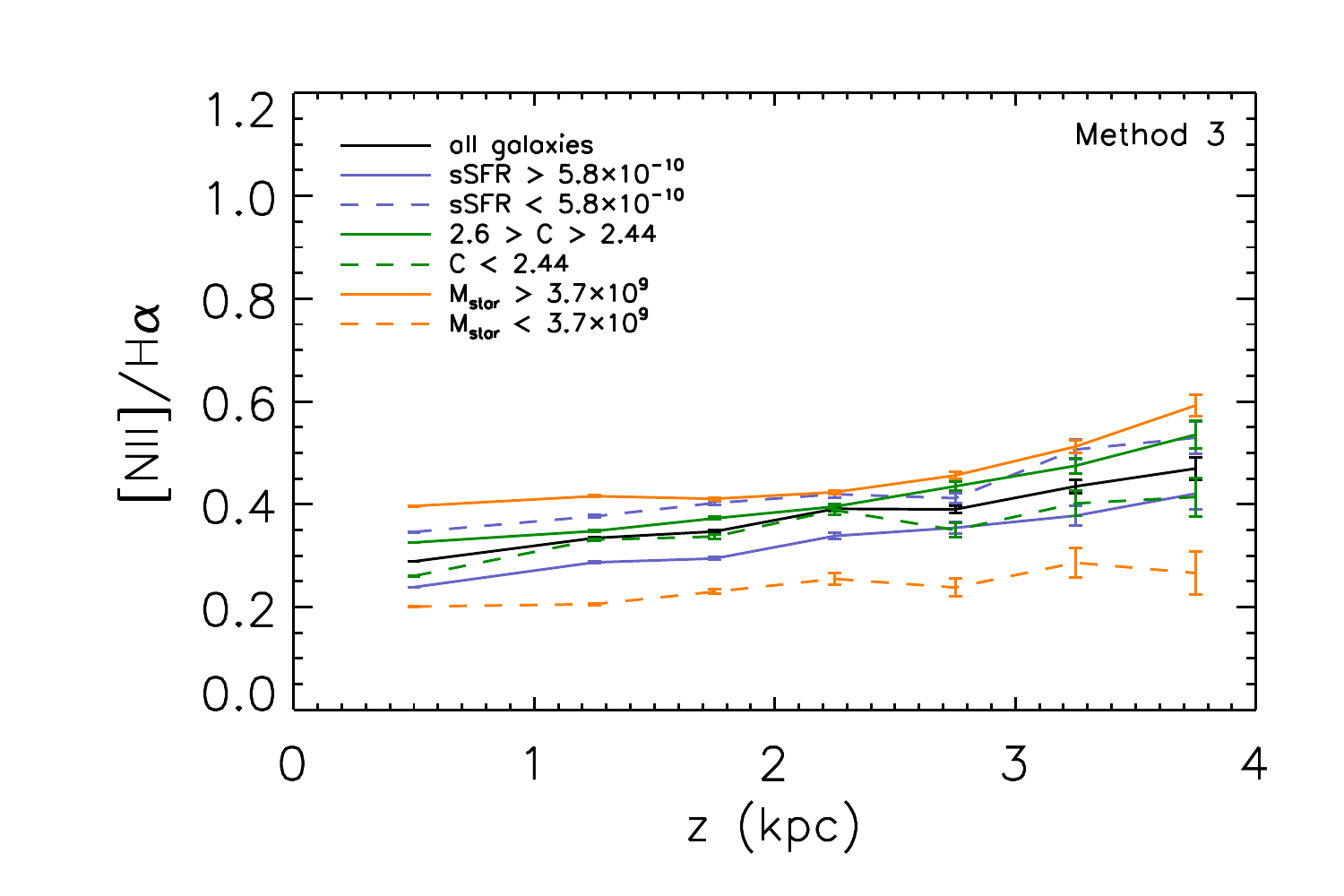}
   \caption{Ratio of emission lines \nii\ and H$\alpha$ as a function of $z$ for various subsamples with the three different stacking methods, \textit{top} panel is with Method 1, \textit{middle} with Method 2, and the \textit{bottom} is with Method 3.}
\label{nII_ha}
    \end{figure}

We have also plotted only the changes in the \nii/H$\alpha$ ratio with distance for the different samples, shown in Fig \ref{nII_ha}.  There is no significant difference amongst the different stacking methods.  \nii/H$\alpha$ can be a temperature indicator, assuming the relative abundances of nitrogen and hydrogen are constant.  This assumption might not be valid in the halo, where for resolved populations studies of very nearby galaxies a range of metalicity gradients is seen \citep{2016MNRAS.457.1419M}. For most of the subsamples this ratio slightly increases by 0.1 to 0.2 with distance, except for M$_{star}$ subsamples.  The high M$_{star}$ sample increases from $\sim0.45$ to 0.8 in Method 1 and 2 and to 0.6 in Method 3.  The ratio in the low M$_{star}$ sample stays constant in all three methods.  This is consistent with Fig \ref{bpt_mstar} where the low M$_{star}$ sample at all distances stays within the star formation area of the BPT diagram.

For more details, Table \ref{em_values} lists the surface brightnesses of all the bright emission lines using the three stacking methods for each minor axis bin for all the samples (excluding the large-$z$ sample).


\section{Conclusion}

Studying eDIG around nearby galaxies is a challenging task due to its naturally low surface brightness.  Previous spectroscopic studies usually observed individual, very nearby galaxies with long integration times to detect eDIG out to a few kpc.  Using the MaNGA survey and with the technique of stacking spectra across several dozen galaxies, we can observe the eDIG into the halo and study the radial trends for an average galaxy population.  We have stacked across the galaxies using three different methods for normalizing the distance from the midplane, by minor axis effective radii ($b_e$), major axis effective radii (R$_e$), and physical distance (kpc).  In most cases the three methods were in agreement.  In this current paper we have shown that:

$\quad\bullet$  We can clearly detect emission lines out to $\sim$\;4\;kpc above the disk with our full sample of 49 galaxies (Fig \ref{zoom_36g} and \ref{em_36g}).  The emission line ratios (Fig \ref{emrat_36g}) are consistent with what has previously been observed in individual nearby galaxies including the Milky Way.  

$\quad\bullet$  With only 16 galaxies observed to larger radii, we can already have some detections for the bright emission lines, namely H$\alpha$ and \oii, and upper limits of the other bright emission lines out to $\sim$\;9\;kpc.  Past $\sim6$\;kpc the \oii\ and H$\alpha$ surface brightnesses are comparable and slightly flatten, suggesting a possible change in the heating source, such as HOLMES, shocks, inflows, etc (Fig. \ref{em_13g}).

$\quad\bullet$  By splitting our sample by different galactic properties, we clearly detect changes to the eDIG in the outer disk and halo.  The emission line surface brightnesses are higher both in the disk and halo for the high sSFR galaxies compared to the low sSFR galaxies (Fig \ref{ha}).  The high M$_{star}$ sample has the highest \nii/H$\alpha$ ratio with the strongest increase with distance, indicating a large temperature gradient.  The high C galaxies also show a slight increase of \nii/H$\alpha$ and the low sSFR sample increases as well (Fig \ref{nII_ha}).  

$\quad\bullet$  In the BPT diagrams, all the split subsamples occupy different areas of the diagram and show different trends with distance (Fig \ref{bpt_ssfr}, \ref{bpt_conc}, and \ref{bpt_mstar}).  The low M$_{star}$, low C, and high sSFR tend to lie near the starforming region of the BPT diagram.  The other subsamples' outer disk and halo values are between the K03 and K01 lines in the \nii/H$\alpha$ diagram, and past K01 in the \sii/H$\alpha$ diagram.  This shows that eDIG properties depend on galactic properties: sSFR, bulge to disk ratio, and stellar mass.


As the MaNGA observations continue, the sample size will keep increasing, allowing for further study of outskirts of galaxies and eDIG.  By the end of the MaNGA survey, with $\sim300$ galaxies out to 4\;kpc and $\sim100$ galaxies out to 9\;kpc, we can better determine the dependence of the eDIG on galactic properties.  In combination with models we can improve our understanding of the possible additional heating source(s) and the extent of the eDIG. 

\begin{table*}
\caption{Galaxy properties \label{gal_prop}}
\centering
\begin{tabular}{cccccccccc}
\hline\hline
\noalign{\smallskip}
Plate & IFU design& RA & dec & redshift & R$_e$ & b/a & M$_{star}$ & C & sSFR\\
 & & & & & (arcsec) & & ($10^9$\;M$_\odot$) & & ($10^{-10}$yr$^{-1}$)  \\
\noalign{\smallskip}
\hline
\noalign{\smallskip}
8082 &  3702 &  49.1086  &   0.3216 & 0.021 &  5.27  &  0.251 &  0.89   &   2.55 &  2.09 \\
8083 &  3703 &  50.8271  &   0.9399 & 0.036 &  2.80  &  0.255 &  3.33   &   2.23 &  2.55\\
8137\tablefootmark{a} & 12701 & 115.1669  &  42.4540 & 0.038 &  6.07  &  0.206 &  3.19   &   2.44 &  5.24\\
8135 & 12705 & 115.4431  &  37.6155 & 0.027 & 10.71  &  0.189 &  3.89   &   2.51 &  3.67 \\
8137 &  3704 & 115.6416  &  44.2159 & 0.042 &  4.12  &  0.247 &  4.74 &   2.55 & 7.25 \\
8137\tablefootmark{a} & 12705 & 116.4754  &  43.5194 & 0.041 &  6.90  &  0.279 &  5.45 &   2.59 &  6.09\\
8138\tablefootmark{a} &  6103 & 117.3182  &  46.2050 & 0.038 &  5.01  &  0.204 &  4.67   &   2.55 & 6.08\\
8140 \tablefootmark{a}& 12702 & 117.9032  &  41.4604 & 0.042 &  4.63  &  0.232 &  5.84  &  2.32 & 5.81  \\
8143 & 12703 & 120.9085  &  43.34305  &  0.015  &  17.42 &  0.16 & 1.42 & 2.43 & 6.11 \\
8249\tablefootmark{a} & 12701 & 136.1563  &  44.8747 & 0.035 &  4.02  &  0.168 & 0.90 & 2.54 & 6.18 \\ 
8247\tablefootmark{a} & 12705 & 137.0681  &  41.6392 & 0.041 &  6.98  &  0.150 &  4.66   &   2.51 & 3.76 \\
8250 & 12705 & 140.3988  &  43.2572 & 0.016 & 10.72  &  0.167 &  0.67   &   2.41 & 10.2 \\
8252 &  9101 & 144.6924  &  48.5629 & 0.025 &  8.19  &  0.200 &  5.95   &   2.53 &  7.71\\
8253 & 12704 & 159.1533  &  43.5068 & 0.025 &  9.54  &  0.164 &  7.52  &   2.51 & 1.62 \\
8254\tablefootmark{a} & 12702 & 163.5181  &  43.5328 & 0.037 &  5.94  &  0.175 &  2.62   &  2.55 & 4.11  \\
8256 & 12701 & 164.5852  &  40.7882 & 0.026 &  4.81  &  0.221 &  2.17  &   2.41 &  7.76\\
8448 &  6102 & 165.1879  &  22.28775  &  0.023  &  6.43  &  0.22 & 0.92 & 2.55 & 12.4 \\
8257 &  9101 & 165.4536  &  44.8138 & 0.025 &  8.28  &  0.260 &  4.73  &   2.45 & 6.33 \\
8257 & 12702 & 166.4033  &  46.1736 & 0.025 & 13.94  &  0.175 &  2.93   &   2.34 &  9.68\\
8451 & 12705 & 166.4771  &  41.0632 & 0.047 &  8.11  &  0.235 &  10.2  &   2.36 &   5.39\\
8448 & 12704 & 166.7397  &  22.83580  &  0.023  &  9.97  &  0.15 & 4.49 & 2.41 & 2.65  \\
8257 & 12705 & 167.0346  &  45.9846 & 0.036 & 17.54  &  0.274 &  27.9  &   2.28 & 0.79 \\
8466 \tablefootmark{a}& 12703 & 171.5091  &  45.4339 & 0.034 &  6.61  &  0.150 &  1.37   &   2.57 &  6.47\\
8259 & 12702 & 178.5063  &  44.6423 & 0.024 & 11.11  &  0.271 &  6.25  &   2.48 &   4.07 \\
8259 & 12703 & 180.9426  &  43.9845 & 0.023 & 11.72  &  0.154 &  1.86   &   2.36 &  5.09\\ 
8263 & 12702 & 186.0307  &  45.4345 & 0.024 & 12.77  &  0.150 &  6.34 &   2.53 & 2.03 \\
8341 & 12701 & 189.0321  &  46.6433 & 0.024 &  9.28  &  0.226 &  3.15 &   2.36 &  6.16\\
8317\tablefootmark{a} & 6101  & 191.6746  &  43.3090 & 0.040 &  5.37  &  0.246 & 3.73 & 2.27 & 5.73 \\
8465 &  9101 & 197.5807  &  47.1241 & 0.024 &  8.27  &  0.199 &  5.11  &   2.09 &  3.99\\
8330 & 12704 & 203.6247  &  38.29412  &  0.026  &  12.80 &  0.23 & 4.63 & 2.31 & 6.63  \\
7495 &  3703 & 205.2433  &  27.7263 & 0.029 &  5.08  &  0.207 &  1.80   &   2.25 &  9.56\\ 
8325 &  3702 & 209.8321  &  47.9562 & 0.015 &  7.84  &  0.208 &  0.79   &   2.53 &  4.94\\
8329\tablefootmark{a} & 12703 & 213.3654  &  43.9138 & 0.040 &  5.67  &  0.276 &  3.23   &   2.35 &  8.31\\
8335 & 12703 & 216.8768  &  40.9637 & 0.018 & 15.49  &  0.150 &  3.47   &   1.68 &  1.49\\
8552\tablefootmark{a} & 12702 & 227.9284  &  43.9704 & 0.028 &  7.74  &  0.193 &  1.11   &   2.48 &  47.8\\
7443 & 12704 & 232.4611  &  42.6290 & 0.019 & 21.51  &  0.245 &  8.67  &   2.57 & 2.48 \\
8551 & 12704 & 233.3194  &  45.6985 & 0.029 & 10.02  &  0.293 &  20.1  &   2.60 & 3.85 \\
8485 &  3704 & 235.5806  &  47.94353  &  0.037  &  5.09  &  0.20 & 2.53 & 2.34 & 7.39  \\
8481 & 12702 & 237.3048  &  55.08883  &  0.047  &  9.39  &  0.16 & 12.1 & 2.39 & 6.17 \\
8486 & 12704 & 238.2618  &  46.7680 & 0.020 & 10.45  &  0.299 &  1.19  &   2.41 &  3.29 \\
8603 &  6103 & 247.8004  &  40.42187  &  0.027  &  5.69  &  0.26 & 0.93 & 2.58 & 52.1  \\ 
8484 &  6101 & 248.0557  &  44.40330  &  0.031  &  10.23 &  0.24 & 15.4 & 2.57 & 0.99 \\
7991 &  6103 & 257.8338  &  56.9913 & 0.031 &  5.56  &  0.259 &  2.65   &   2.53 &  6.42\\
7962\tablefootmark{a} & 12705 & 259.0830  &  26.8502 & 0.048 &  6.18  &  0.237 & 9.59 & 2.31 & 11.6 \\
7990 &  9101 & 259.7555  &  57.17350  &  0.028  &  7.81  &  0.29 & 6.40 & 2.52 & 1.50  \\
7815 &  6101 & 316.5416  &  10.3454 & 0.017 &  6.41  &  0.262 &  0.54 &   2.42 &  11.1\\
8618 & 12701 & 317.9796  &  11.37945  &  0.018  &  18.80 &  0.21 & 2.36 & 2.41 & 1.27 \\
7975 &  3701 & 324.1525  &  10.5067 & 0.040 &  5.07  &  0.293 &  4.18   &   2.36 &  8.93\\
7977 & 12704 & 332.4183  &  13.6358 & 0.027 & 13.00  &  0.249 &  18.4  &   2.47 &  3.88\\
\hline
\noalign{\smallskip}
8552\tablefootmark{a} & 6102  & 227.1277  &  42.8667 & 0.040 &  3.48  &  0.338 & 3.69 & 2.54 & 8.35 \\
8483\tablefootmark{a} & 9102  & 248.3978  &  48.3798 & 0.039 &  4.74  &  0.391 & 2.70 & 2.10 & 7.30 \\
8606\tablefootmark{a} & 6101  & 254.4476  &  37.6877 & 0.042 &  4.14  &  0.393 & 5.60 & 2.28 & 5.92 \\
8618\tablefootmark{a} & 9102  & 319.2715  &   9.9723 & 0.043 &  5.27  &  0.326 & 9.52 & 2.15 & 7.77 \\
\noalign{\smallskip}
\hline
\end{tabular}
\tablefoot{Plate, IFU design number, Right Ascension (RA) and declination (dec) from J2000, redshift, major axis R$_e$ (arcseconds), b/a, stellar mass (M$_{star}$, $10^9$\;M$_\odot$), concentration index (C), and the specific star formation rate (sSFR, $10^{-10}$yr$^{-1}$) for each galaxy. The galaxies listed below the line were used in the large-$z$ sample but not in the full sample.}
\tablefoottext{a}{Galaxy was part of Large-$z$ stack}
\end{table*}

\begin{table*}
\caption{Minor axis bin properties\label{bin_info}}
\centering
\begin{tabular}{ccccccc}
\hline\hline
\noalign{\smallskip}
Sample & Minor axis bin ($b_e$) & Num of galaxies & Num of fibers & SB continuum\tablefootmark{a}
& S/N blue\tablefootmark{b} & S/N H$\alpha$\tablefootmark{c} \\
\noalign{\smallskip}
\hline
\noalign{\smallskip}
Full sample & 2.0-2.5 & 49 & 2981 & 22.7 & 40.31 & 72.35\\
Full sample & 2.5-3.0 & 49 & 2736 & 23.3 & 22.42 & 41.41\\
Full sample & 3.0-3.5 & 49 & 2505 & 23.9 & 13.61 & 27.34\\
Full sample & 3.5-4.0 & 49 & 2281 & 24.4 & 8.31 & 18.99\\
Low sSFR & 2.0-2.5 & 24 & 1845 & 22.9 & 27.68 & 47.73\\
Low sSFR & 2.5-3.0 & 24 & 1614 & 23.5 & 15.52 & 27.38\\
Low sSFR & 3.0-3.5 & 24 & 1339 & 24.0 & 9.90 & 18.03\\
Low sSFR & 3.5-4.0 & 24 & 1158 & 24.4 & 5.88 & 11.82\\
High sSFR & 2.0-2.5 & 24 & 1119 & 22.4 & 29.13 & 55.26\\
High sSFR & 2.5-3.0 & 24 & 1097 & 23.2 & 15.55 & 31.41\\
High sSFR & 3.0-3.5 & 24 & 1136 & 23.8 & 8.70 & 19.93\\
High sSFR & 3.5-4.0 & 24 & 1106 & 24.3 & 5.64 & 14.81\\
Low C & 2.0-2.5 & 24 & 1633 & 23.0 & 22.32 & 38.33\\
Low C & 2.5-3.0 & 24 & 1539 & 23.7 & 12.04 & 21.32\\
Low C & 3.0-3.5 & 24 & 1398 & 24.4 & 7.01 & 14.46\\
Low C & 3.5-4.0 & 24 & 1206 & 24.9 & 4.07 & 8.39\\
High C & 2.0-2.5 & 24 & 1325 & 22.5 & 35.23 & 63.15\\
High C & 2.5-3.0 & 24 & 1162 & 23.0 & 19.62 & 37.31\\
High C & 3.0-3.5 & 24 & 1085 & 23.5 & 12.08 & 24.47\\
High C & 3.5-4.0 & 24 & 1039 & 24.0 & 7.54 & 17.54\\
Low M$_{star}$ & 2.0-2.5 & 24 & 1317 & 23.2 & 23.66 & 43.59\\
Low M$_{star}$ & 2.5-3.0 & 24 & 1258 & 23.9 & 13.09 & 26.82\\
Low M$_{star}$ & 3.0-3.5 & 24 & 1267 & 24.4 & 7.70 & 17.31\\
Low M$_{star}$ & 3.5-4.0 & 24 & 1112 & 24.8 & 5.04 & 11.55\\
High M$_{star}$ & 2.0-2.5 & 24 & 1639 & 22.4 & 32.43 & 56.42\\
High M$_{star}$ & 2.5-3.0 & 24 & 1441 & 23.0 & 17.90 & 31.54\\
High M$_{star}$ & 3.0-3.5 & 24 & 1216 & 23.5 & 10.98 & 21.29\\
High M$_{star}$ & 3.5-4.0 & 24 & 1135 & 24.0 & 6.48 & 14.69\\
Large-$z$ & 4.0-4.5\tablefootmark{d} & 16 & 382 & 23.2 & 4.69 & 11.38\\
Large-$z$ & 4.5-5.0\tablefootmark{d} & 16 & 331 & 23.8 & 2.63 & 7.08\\
Large-$z$ & 5.0-6.0\tablefootmark{d} & 16 & 760 & 24.3 & 2.38 & 8.06\\
Large-$z$ & 6.0-7.0\tablefootmark{d} & 16 & 700 & 25.3 & 0.71 & 3.71\\
Large-$z$ & 7.0-9.0\tablefootmark{d} & 16 & 1351 & 26.0 & 0.26 & 3.75\\
Large-$z$ & 7.0-10.0\tablefootmark{d} & 16 & 1871 & 26.2 & 0.14 & 4.32\\

\noalign{\smallskip}
\hline
\end{tabular}
\tablefoot{Properties of the various outermost minor axis bins for the different samples.  Provided here are the number of galaxies included in each sample
as well as the number of fibers/spectra stacked, the S/N in the blue band, and the S/N for H$\alpha$ for each bin.}
\tablefoottext{a}{Estimate of the surface brightness of the continuum in mag$_{AB}$\;arcsec$^{-2}$ between 4000 and 5500 $\AA$}
\tablefoottext{b}{Median S/N between 4000 and 5500 $\AA$}
\tablefoottext{c}{Median S/N of H$\alpha$ region between 6558 and 6572 $\AA$}
\tablefoottext{d}{Units are in kpc and not $b_e$}
\end{table*}


\begin{acknowledgements}
Funding for the Sloan Digital Sky Survey IV has been provided by
the Alfred P. Sloan Foundation, the U.S. Department of Energy Office of
Science, and the Participating Institutions. SDSS-IV acknowledges
support and resources from the Center for High-Performance Computing at
the University of Utah. The SDSS web site is www.sdss.org.

SDSS-IV is managed by the Astrophysical Research Consortium for the 
Participating Institutions of the SDSS Collaboration including the 
Brazilian Participation Group, the Carnegie Institution for Science, 
Carnegie Mellon University, the Chilean Participation Group, the French Participation Group, Harvard-Smithsonian Center for Astrophysics, 
Instituto de Astrof\'isica de Canarias, The Johns Hopkins University, 
Kavli Institute for the Physics and Mathematics of the Universe (IPMU) / 
University of Tokyo, Lawrence Berkeley National Laboratory, 
Leibniz Institut f\"ur Astrophysik Potsdam (AIP),  
Max-Planck-Institut f\"ur Astronomie (MPIA Heidelberg), 
Max-Planck-Institut f\"ur Astrophysik (MPA Garching), 
Max-Planck-Institut f\"ur Extraterrestrische Physik (MPE), 
National Astronomical Observatory of China, New Mexico State University, 
New York University, University of Notre Dame, 
Observat\'ario Nacional / MCTI, The Ohio State University, 
Pennsylvania State University, Shanghai Astronomical Observatory, 
United Kingdom Participation Group,
Universidad Nacional Aut\'onoma de M\'exico, University of Arizona, 
University of Colorado Boulder, University of Oxford, University of Portsmouth, 
University of Utah, University of Virginia, University of Washington, University of Wisconsin, 
Vanderbilt University, and Yale University.

D.B. acknowledges support from grant RSF 14-50-00043.  MAB acknowledges NSF-AST/1517006.
\end{acknowledgements}

\bibliographystyle{aa}
\bibliography{edig_v5}

\clearpage
\begin{appendix}

\section{Inclination Effect}
  \begin{figure}[!ht]
   \centering    
\includegraphics[width=0.49\textwidth]{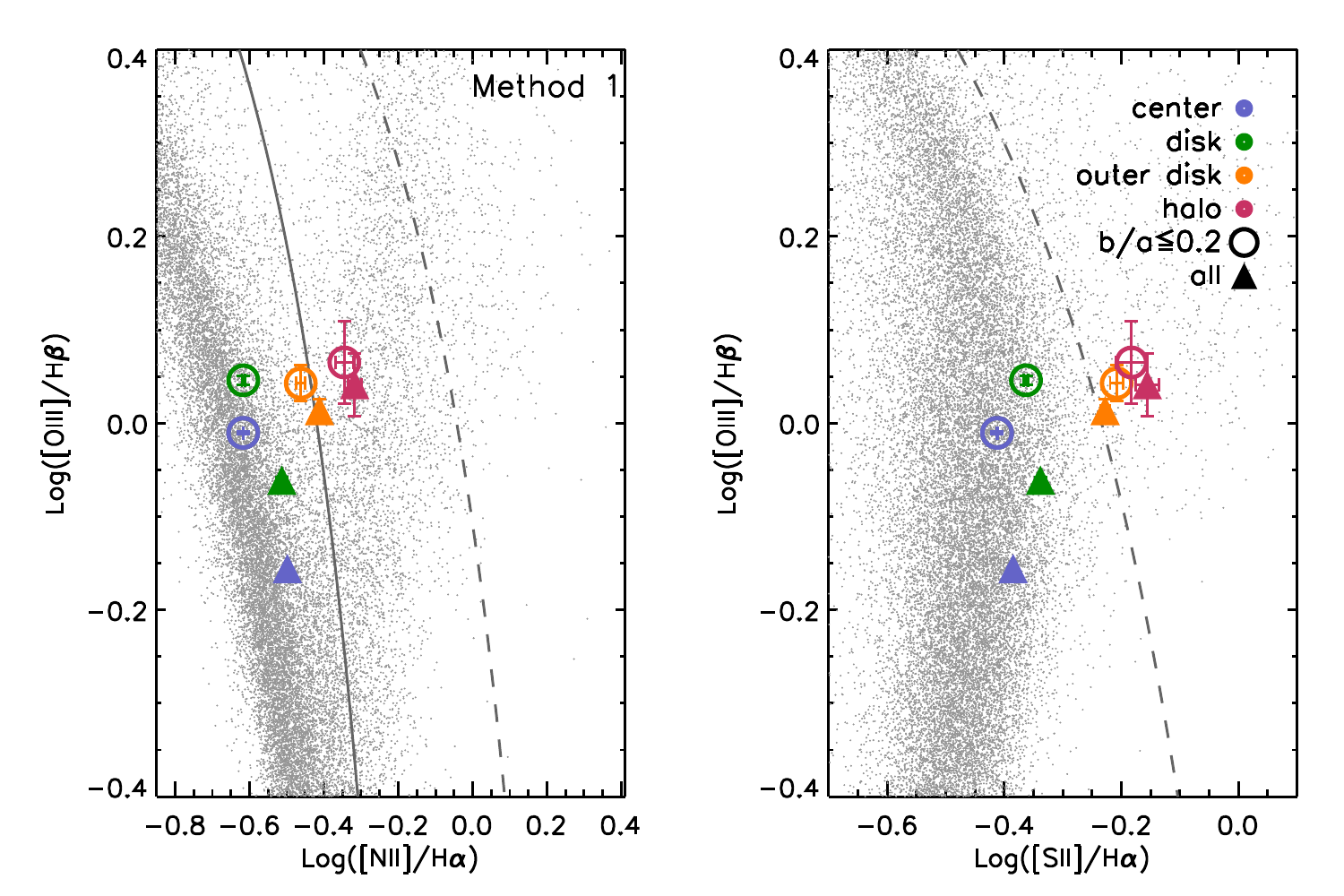}
\includegraphics[width=0.49\textwidth]{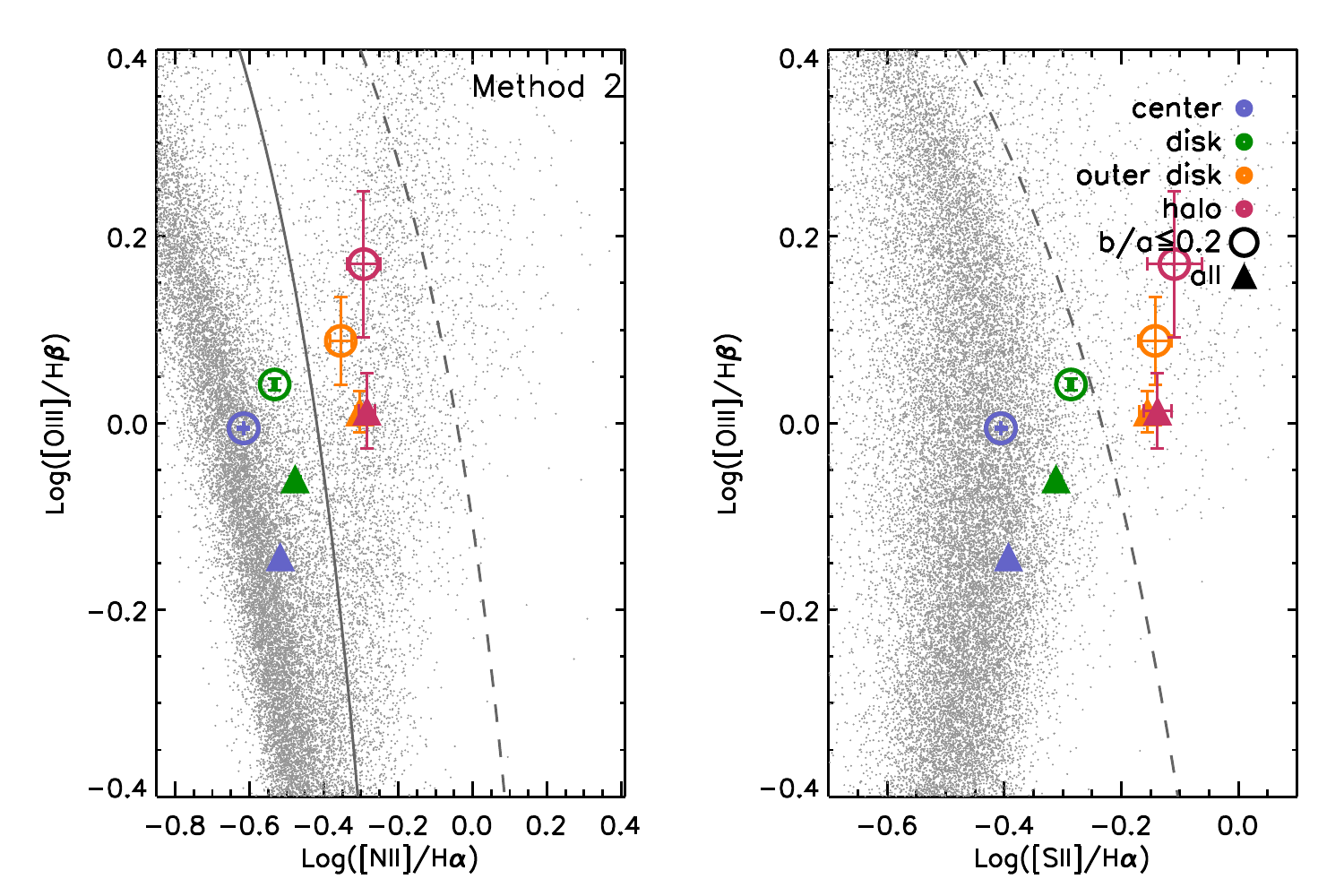}
\includegraphics[width=0.49\textwidth]{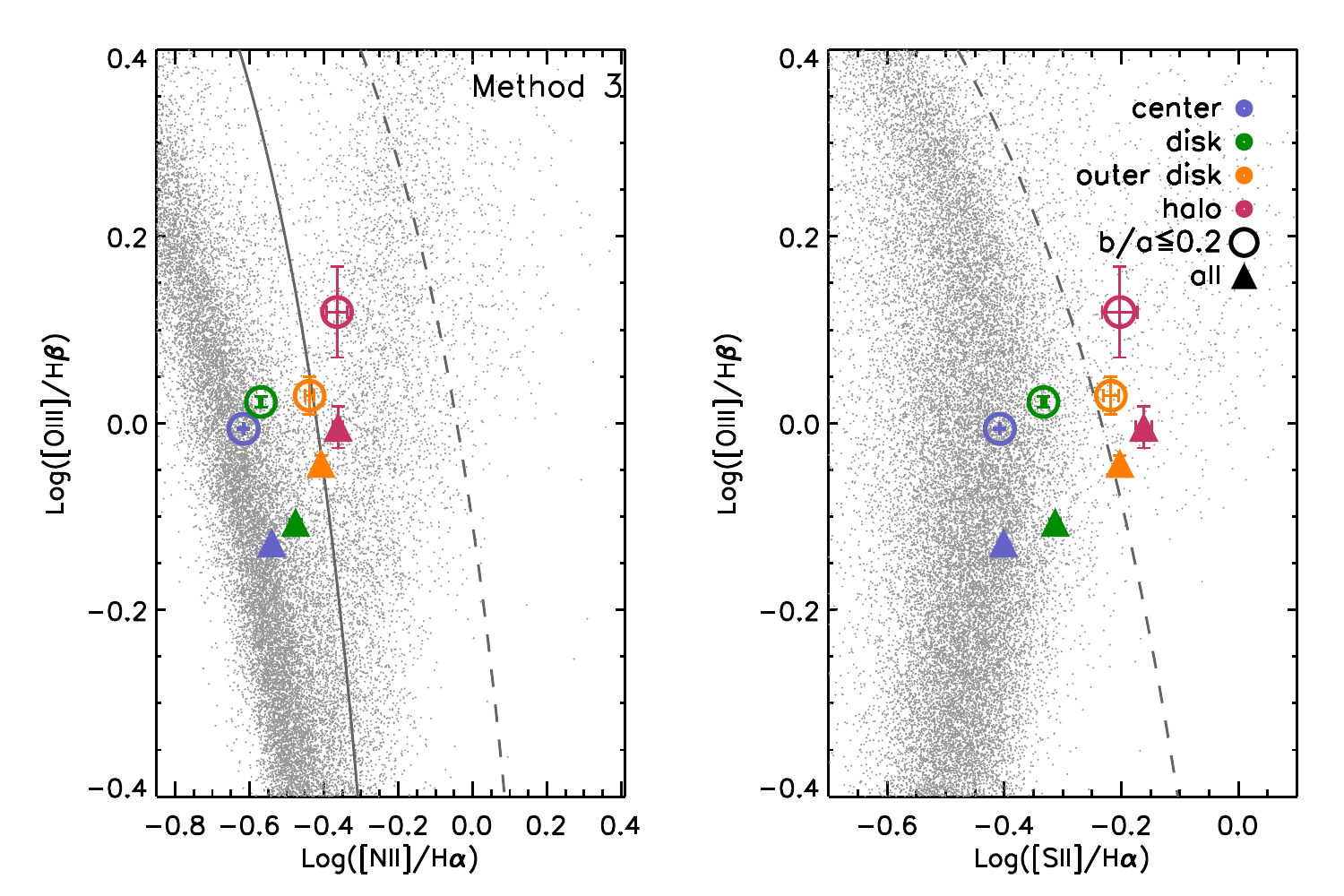}
   \caption{BPT diagrams comparing the ratios from the full sample (filled triangles) with galaxies in the sample that have a b/a$\leq$0.2 (open circles).  \textit{top} panel is with Method 1, \textit{middle} with Method 2, and the \textit{bottom} is with Method 3.}
\label{bpt_lba}
    \end{figure}
    
    The restriction on inclination for this analysis is that each galaxy must have a b/a<0.3.  Thus, many galaxies in our sample are not truly edge-on (see Fig \ref{mosaic} and Table \ref{gal_prop}).  When binning vertically above and below the galaxy midplane we are most likely stacking different areas from the each galaxy in the sample.  In other words for a highly inclined galaxy, 2 kpc above the midplane may already be in the halo, but for a less inclined galaxy would still be in the outer disk.  To see how this mixing affects our results, we have stacked galaxies with b/a$\leq0.2$ and compared them with the full sample (which has b/a<0.3).  We chose b/a=0.2 as the cutoff because this is typically considered to be the intrinsic thickness of a galaxy when converting between b/a and inclination \citep{1926ApJ....64..321H}.  Therefore, galaxies with b/a$leq0.2$ should be edge-on systems.  There are only 18 galaxies in this sample, which is why we can only do this comparison for with the full sample and not the split subsamples or large-$z$ sample where the S/N would be too low in the outskirts.  The full and near edge-on samples have similar median sSFR, C, and M$_{star}$, so the differences seen between then are from different cuts in b/a.  We have plotted this comparison on the BPT diagrams for the three stacking methods in Fig \ref{bpt_lba}.  We define center, disk, outer disk and halo for Method 1 as 0.0-1.0, 1.0-1.5, 2.0-2.5, and 3.0-3.5\;$b_e$, Method 2 as 0.0-0.2, 0.2-0.4, 0.4-0.6, and 0.8-1.0\;R$_e$, and for Method 3 as 0.0-1.0, 1.0-1.5, 2.0-2.5, and 3.0-3.5\;kpc, respectively.  Qualitatively the trend with distance from the midplane of the stacks' emission line ratios to increase with distance and move away from the star forming region of the BPT diagrams is similar between the two samples.  The main difference is an offset to lower \oiii/H$\beta$ for the full sample (filled triangles) compared to the near edge-on sample (open circles).  The amount of this offset changes for each of the stacking methods.  Method 1 in the outskirts has less of an offset compared to the other two methods, which is why for the full and split samples we show the BPT diagrams for Method 1.  For this analysis, since the qualitative trends are same, our conclusions on the properties of eDIG in MaNGA galaxies are robust to the presence of galaxies that are not truly edge-on. 

\section{Emission Line Surface Brightnesses}

We provide a table with the emission line surface brightness values for each bin and for the three stacking methods of the full sample and the split subsamples (Table \ref{em_values}).  The values are for the emission lines \oii$\lambda$3729, H$\beta$, \oiii$\lambda$5007, H$\alpha$, \nii$\lambda$6584, \sii$\lambda$6717, and \sii$\lambda$6731 in units of $10^{36}$\;erg\;s$^{-1}\;$kpc$^{-2}$.  The errors are from the spectral fitting. 

\onecolumn
\begin{longtab}
\begin{longtable}{ccccccccc}
\caption{Emission line surface brightnesses\label{em_values}} \\
\hline\hline
Sample & Minor axis bin & \oii$\lambda$3729 & H$\beta$ & \oiii$\lambda$5007 & H$\alpha$ & \nii$\lambda$6584 & \sii$\lambda$6717 & \sii$\lambda$6731 \\
\hline
\endfirsthead
\caption{continued.}\\
\hline\hline
Sample & Minor axis bin & \oii$\lambda$3729 & H$\beta$ & \oiii$\lambda$5007 & H$\alpha$ & \nii$\lambda$6584 & \sii$\lambda$6717 & \sii$\lambda$6731 \\
\hline
\endhead
\hline
\endfoot
\hline\hline
\noalign{\smallskip}
Full & 0.0-1.0$b_e$  & 192.6$\pm$0.7 & 175.3$\pm$0.3 & 124.0$\pm$0.3 & 637.5$\pm$0.4 & 202.6$\pm$0.3 & 163.2$\pm$0.3 & 102.0$\pm$0.3 \\
Full & 1.0-1.5$b_e$  & 116.5$\pm$0.8 & 82.2$\pm$0.4 & 71.5$\pm$0.4 & 276.9$\pm$0.4 & 84.9$\pm$0.4 & 75.6$\pm$0.4 & 51.5$\pm$0.4 \\
Full & 1.5-2.0$b_e$  & 64.7$\pm$0.8 & 38.1$\pm$0.4 & 37.4$\pm$0.4 & 108.9$\pm$0.4 & 40.0$\pm$0.3 & 36.0$\pm$0.4 & 23.8$\pm$0.4 \\
Full & 2.0-2.5$b_e$  & 34.2$\pm$0.8 & 17.4$\pm$0.4 & 18.0$\pm$0.4 & 49.7$\pm$0.3 & 19.3$\pm$0.3 & 17.4$\pm$0.4 & 12.1$\pm$0.4 \\
Full & 2.5-3.0$b_e$  & 21.9$\pm$0.8 & 10.5$\pm$0.4 & 9.6$\pm$0.4 & 27.4$\pm$0.4 & 11.8$\pm$0.3 & 11.3$\pm$0.4 & 7.7$\pm$0.4 \\
Full & 3.0-3.5$b_e$  & 15.7$\pm$0.8 & 6.8$\pm$0.4 & 7.5$\pm$0.4 & 19.0$\pm$0.4 & 9.2$\pm$0.3 & 7.7$\pm$0.4 & 5.5$\pm$0.4 \\
Full & 3.5-4.0$b_e$  & 12.6$\pm$0.9 & 6.1$\pm$0.4 & 5.0$\pm$0.4 & 15.2$\pm$0.4 & 7.3$\pm$0.4 & 6.5$\pm$0.4 & 4.1$\pm$0.4 \\
Full & 0.0-1.0kpc  & 220.2$\pm$0.7 & 193.1$\pm$0.4 & 145.7$\pm$0.4 & 718.2$\pm$0.4 & 207.3$\pm$0.3 & 177.9$\pm$0.4 & 110.3$\pm$0.3 \\
Full & 1.0-1.5kpc  & 126.9$\pm$0.9 & 95.5$\pm$0.5 & 74.8$\pm$0.4 & 314.2$\pm$0.5 & 105.0$\pm$0.4 & 94.5$\pm$0.4 & 58.4$\pm$0.4 \\
Full & 1.5-2.0kpc  & 82.3$\pm$0.8 & 54.4$\pm$0.4 & 45.9$\pm$0.4 & 173.8$\pm$0.4 & 60.3$\pm$0.4 & 55.1$\pm$0.4 & 34.1$\pm$0.4 \\
Full & 2.0-2.5kpc  & 53.3$\pm$0.8 & 32.4$\pm$0.4 & 29.4$\pm$0.4 & 89.0$\pm$0.4 & 34.8$\pm$0.3 & 33.8$\pm$0.4 & 22.1$\pm$0.4 \\
Full & 2.5-3.0kpc  & 32.4$\pm$0.8 & 17.7$\pm$0.4 & 16.2$\pm$0.4 & 50.9$\pm$0.4 & 19.9$\pm$0.3 & 19.2$\pm$0.4 & 11.7$\pm$0.4 \\
Full & 3.0-3.5kpc  & 21.5$\pm$0.8 & 10.7$\pm$0.4 & 10.6$\pm$0.4 & 29.0$\pm$0.4 & 12.6$\pm$0.3 & 11.6$\pm$0.4 & 8.3$\pm$0.4 \\
Full & 3.5-4.0kpc  & 15.5$\pm$0.9 & 7.1$\pm$0.4 & 7.0$\pm$0.4 & 18.0$\pm$0.4 & 8.4$\pm$0.4 & 8.0$\pm$0.4 & 4.8$\pm$0.4 \\
Full & 0.0-0.2R$_e$  & 194.3$\pm$0.7 & 174.8$\pm$0.3 & 127.4$\pm$0.3 & 648.2$\pm$0.4 & 197.2$\pm$0.3 & 164.1$\pm$0.3 & 100.9$\pm$0.3 \\
Full & 0.2-0.4R$_e$  & 94.1$\pm$0.6 & 64.1$\pm$0.3 & 55.9$\pm$0.3 & 203.1$\pm$0.3 & 67.6$\pm$0.2 & 59.4$\pm$0.3 & 39.7$\pm$0.3 \\
Full & 0.4-0.6R$_e$  & 34.2$\pm$0.6 & 17.7$\pm$0.3 & 17.1$\pm$0.3 & 46.6$\pm$0.3 & 20.7$\pm$0.2 & 17.6$\pm$0.3 & 12.2$\pm$0.3 \\
Full & 0.6-0.8R$_e$  & 15.8$\pm$0.6 & 7.6$\pm$0.3 & 7.8$\pm$0.3 & 19.1$\pm$0.3 & 9.5$\pm$0.2 & 7.6$\pm$0.3 & 5.7$\pm$0.3 \\
Full & 0.8-1.0R$_e$  & 11.0$\pm$0.7 & 4.8$\pm$0.3 & 4.9$\pm$0.3 & 12.6$\pm$0.3 & 6.5$\pm$0.3 & 5.2$\pm$0.3 & 3.9$\pm$0.3 \\
Low sSFR & 0.0-1.0$b_e$  & 126.5$\pm$0.8 & 121.9$\pm$0.4 & 72.9$\pm$0.4 & 463.7$\pm$0.4 & 172.9$\pm$0.4 & 120.0$\pm$0.4 & 80.8$\pm$0.4 \\
Low sSFR & 1.0-1.5$b_e$  & 72.3$\pm$1.1 & 51.1$\pm$0.5 & 42.3$\pm$0.5 & 176.4$\pm$0.5 & 67.7$\pm$0.5 & 50.4$\pm$0.5 & 39.8$\pm$0.5 \\
Low sSFR & 1.5-2.0$b_e$  & 42.5$\pm$1.1 & 24.2$\pm$0.5 & 23.1$\pm$0.5 & 69.7$\pm$0.5 & 34.3$\pm$0.4 & 24.2$\pm$0.5 & 20.1$\pm$0.5 \\
Low sSFR & 2.0-2.5$b_e$  & 23.4$\pm$1.1 & 11.6$\pm$0.5 & 13.0$\pm$0.5 & 32.8$\pm$0.5 & 18.1$\pm$0.4 & 11.1$\pm$0.5 & 10.3$\pm$0.5 \\
Low sSFR & 2.5-3.0$b_e$  & 16.1$\pm$1.1 & 6.8$\pm$0.5 & 6.9$\pm$0.5 & 18.5$\pm$0.5 & 10.5$\pm$0.4 & 7.5$\pm$0.5 & 6.6$\pm$0.5 \\
Low sSFR & 3.0-3.5$b_e$  & 11.0$\pm$1.2 & 4.1$\pm$0.6 & 6.2$\pm$0.5 & 14.0$\pm$0.5 & 9.2$\pm$0.5 & 5.1$\pm$0.5 & 4.7$\pm$0.5 \\
Low sSFR & 3.5-4.0$b_e$  & 8.1$\pm$1.3 & 4.0$\pm$0.6 & 3.4$\pm$0.5 & 10.7$\pm$0.5 & 6.9$\pm$0.5 & 4.5$\pm$0.6 & 4.6$\pm$0.6 \\
Low sSFR & 0.0-1.0kpc  & 138.3$\pm$0.9 & 129.7$\pm$0.5 & 81.1$\pm$0.5 & 502.8$\pm$0.5 & 174.0$\pm$0.4 & 132.8$\pm$0.5 & 84.4$\pm$0.5 \\
Low sSFR & 1.0-1.5kpc  & 88.8$\pm$1.2 & 68.2$\pm$0.6 & 45.8$\pm$0.6 & 243.7$\pm$0.6 & 91.7$\pm$0.5 & 72.1$\pm$0.6 & 49.2$\pm$0.6 \\
Low sSFR & 1.5-2.0kpc  & 64.5$\pm$1.2 & 44.5$\pm$0.6 & 32.6$\pm$0.5 & 150.6$\pm$0.6 & 60.6$\pm$0.5 & 48.1$\pm$0.6 & 33.4$\pm$0.5 \\
Low sSFR & 2.0-2.5kpc  & 46.1$\pm$1.2 & 28.8$\pm$0.6 & 23.3$\pm$0.5 & 87.9$\pm$0.5 & 36.9$\pm$0.5 & 31.4$\pm$0.5 & 23.3$\pm$0.5 \\
Low sSFR & 2.5-3.0kpc  & 28.9$\pm$1.1 & 16.2$\pm$0.5 & 13.6$\pm$0.5 & 49.1$\pm$0.5 & 20.2$\pm$0.5 & 18.2$\pm$0.5 & 12.2$\pm$0.5 \\
Low sSFR & 3.0-3.5kpc  & 20.3$\pm$1.2 & 10.0$\pm$0.6 & 9.3$\pm$0.5 & 29.1$\pm$0.5 & 14.8$\pm$0.5 & 11.4$\pm$0.6 & 9.6$\pm$0.5 \\
Low sSFR & 3.5-4.0kpc  & 15.9$\pm$1.2 & 5.8$\pm$0.6 & 6.5$\pm$0.5 & 17.3$\pm$0.5 & 9.1$\pm$0.5 & 6.7$\pm$0.5 & 5.2$\pm$0.5 \\
Low sSFR & 0.0-0.2R$_e$  & 125.0$\pm$0.8 & 120.2$\pm$0.4 & 71.5$\pm$0.4 & 465.3$\pm$0.5 & 169.5$\pm$0.4 & 121.6$\pm$0.4 & 80.6$\pm$0.4 \\
Low sSFR & 0.2-0.4R$_e$  & 64.3$\pm$0.8 & 43.0$\pm$0.4 & 34.8$\pm$0.4 & 141.3$\pm$0.4 & 58.4$\pm$0.3 & 43.9$\pm$0.4 & 32.0$\pm$0.4 \\
Low sSFR & 0.4-0.6R$_e$  & 25.5$\pm$0.8 & 12.4$\pm$0.4 & 13.0$\pm$0.3 & 34.1$\pm$0.3 & 19.1$\pm$0.3 & 13.3$\pm$0.4 & 10.3$\pm$0.4 \\
Low sSFR & 0.6-0.8R$_e$  & 11.6$\pm$0.8 & 5.4$\pm$0.4 & 6.4$\pm$0.4 & 14.4$\pm$0.4 & 9.1$\pm$0.3 & 5.4$\pm$0.4 & 4.8$\pm$0.4 \\
Low sSFR & 0.8-1.0R$_e$  & 7.9$\pm$1.0 & 2.7$\pm$0.5 & 3.3$\pm$0.4 & 9.3$\pm$0.4 & 6.4$\pm$0.4 & 3.6$\pm$0.5 & 3.9$\pm$0.5 \\
High sSFR & 0.0-1.0$b_e$  & 348.7$\pm$1.2 & 288.7$\pm$0.6 & 240.9$\pm$0.6 & 1032.1$\pm$0.7 & 258.3$\pm$0.5 & 244.8$\pm$0.6 & 143.0$\pm$0.5 \\
High sSFR & 1.0-1.5$b_e$  & 210.3$\pm$1.4 & 145.9$\pm$0.7 & 139.2$\pm$0.7 & 504.5$\pm$0.8 & 119.1$\pm$0.6 & 124.7$\pm$0.7 & 77.1$\pm$0.6 \\
High sSFR & 1.5-2.0$b_e$  & 112.2$\pm$1.3 & 66.3$\pm$0.6 & 68.5$\pm$0.6 & 190.0$\pm$0.6 & 51.7$\pm$0.5 & 57.5$\pm$0.6 & 31.6$\pm$0.6 \\
High sSFR & 2.0-2.5$b_e$  & 54.2$\pm$1.2 & 27.7$\pm$0.6 & 28.2$\pm$0.6 & 85.6$\pm$0.6 & 22.0$\pm$0.5 & 28.1$\pm$0.6 & 15.9$\pm$0.6 \\
High sSFR & 2.5-3.0$b_e$  & 29.9$\pm$1.2 & 16.1$\pm$0.6 & 13.8$\pm$0.6 & 43.2$\pm$0.6 & 13.9$\pm$0.5 & 16.6$\pm$0.6 & 9.3$\pm$0.6 \\
High sSFR & 3.0-3.5$b_e$  & 20.2$\pm$1.2 & 9.4$\pm$0.6 & 8.9$\pm$0.5 & 24.9$\pm$0.5 & 8.5$\pm$0.5 & 10.1$\pm$0.6 & 6.4$\pm$0.6 \\
High sSFR & 3.5-4.0$b_e$  & 16.6$\pm$1.2 & 8.1$\pm$0.6 & 6.8$\pm$0.6 & 20.3$\pm$0.6 & 7.3$\pm$0.5 & 8.4$\pm$0.6 & 3.6$\pm$0.6 \\
High sSFR & 0.0-1.0kpc  & 348.4$\pm$1.2 & 280.8$\pm$0.6 & 242.3$\pm$0.6 & 1024.6$\pm$0.7 & 244.4$\pm$0.5 & 230.9$\pm$0.5 & 138.7$\pm$0.5 \\
High sSFR & 1.0-1.5kpc  & 185.1$\pm$1.4 & 130.2$\pm$0.7 & 115.9$\pm$0.7 & 409.2$\pm$0.7 & 117.3$\pm$0.6 & 119.6$\pm$0.7 & 67.0$\pm$0.6 \\
High sSFR & 1.5-2.0kpc  & 106.7$\pm$1.3 & 63.7$\pm$0.6 & 60.4$\pm$0.6 & 191.8$\pm$0.6 & 56.5$\pm$0.5 & 58.9$\pm$0.6 & 31.9$\pm$0.6 \\
High sSFR & 2.0-2.5kpc  & 61.0$\pm$1.2 & 34.4$\pm$0.6 & 34.5$\pm$0.6 & 87.8$\pm$0.6 & 29.7$\pm$0.5 & 33.3$\pm$0.6 & 18.5$\pm$0.6 \\
High sSFR & 2.5-3.0kpc  & 35.7$\pm$1.2 & 18.1$\pm$0.6 & 18.0$\pm$0.6 & 49.3$\pm$0.5 & 17.4$\pm$0.5 & 17.7$\pm$0.6 & 9.2$\pm$0.5 \\
High sSFR & 3.0-3.5kpc  & 22.7$\pm$1.2 & 11.5$\pm$0.6 & 11.5$\pm$0.5 & 28.3$\pm$0.5 & 10.7$\pm$0.5 & 10.6$\pm$0.6 & 6.3$\pm$0.5 \\
High sSFR & 3.5-4.0kpc  & 14.5$\pm$1.2 & 8.9$\pm$0.6 & 7.6$\pm$0.6 & 19.0$\pm$0.6 & 8.0$\pm$0.5 & 8.9$\pm$0.6 & 4.5$\pm$0.6 \\
High sSFR & 0.0-0.2R$_e$  & 343.4$\pm$1.2 & 281.5$\pm$0.6 & 244.6$\pm$0.6 & 1021.1$\pm$0.7 & 242.9$\pm$0.5 & 236.3$\pm$0.6 & 134.7$\pm$0.5 \\
High sSFR & 0.2-0.4R$_e$  & 157.1$\pm$1.0 & 104.5$\pm$0.5 & 101.7$\pm$0.5 & 330.1$\pm$0.5 & 86.0$\pm$0.4 & 90.8$\pm$0.5 & 55.5$\pm$0.4 \\
High sSFR & 0.4-0.6R$_e$  & 48.3$\pm$0.8 & 26.2$\pm$0.4 & 24.7$\pm$0.4 & 69.8$\pm$0.4 & 23.6$\pm$0.4 & 25.6$\pm$0.4 & 15.5$\pm$0.4 \\
High sSFR & 0.6-0.8R$_e$  & 20.9$\pm$0.9 & 10.4$\pm$0.4 & 9.9$\pm$0.4 & 25.8$\pm$0.4 & 9.8$\pm$0.4 & 10.5$\pm$0.4 & 6.8$\pm$0.4 \\
High sSFR & 0.8-1.0R$_e$  & 13.8$\pm$0.9 & 7.0$\pm$0.5 & 6.8$\pm$0.4 & 16.1$\pm$0.4 & 6.3$\pm$0.4 & 6.9$\pm$0.5 & 3.9$\pm$0.5 \\
Low C & 0.0-1.0$b_e$  & 210.8$\pm$0.9 & 168.4$\pm$0.4 & 141.5$\pm$0.4 & 558.7$\pm$0.5 & 164.6$\pm$0.4 & 149.3$\pm$0.4 & 90.1$\pm$0.4 \\
Low C & 1.0-1.5$b_e$  & 120.2$\pm$1.1 & 74.8$\pm$0.6 & 75.0$\pm$0.5 & 229.7$\pm$0.6 & 65.8$\pm$0.5 & 67.2$\pm$0.5 & 43.0$\pm$0.5 \\
Low C & 1.5-2.0$b_e$  & 68.7$\pm$1.1 & 38.1$\pm$0.5 & 41.8$\pm$0.5 & 102.0$\pm$0.5 & 32.5$\pm$0.4 & 34.7$\pm$0.5 & 22.7$\pm$0.5 \\
Low C & 2.0-2.5$b_e$  & 31.1$\pm$1.0 & 15.0$\pm$0.5 & 16.8$\pm$0.5 & 40.6$\pm$0.5 & 13.5$\pm$0.4 & 14.1$\pm$0.5 & 9.8$\pm$0.5 \\
Low C & 2.5-3.0$b_e$  & 19.2$\pm$1.1 & 8.8$\pm$0.5 & 7.8$\pm$0.5 & 20.9$\pm$0.5 & 7.5$\pm$0.4 & 9.0$\pm$0.5 & 5.8$\pm$0.5 \\
Low C & 3.0-3.5$b_e$  & 12.7$\pm$1.1 & 5.9$\pm$0.5 & 6.0$\pm$0.5 & 15.2$\pm$0.5 & 5.5$\pm$0.4 & 6.5$\pm$0.5 & 4.0$\pm$0.5 \\
Low C & 3.5-4.0$b_e$  & 12.6$\pm$1.2 & 4.9$\pm$0.6 & 3.7$\pm$0.5 & 12.3$\pm$0.5 & 4.8$\pm$0.5 & 6.1$\pm$0.5 & 2.6$\pm$0.5 \\
Low C & 0.0-1.0kpc  & 244.7$\pm$1.0 & 184.4$\pm$0.5 & 163.5$\pm$0.5 & 636.9$\pm$0.5 & 165.9$\pm$0.4 & 164.8$\pm$0.5 & 98.6$\pm$0.4 \\
Low C & 1.0-1.5kpc  & 118.9$\pm$1.2 & 77.1$\pm$0.6 & 70.5$\pm$0.6 & 221.9$\pm$0.6 & 73.4$\pm$0.5 & 78.9$\pm$0.6 & 43.2$\pm$0.5 \\
Low C & 1.5-2.0kpc  & 66.3$\pm$1.1 & 41.9$\pm$0.5 & 37.7$\pm$0.5 & 117.5$\pm$0.5 & 39.6$\pm$0.5 & 43.9$\pm$0.5 & 25.8$\pm$0.5 \\
Low C & 2.0-2.5kpc  & 42.0$\pm$1.1 & 25.5$\pm$0.5 & 23.8$\pm$0.5 & 57.7$\pm$0.5 & 22.4$\pm$0.4 & 25.1$\pm$0.5 & 15.6$\pm$0.5 \\
Low C & 2.5-3.0kpc  & 27.1$\pm$1.1 & 12.8$\pm$0.5 & 12.3$\pm$0.5 & 35.2$\pm$0.5 & 12.3$\pm$0.4 & 14.3$\pm$0.5 & 7.6$\pm$0.5 \\
Low C & 3.0-3.5kpc  & 18.1$\pm$1.1 & 8.6$\pm$0.5 & 7.5$\pm$0.5 & 19.4$\pm$0.5 & 7.8$\pm$0.4 & 8.7$\pm$0.5 & 5.7$\pm$0.5 \\
Low C & 3.5-4.0kpc  & 12.3$\pm$1.1 & 6.4$\pm$0.5 & 4.2$\pm$0.5 & 12.9$\pm$0.5 & 5.4$\pm$0.4 & 5.9$\pm$0.5 & 2.4$\pm$0.5 \\
Low C & 0.0-0.2R$_e$  & 210.1$\pm$0.9 & 165.3$\pm$0.4 & 142.7$\pm$0.4 & 568.6$\pm$0.5 & 156.3$\pm$0.4 & 148.3$\pm$0.4 & 86.1$\pm$0.4 \\
Low C & 0.2-0.4R$_e$  & 85.9$\pm$0.8 & 51.0$\pm$0.4 & 51.5$\pm$0.4 & 147.7$\pm$0.4 & 47.4$\pm$0.3 & 47.2$\pm$0.4 & 29.9$\pm$0.4 \\
Low C & 0.4-0.6R$_e$  & 29.4$\pm$0.8 & 14.7$\pm$0.4 & 13.9$\pm$0.3 & 34.7$\pm$0.3 & 14.3$\pm$0.3 & 13.6$\pm$0.4 & 9.6$\pm$0.4 \\
Low C & 0.6-0.8R$_e$  & 12.6$\pm$0.8 & 6.2$\pm$0.4 & 5.9$\pm$0.4 & 13.9$\pm$0.4 & 5.9$\pm$0.3 & 6.1$\pm$0.4 & 4.4$\pm$0.4 \\
Low C & 0.8-1.0R$_e$  & 10.3$\pm$0.9 & 4.4$\pm$0.5 & 3.9$\pm$0.4 & 9.4$\pm$0.4 & 4.2$\pm$0.4 & 5.0$\pm$0.5 & 2.7$\pm$0.4 \\
High C & 0.0-1.0$b_e$  & 170.7$\pm$1.0 & 178.5$\pm$0.5 & 97.1$\pm$0.5 & 733.6$\pm$0.6 & 255.4$\pm$0.5 & 170.4$\pm$0.5 & 111.6$\pm$0.5 \\
High C & 1.0-1.5$b_e$  & 112.6$\pm$1.3 & 88.2$\pm$0.6 & 65.0$\pm$0.6 & 328.3$\pm$0.6 & 109.9$\pm$0.5 & 81.4$\pm$0.6 & 59.9$\pm$0.6 \\
High C & 1.5-2.0$b_e$  & 60.3$\pm$1.2 & 36.5$\pm$0.6 & 31.7$\pm$0.5 & 111.4$\pm$0.6 & 49.8$\pm$0.5 & 35.5$\pm$0.6 & 23.7$\pm$0.5 \\
High C & 2.0-2.5$b_e$  & 37.7$\pm$1.1 & 20.5$\pm$0.6 & 19.4$\pm$0.5 & 61.8$\pm$0.5 & 26.9$\pm$0.5 & 20.7$\pm$0.5 & 14.6$\pm$0.5 \\
High C & 2.5-3.0$b_e$  & 25.2$\pm$1.2 & 12.5$\pm$0.6 & 12.2$\pm$0.6 & 35.8$\pm$0.6 & 18.3$\pm$0.5 & 13.7$\pm$0.6 & 9.7$\pm$0.6 \\
High C & 3.0-3.5$b_e$  & 18.6$\pm$1.2 & 7.8$\pm$0.6 & 9.5$\pm$0.6 & 23.6$\pm$0.6 & 14.5$\pm$0.5 & 8.9$\pm$0.6 & 7.4$\pm$0.6 \\
High C & 3.5-4.0$b_e$  & 12.2$\pm$1.3 & 7.6$\pm$0.6 & 6.6$\pm$0.6 & 18.8$\pm$0.6 & 10.4$\pm$0.5 & 6.8$\pm$0.6 & 5.9$\pm$0.6 \\
High C & 0.0-1.0kpc  & 189.5$\pm$1.1 & 201.1$\pm$0.6 & 119.1$\pm$0.6 & 833.8$\pm$0.7 & 271.4$\pm$0.5 & 187.4$\pm$0.6 & 121.2$\pm$0.6 \\
High C & 1.0-1.5kpc  & 135.3$\pm$1.4 & 115.9$\pm$0.7 & 77.5$\pm$0.7 & 436.7$\pm$0.8 & 151.9$\pm$0.7 & 108.8$\pm$0.7 & 74.5$\pm$0.7 \\
High C & 1.5-2.0kpc  & 101.5$\pm$1.3 & 69.6$\pm$0.7 & 56.2$\pm$0.6 & 244.1$\pm$0.7 & 90.9$\pm$0.6 & 68.2$\pm$0.6 & 44.2$\pm$0.6 \\
High C & 2.0-2.5kpc  & 66.5$\pm$1.3 & 40.6$\pm$0.6 & 36.2$\pm$0.6 & 129.8$\pm$0.6 & 51.3$\pm$0.6 & 44.1$\pm$0.6 & 29.8$\pm$0.6 \\
High C & 2.5-3.0kpc  & 39.0$\pm$1.2 & 23.1$\pm$0.6 & 21.1$\pm$0.6 & 69.2$\pm$0.6 & 30.1$\pm$0.5 & 24.6$\pm$0.6 & 16.7$\pm$0.6 \\
High C & 3.0-3.5kpc  & 25.4$\pm$1.3 & 13.6$\pm$0.6 & 14.5$\pm$0.6 & 41.0$\pm$0.6 & 19.5$\pm$0.5 & 14.9$\pm$0.6 & 11.5$\pm$0.6 \\
High C & 3.5-4.0kpc  & 19.5$\pm$1.4 & 8.1$\pm$0.7 & 10.4$\pm$0.6 & 24.4$\pm$0.6 & 13.1$\pm$0.6 & 10.7$\pm$0.7 & 8.0$\pm$0.6 \\
High C & 0.0-0.2R$_e$  & 173.0$\pm$1.1 & 184.1$\pm$0.6 & 100.9$\pm$0.5 & 759.7$\pm$0.6 & 260.0$\pm$0.5 & 175.0$\pm$0.5 & 115.7$\pm$0.5 \\
High C & 0.2-0.4R$_e$  & 104.5$\pm$0.9 & 78.4$\pm$0.5 & 60.5$\pm$0.4 & 278.1$\pm$0.5 & 96.7$\pm$0.4 & 72.5$\pm$0.4 & 50.8$\pm$0.4 \\
High C & 0.4-0.6R$_e$  & 40.6$\pm$0.9 & 21.5$\pm$0.4 & 21.4$\pm$0.4 & 64.6$\pm$0.4 & 29.5$\pm$0.4 & 22.1$\pm$0.4 & 14.8$\pm$0.4 \\
High C & 0.6-0.8R$_e$  & 20.0$\pm$0.9 & 9.6$\pm$0.4 & 10.7$\pm$0.4 & 27.6$\pm$0.4 & 14.8$\pm$0.4 & 9.5$\pm$0.4 & 7.3$\pm$0.4 \\
High C & 0.8-1.0R$_e$  & 11.7$\pm$1.0 & 4.9$\pm$0.5 & 6.0$\pm$0.4 & 15.5$\pm$0.4 & 9.2$\pm$0.4 & 5.5$\pm$0.5 & 5.1$\pm$0.5 \\
Low M$_{star}$ & 0.0-1.0$b_e$  & 276.3$\pm$1.0 & 190.4$\pm$0.5 & 192.3$\pm$0.5 & 665.2$\pm$0.6 & 139.7$\pm$0.4 & 149.3$\pm$0.5 & 83.2$\pm$0.4 \\
Low M$_{star}$ & 1.0-1.5$b_e$  & 156.1$\pm$1.3 & 90.6$\pm$0.6 & 98.9$\pm$0.6 & 305.7$\pm$0.6 & 55.6$\pm$0.5 & 71.8$\pm$0.6 & 41.9$\pm$0.5 \\
Low M$_{star}$ & 1.5-2.0$b_e$  & 72.1$\pm$1.2 & 38.7$\pm$0.6 & 42.0$\pm$0.5 & 114.2$\pm$0.5 & 23.8$\pm$0.5 & 31.5$\pm$0.5 & 17.6$\pm$0.5 \\
Low M$_{star}$ & 2.0-2.5$b_e$  & 38.0$\pm$1.1 & 18.0$\pm$0.5 & 19.0$\pm$0.5 & 55.0$\pm$0.5 & 11.0$\pm$0.4 & 15.6$\pm$0.5 & 10.0$\pm$0.5 \\
Low M$_{star}$ & 2.5-3.0$b_e$  & 22.0$\pm$1.1 & 11.4$\pm$0.5 & 8.9$\pm$0.5 & 28.6$\pm$0.5 & 5.8$\pm$0.5 & 9.6$\pm$0.5 & 5.4$\pm$0.5 \\
Low M$_{star}$ & 3.0-3.5$b_e$  & 15.5$\pm$1.2 & 7.0$\pm$0.5 & 7.0$\pm$0.5 & 19.5$\pm$0.5 & 4.8$\pm$0.5 & 6.6$\pm$0.5 & 4.7$\pm$0.5 \\
Low M$_{star}$ & 3.5-4.0$b_e$  & 13.6$\pm$1.2 & 6.0$\pm$0.6 & 4.2$\pm$0.5 & 15.8$\pm$0.5 & 3.7$\pm$0.5 & 6.1$\pm$0.6 & 3.1$\pm$0.5 \\
Low M$_{star}$ & 0.0-1.0kpc  & 293.6$\pm$1.0 & 198.5$\pm$0.5 & 202.4$\pm$0.5 & 706.3$\pm$0.6 & 141.9$\pm$0.4 & 151.9$\pm$0.4 & 90.5$\pm$0.4 \\
Low M$_{star}$ & 1.0-1.5kpc  & 121.1$\pm$1.2 & 69.4$\pm$0.6 & 70.4$\pm$0.6 & 225.2$\pm$0.6 & 46.3$\pm$0.5 & 59.3$\pm$0.5 & 32.9$\pm$0.5 \\
Low M$_{star}$ & 1.5-2.0kpc  & 59.9$\pm$1.1 & 31.6$\pm$0.5 & 29.4$\pm$0.5 & 97.5$\pm$0.5 & 22.4$\pm$0.4 & 27.4$\pm$0.5 & 16.4$\pm$0.5 \\
Low M$_{star}$ & 2.0-2.5kpc  & 32.7$\pm$1.1 & 15.4$\pm$0.5 & 14.3$\pm$0.5 & 41.5$\pm$0.5 & 10.6$\pm$0.4 & 13.6$\pm$0.5 & 8.8$\pm$0.5 \\
Low M$_{star}$ & 2.5-3.0kpc  & 20.3$\pm$1.1 & 9.0$\pm$0.5 & 8.4$\pm$0.5 & 25.6$\pm$0.5 & 6.1$\pm$0.4 & 8.2$\pm$0.5 & 4.6$\pm$0.5 \\
Low M$_{star}$ & 3.0-3.5kpc  & 15.1$\pm$1.1 & 6.0$\pm$0.5 & 6.0$\pm$0.5 & 16.0$\pm$0.5 & 4.6$\pm$0.4 & 5.3$\pm$0.5 & 3.8$\pm$0.5 \\
Low M$_{star}$ & 3.5-4.0kpc  & 8.0$\pm$1.1 & 4.1$\pm$0.5 & 3.5$\pm$0.5 & 10.9$\pm$0.5 & 2.9$\pm$0.4 & 3.1$\pm$0.5 & 1.4$\pm$0.5 \\
Low M$_{star}$ & 0.0-0.2R$_e$  & 270.6$\pm$1.0 & 184.3$\pm$0.5 & 190.4$\pm$0.5 & 653.0$\pm$0.5 & 130.9$\pm$0.4 & 145.0$\pm$0.4 & 80.1$\pm$0.4 \\
Low M$_{star}$ & 0.2-0.4R$_e$  & 96.9$\pm$0.8 & 53.4$\pm$0.4 & 59.2$\pm$0.4 & 170.5$\pm$0.4 & 32.4$\pm$0.3 & 42.7$\pm$0.4 & 25.2$\pm$0.3 \\
Low M$_{star}$ & 0.4-0.6R$_e$  & 27.9$\pm$0.8 & 13.0$\pm$0.4 & 11.9$\pm$0.4 & 36.3$\pm$0.3 & 7.4$\pm$0.3 & 11.4$\pm$0.4 & 6.9$\pm$0.3 \\
Low M$_{star}$ & 0.6-0.8R$_e$  & 14.1$\pm$0.8 & 6.9$\pm$0.4 & 5.9$\pm$0.4 & 16.5$\pm$0.4 & 4.0$\pm$0.3 & 5.8$\pm$0.4 & 3.8$\pm$0.4 \\
Low M$_{star}$ & 0.8-1.0R$_e$  & 11.4$\pm$1.1 & 6.2$\pm$0.5 & 4.4$\pm$0.5 & 14.1$\pm$0.5 & 3.6$\pm$0.4 & 5.1$\pm$0.5 & 3.4$\pm$0.5 \\
High M$_{star}$ & 0.0-1.0$b_e$  & 153.4$\pm$0.9 & 162.9$\pm$0.5 & 82.1$\pm$0.4 & 638.7$\pm$0.5 & 259.3$\pm$0.4 & 164.1$\pm$0.5 & 111.1$\pm$0.4 \\
High M$_{star}$ & 1.0-1.5$b_e$  & 95.6$\pm$1.2 & 77.3$\pm$0.6 & 57.2$\pm$0.5 & 266.8$\pm$0.6 & 109.0$\pm$0.5 & 76.5$\pm$0.5 & 58.0$\pm$0.5 \\
High M$_{star}$ & 1.5-2.0$b_e$  & 60.1$\pm$1.1 & 36.8$\pm$0.5 & 34.2$\pm$0.5 & 109.0$\pm$0.5 & 55.7$\pm$0.5 & 38.6$\pm$0.5 & 28.3$\pm$0.5 \\
High M$_{star}$ & 2.0-2.5$b_e$  & 31.3$\pm$1.1 & 16.4$\pm$0.5 & 17.2$\pm$0.5 & 47.2$\pm$0.5 & 27.5$\pm$0.5 & 18.4$\pm$0.5 & 13.8$\pm$0.5 \\
High M$_{star}$ & 2.5-3.0$b_e$  & 21.9$\pm$1.1 & 9.3$\pm$0.6 & 10.1$\pm$0.5 & 26.7$\pm$0.5 & 18.1$\pm$0.5 & 12.5$\pm$0.5 & 9.5$\pm$0.5 \\
High M$_{star}$ & 3.0-3.5$b_e$  & 15.6$\pm$1.2 & 6.5$\pm$0.6 & 8.1$\pm$0.5 & 18.9$\pm$0.5 & 14.4$\pm$0.5 & 8.7$\pm$0.6 & 6.4$\pm$0.6 \\
High M$_{star}$ & 3.5-4.0$b_e$  & 11.0$\pm$1.3 & 5.8$\pm$0.6 & 6.0$\pm$0.6 & 14.5$\pm$0.6 & 10.8$\pm$0.5 & 6.7$\pm$0.6 & 4.9$\pm$0.6 \\
High M$_{star}$ & 0.0-1.0kpc  & 161.1$\pm$1.1 & 182.9$\pm$0.6 & 84.2$\pm$0.5 & 756.0$\pm$0.7 & 299.6$\pm$0.6 & 193.8$\pm$0.6 & 125.4$\pm$0.6 \\
High M$_{star}$ & 1.0-1.5kpc  & 131.4$\pm$1.4 & 120.3$\pm$0.8 & 71.5$\pm$0.7 & 432.9$\pm$0.8 & 180.0$\pm$0.7 & 127.4$\pm$0.7 & 83.4$\pm$0.7 \\
High M$_{star}$ & 1.5-2.0kpc  & 106.8$\pm$1.3 & 80.4$\pm$0.7 & 58.6$\pm$0.7 & 281.5$\pm$0.7 & 115.6$\pm$0.6 & 86.8$\pm$0.7 & 56.9$\pm$0.7 \\
High M$_{star}$ & 2.0-2.5kpc  & 77.2$\pm$1.3 & 52.1$\pm$0.7 & 42.6$\pm$0.6 & 165.0$\pm$0.6 & 69.9$\pm$0.6 & 56.1$\pm$0.6 & 39.7$\pm$0.6 \\
High M$_{star}$ & 2.5-3.0kpc  & 48.2$\pm$1.2 & 29.5$\pm$0.6 & 26.0$\pm$0.6 & 87.6$\pm$0.6 & 40.0$\pm$0.5 & 32.7$\pm$0.6 & 21.8$\pm$0.6 \\
High M$_{star}$ & 3.0-3.5kpc  & 30.6$\pm$1.3 & 17.9$\pm$0.6 & 17.8$\pm$0.6 & 51.9$\pm$0.6 & 26.6$\pm$0.5 & 19.9$\pm$0.6 & 15.3$\pm$0.6 \\
High M$_{star}$ & 3.5-4.0kpc  & 23.6$\pm$1.3 & 11.0$\pm$0.6 & 12.1$\pm$0.6 & 30.4$\pm$0.6 & 18.0$\pm$0.5 & 14.1$\pm$0.6 & 10.1$\pm$0.6 \\
High M$_{star}$ & 0.0-0.2R$_e$  & 153.9$\pm$0.9 & 163.8$\pm$0.5 & 80.1$\pm$0.5 & 658.0$\pm$0.6 & 264.7$\pm$0.5 & 169.1$\pm$0.5 & 112.9$\pm$0.5 \\
High M$_{star}$ & 0.2-0.4R$_e$  & 92.3$\pm$0.9 & 71.2$\pm$0.4 & 52.4$\pm$0.4 & 235.1$\pm$0.4 & 101.6$\pm$0.4 & 71.9$\pm$0.4 & 51.3$\pm$0.4 \\
High M$_{star}$ & 0.4-0.6R$_e$  & 38.7$\pm$0.8 & 21.0$\pm$0.4 & 21.7$\pm$0.4 & 56.7$\pm$0.4 & 34.1$\pm$0.4 & 22.6$\pm$0.4 & 16.9$\pm$0.4 \\
High M$_{star}$ & 0.6-0.8R$_e$  & 17.1$\pm$0.8 & 7.9$\pm$0.4 & 9.6$\pm$0.4 & 21.8$\pm$0.4 & 15.5$\pm$0.4 & 9.3$\pm$0.4 & 7.4$\pm$0.4 \\
High M$_{star}$ & 0.8-1.0R$_e$  & 10.6$\pm$0.9 & 3.8$\pm$0.5 & 5.3$\pm$0.4 & 11.7$\pm$0.4 & 9.0$\pm$0.4 & 5.2$\pm$0.4 & 4.2$\pm$0.4 \\

\hline
\end{longtable}
\tablefoot{All of the emission line surface brightnesses are in $10^{36}$\;erg\;s$^{-1}\;$kpc$^{-2}$ and the errors are from the spectral fitting analysis.}
\end{longtab}

\twocolumn

\end{appendix}
\end{document}